\documentclass[twocolumn]{aastex631}

\usepackage{graphicx}	
\usepackage{rotating}
\usepackage{amsmath}	
\usepackage{amssymb}	
\usepackage{xspace}	
\usepackage{color}
\usepackage{tablefootnote}
\usepackage{fleqn,natbib}
\usepackage{threeparttable}
\usepackage{xcolor}
\usepackage{dcolumn}
\usepackage{float}
\usepackage{multirow,longtable,supertabular}
\usepackage{hyperref}

\newcommand{\AstroSat}{{\em AstroSat}\xspace}
\newcommand{\fermi}{{\em Fermi}\xspace}
\newcommand{\kw}{{\em Konus}-Wind\xspace}

\newcommand{\keV}{{\rm keV}\xspace}

\newcommand{\s}{{\rm ~s}\xspace}
\newcommand{\swift}{{\em Swift}\xspace}
\newcommand{\tninty}{{$T_{\rm 90}$}\xspace}

\newcommand{\Ep}{$E_{\rm p}$\xspace}
\newcommand{\sw}[1]{\texttt{#1}}

\received{May 5, 2024}
\revised{June 14, 2024}
\accepted{\today}

\graphicspath{{./}{figures/}}

\begin{document}

\title{A detailed time-resolved and energy-resolved spectro-polarimetric study of bright GRBs detected by \AstroSat CZTI in its first year of operation}

\correspondingauthor{Rahul Gupta}
\email{rahulbhu.c157@gmail.com, rahul.gupta@nasa.gov, shashi@aries.res.in}

\author[0000-0003-4905-7801]{Rahul Gupta}
\affiliation{Astrophysics Science Division, NASA Goddard Space Flight Center, Mail Code 661, Greenbelt, MD 20771, USA}
\affiliation{NASA Postdoctoral Program Fellow}
\affiliation{Aryabhatta Research Institute of Observational Sciences (ARIES), Manora Peak, Nainital-263002, India}
\author{S. B. Pandey}
\affiliation{Aryabhatta Research Institute of Observational Sciences (ARIES), Manora Peak, Nainital-263002, India}
\author[0000-0001-6621-259X]{S. Gupta}
\affiliation{Bhabha Atomic Research Center, Mumbai, Maharashtra-400094, India}
\affiliation{Homi Bhabha National Institute, Mumbai, Maharashtra-400094, India}
\author[0000-0001-9856-1866]{T. Chattopadhayay}
\affiliation{Kavli Institute of Particle Astrophysics and Cosmology, Stanford University, 452 Lomita Mall, Stanford, CA 94305, USA}
\author[0000-0003-3352-3142]{D. Bhattacharya}
\affiliation{Department of Physics, Ashoka University, Sonipat, Haryana-131029, India}
\author[0000-0002-6112-7609]{V. Bhalerao}
\affiliation{Indian Institute of Technology Bombay Powai, Mumbai, Maharashtra 400076, India}
\author[0000-0003-2999-3563]{A. J. Castro-Tirado}
\affiliation{Instituto de Astrof\'isica de Andaluc\'ia (IAA-CSIC), Glorieta de la Astronom\'ia s/n, E-18008, Granada, Spain}
\affiliation{Departamento de Ingenier\'ia de Sistemas y Autom\'atica, Escuela de Ingenier\'ias, Universidad de M\'alaga, C\/. Dr. Ortiz Ramos s\/n, E-29071, M\'alaga, Spain}
\author{A. Valeev}
\affiliation{Special Astrophysical Observatory of Russian Academy of Sciences, Nizhniy Arkhyz, Russia}
\author[0000-0003-3164-8056]{A. K. Ror}
\affiliation{Aryabhatta Research Institute of Observational Sciences (ARIES), Manora Peak, Nainital-263002, India}
\author[0000-0002-4394-4138]{V. Sharma}
\affiliation{Astrophysics Science Division, NASA Goddard Space Flight Center, Mail Code 661, Greenbelt, MD 20771, USA}
\affiliation{University of Maryland, Baltimore County, Baltimore, MD 21250, USA}
\affiliation{Center for Research and Exploration in Space Science and Technology, NASA/GSFC, Greenbelt, MD 20771, USA}
\author[0000-0002-4744-9898]{J. Racusin}
\affiliation{Astrophysics Science Division, NASA Goddard Space Flight Center, Mail Code 661, Greenbelt, MD 20771, USA}
\author[0000-0002-9928-0369]{A. Aryan}
\affiliation{Aryabhatta Research Institute of Observational Sciences (ARIES), Manora Peak, Nainital-263002, India}
\affiliation{Graduate Institute of Astronomy, National Central University, 300 Jhongda Road, 32001 Jhongli, Taiwan}
\author[0000-0003-3220-7543]{S. Iyyani}
\affiliation{Indian Institute of Science Education and Research, Thiruvananthapuram, Kerala, India, 695551}
\author[0000-0002-2050-0913]{S. Vadawale}
\affiliation{Physical Research Laboratory Thaltej, Ahmedabad, Gujarat 380009, India}

\begin{abstract}
The radiation mechanism underlying the prompt emission remains unresolved and can be resolved using a systematic and uniform time-resolved spectro-polarimetric study. In this paper, we investigated the spectral, temporal, and polarimetric characteristics of five bright GRBs using archival data from \AstroSat CZTI, \swift BAT, and \fermi GBM. These bright GRBs were detected by CZTI in its first year of operation, and their average polarization characteristics have been published in \cite{2022ApJ...936...12C}. In the present work, we examined the time-resolved (in 100-600 \keV) and energy-resolved polarization measurements of these GRBs with an improved polarimetric technique such as increasing the effective area and bandwidth (by using data from low-gain pixels), using an improved event selection logic to reduce noise in the double events and extend the spectral bandwidth. In addition, we also separately carried out detailed time-resolved spectral analyses of these GRBs using empirical and physical synchrotron models. By these improved time-resolved and energy-resolved spectral and polarimetric studies (not fully coupled spectro-polarimetric fitting), we could pin down the elusive prompt emission mechanism of these GRBs. Our spectro-polarimetric analysis reveals that GRB 160623A, GRB 160703A, and GRB 160821A have Poynting flux-dominated jets. On the other hand, GRB 160325A and GRB 160802A have baryonic-dominated jets with mild magnetization. Furthermore, we observe a rapid change in polarization angle by $\sim$ 90 degrees within the main pulse of very bright GRB 160821A, consistent with our previous results. Our study suggests that the jet composition of GRBs may exhibit a wide range of magnetization, which can be revealed by utilizing spectro-polarimetric investigations of the bright GRBs.    

\end{abstract}

\section{Introduction}

Gamma-ray bursts (GRBs) are among the most energetic and enigmatic phenomena in the Universe. They emit an immense amount of energy in the form of high-energy photons occurring during cataclysmic events such as the collapse of massive stars or the merging of compact objects \citep{2004RvMP...76.1143P, 2015PhR...561....1K}. The exact radiation mechanism driving the prompt emission remains elusive \citep{2004ApJ...613..460B, 2011CRPhy..12..206Z, 2022Galax..10...38B}. Synchrotron emission, typically associated with the radiation emitted as relativistic electrons accelerated in magnetic fields, is commonly believed to underlie the spectral shape of the prompt emission \citep{Uhm:2014, 2019A&A...628A..59O, Tavani:1996, 2020NatAs...4..210Z}. The low energy spectral slope ($\alpha_{\rm pt}$) acts as an indicator tool for understanding the potential radiation physics of GRBs. In scenarios involving fast cooling synchrotron emission, where relativistic electrons rapidly emit all their energy upon acceleration, the theoretically predicted value of $\alpha_{\rm pt}$ is -3/2 \citep{2000ApJ...534L.163G}. However, upon examining the distribution of $\alpha_{\rm pt}$ for numerous GRBs observed with various telescopes such as {\it CGRO}/BATSE and \fermi/GBM, it becomes clear that a substantial number of bursts do not align with the expected characteristics of synchrotron emission \citep{Preece:1998}. This inconsistency suggests the involvement of alternative mechanisms in generating some or all of the emissions. For instance, physical models of photospheric emission have been observed to directly fit the observational data \citep{2017IJMPD..2630018P, 2017SSRv..207...87B, Zeynep:2020arXiv, 2012ApJ...755L...6F}. 
Moreover, thermal photospheric spectra need not strictly adhere to a \sw{Blackbody} distribution; if dissipation takes place just beneath the photosphere, this process could widen the spectrum compared to the standard \sw{Blackbody} spectrum \citep{Beloborodov2017, Ahlgren:2019ApJ, 2005ApJ...628..847R, 2011MNRAS.415.3693R}. Additionally, non-dissipative broadening of the photospheric emission can occur due to high latitude emission, often referred to as the multi-color \sw{Blackbody} effect \citep{2013MNRAS.428.2430L, 2015AdAst2015E..22P, 2019MNRAS.487.5508A}. This effect arises because different parts of the photosphere can have different temperatures, leading to a spectrum that is broader than a single \sw{Blackbody}.

In recent years, significant strides have been made in the study of radiation physics through broadband spectroscopy of prompt emissions. \cite{2017ApJ...846..137O, 2018A&A...616A.138O} performed a joint spectral analysis on a sample of 34 bright bursts observed concurrently by the \swift Burst Alert Telescope (BAT) and the X-ray Telescope (XRT), focusing on prompt gamma-ray emissions. This analysis identified a distinct lower frequency energy break in addition to the typical peak energy break. Notably, the values for $\alpha_{1}$ (photon index below the low energy break) and $\alpha_{2}$ (photon index above the low energy break) aligned with synchrotron theory predictions. This spectral behavior was similarly noted in bright long GRBs observed by \fermi, as discussed by \cite{2018A&A...613A..16R, 2019A&A...625A..60R}, though it was not present in bright short GRBs from \fermi. We also noted comparable spectral characteristics in one of the brightest long-duration GRBs detected by \fermi (GRB 190530A, \citealt{2022MNRAS.511.1694G, 2023arXiv231216265G}). Furthermore, \cite{2019A&A...628A..59O} expanded the analysis to include the optical band and concluded that the synchrotron spectral shape fits well across the spectrum from gamma-ray to optical bands, using a synchrotron physical model. However, it is important to note that the resulting parameters of the spectral fits, such as the bulk Lorentz factor, number density of electrons, and magnetic field strength, showed inconsistencies compared to other GRB prompt emissions analyses. These discrepancies highlight the complexities involved in modeling the emission region of the jet and suggest the need for further investigation to reconcile these differences. These findings further highlight the potential of simultaneous multi-band observations of prompt emissions, from optical to GeV energies, to deepen our understanding of emission mechanisms. However, capturing such simultaneous observations remains a significant challenge due to the extremely short and variable nature of prompt emissions, often concluding before there is time to redirect optical/X-ray instruments to the burst location \citep{2023arXiv231216265G}.

Currently, a major challenge in the spectral analysis of GRBs is the degeneracy among various spectral models. Often, the same dataset can be effectively fitted with different spectral models, all yielding comparably good statistical results \citep{2016MNRAS.456.2157I}. The spectroscopic study of prompt gamma-ray emission of GRBs provides valuable information, yet it alone is inadequate to fully discriminate between various emission models. Consequently, there is a critical need for more constraining observables, such as polarization, for example. \citep{2022JApA...43...37I, Toma:2013arXiv, 2020MNRAS.491.3343G}. 

Polarization measurements offer a reliable means of distinguishing between various potential radiation models of GRBs. This is because different models for prompt emission radiation predict distinct polarization fractions depending on the geometry of the jet. Typically, any asymmetry in the emitting region or viewing geometry results in linearly polarized emission. Synchrotron radiation originating from structured magnetic fields and observed along the jet axis is expected to exhibit a high degree of polarization. Conversely, inverse Compton and photospheric emission typically yield low polarization fractions, except when the jet is observed off-axis \citep{2009ApJ...698.1042T}. Therefore, by conducting polarization measurements for numerous bursts, we can gain tangible insights into the emission mechanisms of GRBs. Therefore, combining polarization measurement with spectroscopy can effectively resolve the degeneracy among different spectral models. Additionally, variations in polarization are crucial as they influence the underlying emission mechanism \citep{2021Galax...9...82G, McConnell:2017NewAR}. The temporal evolution of polarization also serves as a vital tool for comprehending the dynamic nature of the jet. Thus, time-resolved spectro-polarimetric measurements offer valuable information for distinguishing between different GRB models and understanding the radiation mechanisms involved.

Polarization measurements of prompt emission present significant challenges and have yet to be extensively conducted \citep{2021Galax...9...82G}. As of now, such measurements have been attempted for only a limited number of bursts, approximately 40, utilizing instruments such as the Reuven Ramaty High Energy Solar Spectroscopic Imager ($RHESSI$; \citealt{2003Natur.423..415C, 2004MNRAS.350.1288R}), the BATSE Albedo Polarimetry System (BAPS; \citealt{2005A&A...439..245W}), the INTErnational Gamma-Ray Astrophysics Laboratory ($INTEGRAL$; \citealt{2007ApJS..169...75K, 2007A&A...466..895M}), the GAmma-ray burst Polarimeter (GAP; \citealt{2011ApJ...743L..30Y, 2011PASJ...63..625Y, 2012ApJ...758L...1Y}), the Cadmium Zinc Telluride Imager onboard \AstroSat and POLAR \citep{2020A&A...644A.124K, 2019A&A...627A.105B}. However, most analyses have focused only on time and energy-integrated polarization measurements \citep{2021JApA...42..106C}.

Recently, we in \cite{2022ApJ...936...12C} reported the ﬁrst catalog of prompt emission polarization measurements, focusing on twenty bright GRBs observed by Cadmium Zinc Telluride Imager (CZTI) during its first five years of operation. These bursts were selected for their brightness to maximize the number of Compton events available for polarization analysis. The analysis revealed time-integrated polarization measurements in the energy range of 100–600 \keV. Based on the time-integrated polarization analysis, we found that most of these bursts ($\sim$ 75 \%) exhibited a low or zero polarization in the full burst interval (time-resolved and energy-resolved polarization measurement is required to examine if they are intrinsically unpolarized or the polarization angle within the burst is changing over the time) and only about 25 \% of the sample show indications of high linear polarization, including some as high as 71.43\% $\pm$ 26.84\% (GRB 180103A). Such high polarization implies that the mechanism for prompt emission could either be synchrotron radiation within a time-independent ordered magnetic field or Compton drag.

On the other hand, the POLAR instrument was also designed to perform linear polarization measurements of GRBs within an energy range of approximately 50-500 \keV. \cite{2020A&A...644A.124K} analyzed a sample of GRBs detected by POLAR and reported that the time-integrated analysis of the GRBs in their selection is compatible with a low or zero polarization \cite{2022ApJ...936...12C} compared the GRB polarization measurements made by POLAR and \AstroSat. POLAR, with an energy range of 50-500 \keV, is sensitive to lower energies and samples with longer burst durations. In contrast, \AstroSat, sensitive to energies above 100 \keV, samples shorter burst durations. GRB emissions are typically highly structured, and several GRBs, such as GRB 160821A \citep{2019ApJ...882L..10S}, GRB 170114A \citep{2019A&A...627A.105B}, and GRB 100826A \citep{2011ApJ...743L..30Y}, have shown polarization angle changes during bursts. Thus, POLAR's longer sampling duration makes it more likely to detect emissions with varying polarization angles, resulting in lower polarization observations compared to \AstroSat's higher energy, shorter duration sampling. In addition, the discrepancies could also arise from instrument systematics or differences in the GRBs observed by each instrument.

In this paper, we performed the time-resolved and energy-resolved polarization measurements of five bright bursts observed by CZTI in its first year of operation to verify whether the polarization properties are changing for these bursts. Additionally, we also performed a comprehensive time-resolved spectral analysis of those bursts observed by the \fermi mission to constrain their radiation physics. The paper's layout is as follows: In \S~\ref{sample}, we have given the details about our sample for the present study. In \S~\ref{dataanalysis}, we have given the methods of time-averaged, time-resolved, and energy-resolved spectro-polarimetric data analysis. The results and discussion of this work are given in \S~\ref{results} and in \S~\ref{discussion}, respectively. Finally, we have given a summary \& conclusion of this work in \S~\ref{conclusion}.

\section{Sample selection and previous polarization measurements}
\label{sample}

\begin{table*}[!ht]
\caption{List of bright gamma-ray bursts and their properties under investigation in our sample. The reported values of time-integrated polarization fractions (PF) obtained from \protect\cite{2022ApJ...936...12C} are listed in the last column. \AstroSat orbit IDs cited here correspond to those in which the necessary data were telemetered to the ground station. The data include those of Target of Opportunity (ToO), Announcement of Opportunity (AO), and Guaranteed Time (GT) observations. The redshift measurement/host search for the sample was attempted utilizing larger telescopes, such as the 10.4m GTC and the 3.6m DOT (4K $\times$ 4K IMAGER and TANSPEC) to study such transients \citep{2016RMxAC..48...83P}.}
\label{sample_table}
\begin{center}
\begin{scriptsize}
    \begin{tabular}{ |c|c|c|c|c|c|c|}
\hline
Sr. No. & \bf GRB name& Redshift/Host limit (mag)& \tninty  & \bf \AstroSat Orbit ID& Compton Counts & \bf Time-integrated PF (\%) \\
\hline
1& GRB 160325A & --& 42.94 $\pm$ 0.57 & 2652 & 764 & $<$ 45.02\\
2& GRB 160623A & 0.367 & 107.78 $\pm$ 8.69 &3983 & 1714 & $<$ 56.51 \\
3& GRB 160703A & $< $~1.5/$>$ 23 (i)  & 44.40 $\pm$ 2.80 &4135 & 433 & $<$ 62.64\\
4& GRB 160802A & -- &  16.38 $\pm$ 0.36 & 4576 & 1511 & $<$ 51.89\\
5& GRB 160821A & $>$ 23.6 (R) & 43.01 $\pm$ 0.72& 4866 &  2851 &$<$ 33.87 \\
\hline
\end{tabular}
\end{scriptsize}
\end{center}
\end{table*}

For the present work, we selected five GRBs (GRB 160325A, GRB 160623A, GRB 160703A, GRB 160802A, and GRB 160821A) to investigate in-depth the time-resolved and energy-resolved spectral and polarimetric characteristics. These bright bursts were observed by \AstroSat in its first year of operation. These GRBs are selected based on their brightness (fluence values greater than 10$^{-5}$ erg cm$^{-2}$) and their detection in CZTI within certain angles (0-60 and 120-180), where CZTI has good sensitivity for polarization measurements (see section 2 of \citealt{2022ApJ...936...12C} for more information about sample selection). The selected sample of bright bursts (see Figure \ref{TAS_FERMI_dist}) for this study and their time-integrated polarization have been tabulated in Table \ref{sample_table}. Below, we provide brief observations of individual bursts and their previous polarization measurements.

\subsection{GRB 160325A}

GRB 160325A was triggered by \fermi GBM \citep{2009ApJ...702..791M} and LAT \citep{2009ApJ...697.1071A} simultaneously at 06:59:21.51 UT on March 25, 2016 \citep{2016GCN.19224....1R, 2016GCN.19227....1A}. The GBM light curve of GRB 160325A has two separate emission episodes with a total \tninty duration of 43 sec in 50 -300 \keV. The gamma-ray/hard X-ray instruments like \swift BAT \citep{2005SSRv..120..143B, 2016GCN.19222....1S}, \kw \citep{2016GCN.19244....1T}, and \AstroSat \citep{2019ApJ...884..123C} also detected GRB 160325A. Previously, we studied the spectro-polarimetric properties of individual episodes of GRB 160325A and noted that both episodes have different spectral and polarimetric properties. The first episode of GRB 160325A is best fitted using \sw{Cutoff power-law + Blackbody} function and has a low polarization fraction ($<$ 37 \%, an upper limit in 100-380 \keV), suggesting sub-photospheric model as a dominant radiation model for this episode. On the other hand, the second episode of GRB 160325A is best fitted using \sw{Cutoff power-law} function and has high polarization fraction ($>$ 43 \%, a lower limit in 100-380 \keV), suggesting thin shell synchrotron radiation model \citep{2020MNRAS.493.5218S}. Our joint spectro-polarimetric analysis indicates a change in the spectral and polarimetric properties of two episodes of GRB 160325A.

\subsection{GRB 160623A}

GRB 160623A was detected by \fermi GBM at 05:00:34.23 UT on June 23, 2016, \citep{2016GCN.19555....1M}. GRB 160623A was also detected by other GRB triggering instruments \kw \citep{2016GCN.19554....1F}, CALET Gamma-Ray Burst Monitor \citep{2016GCN.19597....1Y}, \fermi LAT \citep{2016GCN.19553....1V}, and \AstroSat \citep{2019ApJ...884..123C}. We noted that \kw, CALET, and \AstroSat detected a bright emission pulse followed by weaker emission phases. However, GBM could not detect the brighter main emission pulse due to the Earth's occultation of the source. GBM detected weaker emission of around 50 sec (see Figure \ref{Prompt_Light_Curves} of the appendix). Recently, we reported the time-averaged PF of ($<$ 56.51 \%, an upper limit) in 100-600 \keV using \AstroSat/CZTI observations \citep{2022ApJ...936...12C}.

For this burst, \swift XRT discovered X-ray afterglow and also observed bright dust-scattered features (radius $\sim$ 3.5 arcmin) around this GRB \citep{2016GCN.19558....1M, 2016GCN.19559....1T, 2017MNRAS.472.1465P}. Many ground-based facilities detected the optical/mm/radio afterglows of GRB 160623A. 

Utilizing the precise localization of the optical afterglow of GRB 160623A, \cite{2016GCN.19708....1M} reported the spectroscopic redshift of the burst ($z$ = 0.367). We also conducted observations of the optical afterglow of GRB 160623A using the 10.4m Gran Telescopio Canarias (GTC) as a part of a larger collaboration. Spectra were gathered at various epochs: on June 25 (1.9 days post-burst) and July 3/4, 2016. We utilized both the R1000B and R2500I grisms, covering the wavelength range of 3800-10000 Å. Analysis of the reddest spectrum (2 x 1200 sec with R2500I) at the afterglow position revealed emission lines of H-alpha and [SII], enabling us to determine a redshift of $z$ = 0.367 (see Figure \ref{Redshift_GRB160623A} of the appendix), which corroborates the value proposed by \cite{2016GCN.19708....1M}. Additionally, the bluest range spectrum (1200 sec) indicated a marginal detection of H-beta, considering the high foreground Galactic extinction along the line of sight. The faint continuum observed in the spectrum from the first epoch extended down to 3800 Å, with no discernible absorption lines present. Based on these observations, we confirmed that this redshift corresponds to the host galaxy of GRB 160623A \citep{2016GCN.19710....1C}.

\subsection{GRB 160703A}

GRB 160703A was detected by \swift BAT at 12:10:05 UT on July 03, 2016 with a \tninty duration of 44.4 $\pm$ 2.8 sec \citep{2016GCN.19645....1C, 2016GCN.19648....1L}. The prompt emission of GRB 160703A was also observed by \kw \citep{2016GCN.19649....1F} and \AstroSat \citep{2016GCN.19739....1B}. Based on \AstroSat CZTI observations of GRB 160703A, we reported the high value of time-averaged PF ($<$ 62.64 \%, an upper limit) in 100-600 \keV \citep{2022ApJ...936...12C}.  

The X-ray and optical counterpart of GRB 160703A was detected by \swift XRT and UVOT instruments \citep{2016GCN.19655....1D, 2016GCN.19656....1H}. The UVOT detected the afterglow of GRB 160703A in all its seven filters, based on this \cite{2016GCN.19656....1H} constrained the redshift of the burst ($z~<$ 1.5). Later follow-up observations using the Giant Metrewave Radio Telescope (GMRT) telescope detected a faint potential radio counterpart of GRB 160703A \citep{2016GCN.19849....1N}.  

\subsection{GRB 160802A}

At 06:13:29.63 UT on August 02, 2016, GRB 160802A was triggered by \fermi GBM with \tninty of 16.4 sec (in 50-300 \keV). GBM provides the localization as RA= 35.29, DEC = +72.69 (J2000) with uncertainty radius of 1 degree \citep{2016GCN.19754....1B}. The GBM light curve of GRB 160802A has two clearly separated emission episodes (see Figure \ref{Prompt_Light_Curves} of the appendix). GRB 160802A was one of the brightest (energy fluence 1.04 $\times$ 10$^{-4}$ erg $\rm s^{-2}$ in 10-1000 \keV) \fermi GBM detected bursts. The burst was independently detected by other gamma-ray detecting satellites/instruments such as \AstroSat CZTI \citep{2016GCN.19782....1B}, \kw \citep{2016GCN.19767....1K}, Lomonosov BDRG \citep{2016GCN.19759....1P}, and CALET Gamma-ray Burst Monitor \citep{2016GCN.19778....1T}.

We in \cite{2018ApJ...862..154C} studied the spectro-polarimetric study of GRB 160802A using joint \fermi and \AstroSat observations. We performed spectral analysis using empirical functions and \sw{XSPEC} software. We noted that the evolution of low-energy photon indices of the \sw{Band} function is harder than those theoretically expected from thin shell synchrotron slow and fast cooling model, indicating photospheric origin. Additionally, we calculated the time-averaged PF = 85 $\pm$ 29 \% using previous polarization tools in 100-300 \keV \citep{2018ApJ...862..154C}. A high value of the time-averaged PF ($<$ 51.89 \%, an upper limit) was also measured 100-600 \keV in \cite{2022ApJ...936...12C} for GRB 160802A using improved polarimetric techniques. Such a high value of PF indicates a synchrotron model if the source was observed on-axis. On the other hand, the photospheric model can also produce such high PF if the source is viewed along the edge. Based on our joint \fermi and \AstroSat spectro-polarimetric observations, we suggested that GRB 160802A might have originated due to subphotospheric dissipation viewed along the edge \citep{2018ApJ...862..154C}.

\subsection{GRB 160821A}

GRB 160821A was detected by \swift BAT and \fermi GBM at 20:34:30 UT on 21 August 2016 \citep{2016GCN.19830....1S, 2016GCN.19835....1S}. The prompt emission of the burst was also discovered independently using \fermi LAT \citep{2016GCN.19831....1M}, \kw \citep{2016GCN.19842....1K}, CALET \citep{2016GCN.19865....1M}, and \AstroSat \citep{2016GCN.19867....1B}. The burst is extremely bright which provides a unique opportunity for detailed spectro-polarimetric analysis using \fermi-\AstroSat observations. We performed the spectro-polarimetric analysis of GRB 160821A and noted a high PF (66$^{+26}_{-27}$ \%) in the time-averaged polarization measurements (in 100-300 \keV). Additionally, the time-resolved polarization measurements give evidence of a change in polarization angle by twice during the entire emission phase of GRB 160821A \citep{2019ApJ...882L..10S}. Recently, for this burst, we reported the time-averaged PF ($<$ 33.87 \%, an upper limit) in 100-600 \keV utilizing the improved polarization measurement tools \citep{2022ApJ...936...12C}.  

\begin{figure*}
\centering
\includegraphics[scale=0.38]{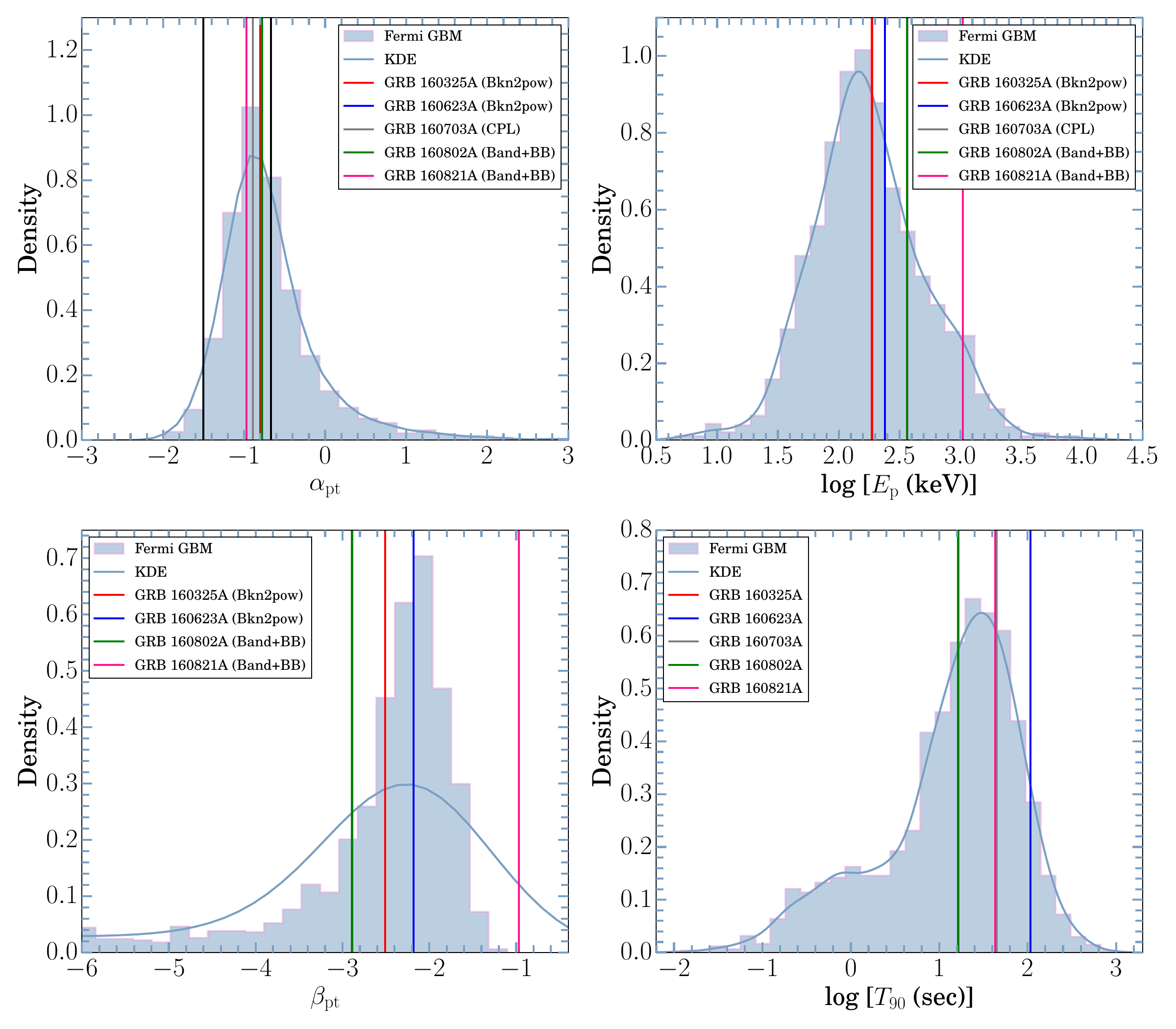}
\includegraphics[scale=0.31]{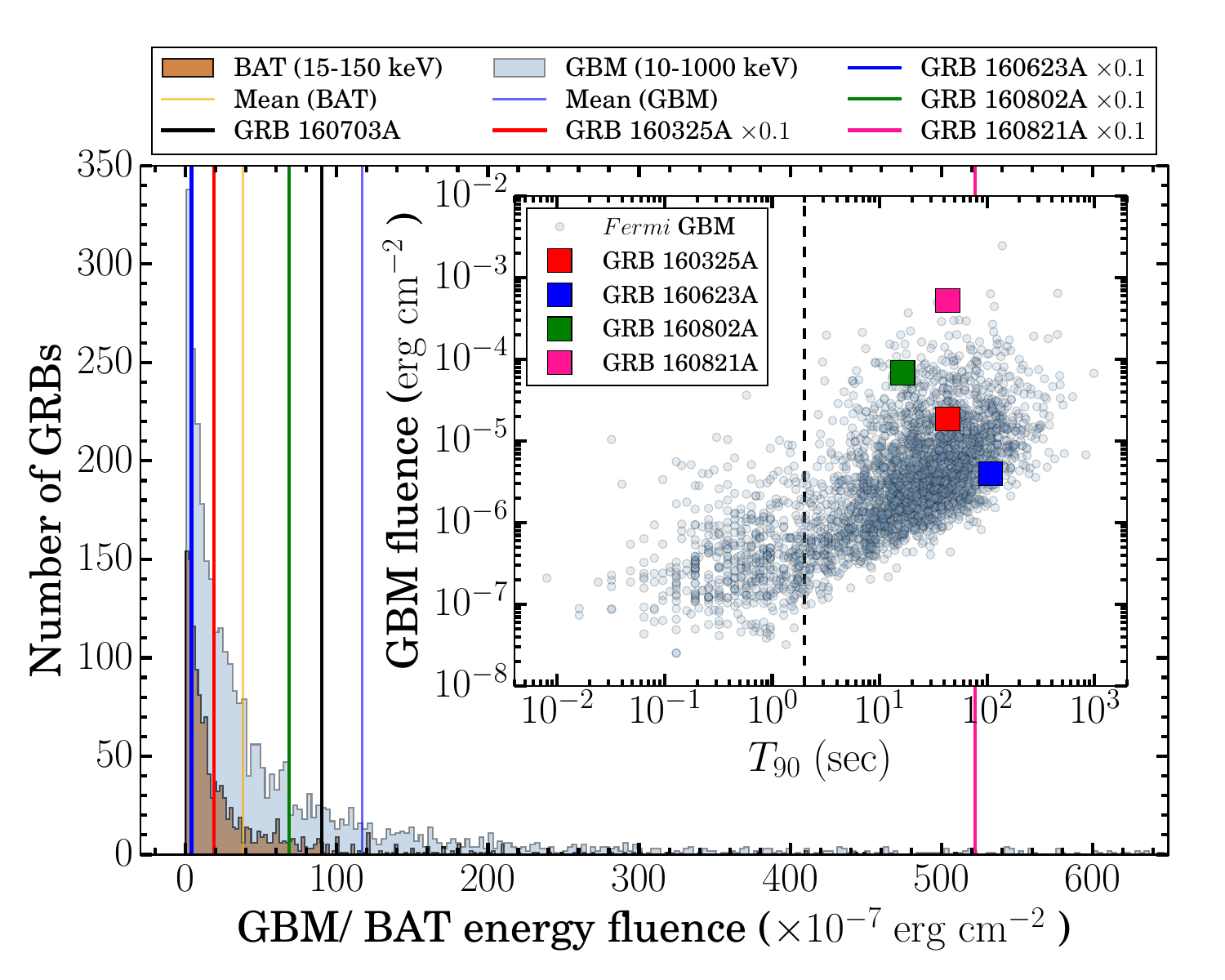}
\includegraphics[scale=0.31]{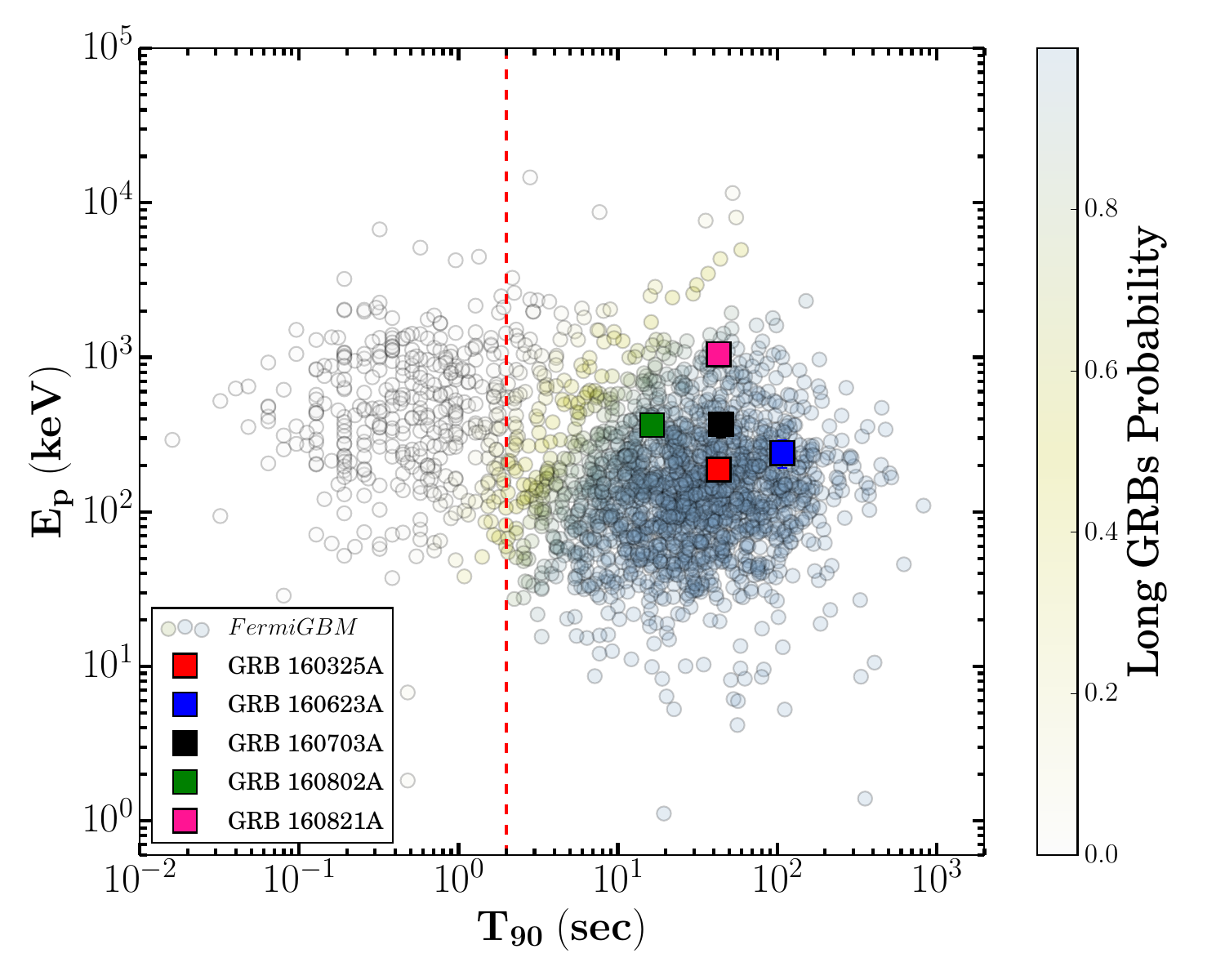}
\caption{{Prompt emission characteristic of the GRBs:} The distributions of basic spectral ($\alpha_{\rm pt}$ (top-left), \Ep (top-right), $\beta_{\rm pt}$ (middle-left)) and temporal (\tninty, middle-right) properties of GBM detected GRBs. The solid black lines correspond to the theoretically predicted values of the low energy photon index from thin-shell synchrotron emission models. The vertical colored lines denote the position of GRBs under study in this paper. The Kernel density estimations (KDE) for all the distributions are shown using grey curves. {Bottom-left:} Histogram of \fermi GBM (light blue) and \swift BAT (orange) energy fluence values. The mean fluence values for the BAT and GBM samples are marked by vertical solid orange and blue lines, respectively. The positions of all five bursts in our sample are marked using vertical-colored lines. The inset plot illustrates the relationship between energy fluence and duration for \fermi GRBs. {Bottom-right}: \Ep-\tninty (harness-duration) plot for \fermi GBM GRBs. The location of five GRBs in our sample is shown using colored squares. The vertical red line represents the threshold for classifying bursts. The figure displays the long and short bursts obtained from the GBM catalog. The probability of long GRBs is represented on the right side of the Y-scale.}
\label{TAS_FERMI_dist}
\end{figure*}

\section{Data analysis}
\label{dataanalysis}
We utilized \AstroSat CZTI data for the polarization measurements of the GRBs in our sample, while \fermi and \swift observations were employed for the spectral analysis of the bursts (see details below). It is crucial to clarify that our analysis does not involve a fully coupled spectro-polarimetric fitting. Despite the absence of a fully coupled spectro-polarimetric analysis, our study provides significant insights into the polarization characteristics and emission mechanisms of the GRBs under investigation.

\subsection{Technique of polarization analysis and improvements}

The \AstroSat CZTI mainly serves as a hard X-ray imaging/spectroscopy detector with a wide field of view. Notably, its ground calibration has revealed polarization measurement capabilities for on-axis sources. Recent experimentation by \cite{2022JATIS...8c8005V} has further validated CZTI's ability to measure off-axis hard X-ray polarization for bright sources such as GRBs. Above 100 \keV, CZTI exhibits a notable probability of Compton scattering. Leveraging the pixilated nature of CZT detectors, it functions as a Compton Polarimeter. Given its distinctive hard X-ray polarization measurement capabilities, the CZTI team has reported polarization measurements of both persistent (such as the Crab pulsar and nebula) and transient (including GRBs) X-ray sources \citep{2016ApJ...833...86R, 2018NatAs...2...50V, 2019ApJ...884..123C}. Despite moderate brightness, energetic transient sources like GRBs are the potential hard X-ray for polarization measurements due to the simultaneous availability of pre and post-burst backgrounds with higher signal-to-noise ratios. For a comprehensive understanding of the prompt emission polarization analysis of GRBs using CZTI data, we in \cite{2022ApJ...936...12C} present detailed techniques. In this study, we present a concise overview of the steps and recent enhancements in the polarization analysis tool for CZTI data.

\begin{itemize}

\item {Selection of Compton events:} To conduct polarization analysis using CZTI data, we initially chose double events detected within a 20 $\mu$s temporal window. Subsequently, we applied Compton criteria, assessing the ratio of energies received on neighboring pixels, to filter out double events resulting from chance coincidence.
    
\item {Creation of Background-Subtracted Azimuthal Angle Distribution:} Compton events were selected within both the GRB emission region and the pre-and post-burst background regions. To define the latter, we excluded instances of spacecraft crossing the South Atlantic Anomaly. Following this, we subtracted the raw azimuthal angle distribution of the GRB emission region from that of the background, resulting in the final background-subtracted azimuthal angle distribution for the GRB.
    
\item {Correction for geometric effects:} Systematic errors stemming from geometric effects and off-axis detection of GRBs impact the background-subtracted azimuthal angle distribution. To address this, we employed the Geant4 toolkit and the \AstroSat mass model to simulate an unpolarized azimuthal angle distribution. This simulation considered the distribution of photons observed from GRB spectra at the same orientation as the \AstroSat spacecraft. Subsequently, we normalized the observed background-subtracted azimuthal angle distribution of the GRB using the simulated unpolarized azimuthal angle distribution.

\item {Calculation of Modulation Amplitude and Polarization Angle:} We employed a sinusoidal function to fit the observed background-subtracted and geometry-corrected azimuthal angle distribution of the GRB. This fitting process enabled us to determine the modulation factor ($\mu$) and polarization angle within the \AstroSat CZTI plane. For the sinusoidal function fitting, we utilized the Markov chain Monte Carlo (MCMC) method.
    
\item {Calculation of Polarization Fraction:} To ascertain the polarization fraction, normalization of the modulation factor ($\mu$) with the simulated modulation amplitude for 100\% polarized radiation $\mu_{100}$ is required. This value is obtained through Geant4 toolkit simulations using the \AstroSat mass model for the same direction and observed spectral parameters. Subsequently, the PF is calculated by normalizing $\mu$ with $\mu_{100}$ for those bursts exhibiting a Bayes factor greater than 2. In instances where the Bayes factor is below 2, we establish a constraint on the polarization fraction by setting a 2$\sigma$ upper limit (refer to \citealt{2022ApJ...936...12C} for further details).
\end{itemize}

Furthermore, we have implemented the following enhancements in the polarization data analysis of the \AstroSat CZTI for this study. This upgraded CZTI pipeline is being utilized for the first time for executing time and energy-resolved polarization measurements of bursts detected by the \AstroSat CZTI.

\subsubsection{Low gain pixels and energy bandwidth} 
Since the launch of the \AstroSat mission, around 20 $\%$ of the CZTI pixels were observed to have electronic gains lower (2$-$4 times) than the laboratory-tested gain values. In the previous studies (e.g., \citealt{2018ApJ...862..154C, 2019ApJ...884..123C, 2019ApJ...874...70C, 2019ApJ...882L..10S, 2020MNRAS.493.5218S, 2022MNRAS.511.1694G}), the sensitive spectroscopic and polarimetric information in 100 - 300 \keV were extracted using the normal-gain pixels only. However, the electronic gain for the low-gain pixels has been constant since the first day of working of CZTI in space; therefore, considering the low-gain pixels after rigorous calibration can extend the energy channels of Compton energy spectra and polarization up to 600 \keV. This new characteristic also makes wider the spectral coverage using single-pixel up to the sub-MeV capacity ($\sim$1 MeV), earlier it was restricted to 150 \keV \citep{2021arXiv210213594C}. Recently, we applied this method for the time-averaged polarization measurement of twenty GRBs detected by the \AstroSat CZTI in its five years of operation \citep{2022ApJ...936...12C}. We are now implementing these improvements for the first time in time-resolved and energy-resolved polarimetric measurements of bright bursts. This new methodology significantly enhances outcomes and extends the energy coverage for prompt emission spectro-polarimetric analysis.

\subsubsection{New event selection logic} 
Hard X-ray detectors are typically sensitive to background noise, such as cosmic rays, owing to their non-focusing nature at these wavelengths. As a result, it is crucial to identify and eliminate such noise events, selecting only those unaffected by background interference for scientific analysis. In the case of CZTI, the previous analysis pipeline for time-resolved spectro-polarimetric studies has incorporated techniques for event selection, but these techniques have some constraints. For example, these algorithms were mainly developed to analyze data from regular X-ray sources where the object flux is significantly lower than the background and thus is not well equipped for transient events like GRBs. \cite{2021JApA...42...37R} re-investigated the features of noise events in CZTI and gave a generalized event choice technique that provides analysis for all types of sources, including GRBs. This algorithm significantly reduces noise levels without considering the source flux dependence. In our current study, focusing on time-resolved and energy-resolved polarimetric analysis using CZTI data, we have used this algorithm, leveraging its improved capability for noise reduction across various source types. 

\subsection{Technique of temporal and spectral analysis}
The temporal profiles of GRBs exhibit distinct characteristics attributed to the erratic behavior of the central engine. To extract temporal information from \fermi GBM data, we employed the Fermi GBM Data Tools \citep{GbmDataTools}. Furthermore, to extract spectra from \fermi GBM data, we employed the \sw{gtburst} tool\footnote{\url{https://fermi.gsfc.nasa.gov/ssc/data/analysis/scitools/gtburst.html}}. For BAT data, both temporal and spectral analyses were carried out using \sw{HEASOFT}, utilizing the most recent BAT calibration files. For detailed insights into the technique employed for BAT data analysis, refer to \cite{2021MNRAS.505.4086G}. It is important to note that \sw{3ML} plugin for the simultaneous fitting of BAT data with data from other instruments, such as \kw, \AstroSat, etc, is not currently available. Consequently, for GRBs observed with BAT, we have relied exclusively on the spectral parameters derived from the \kw instrument, as reported in \cite{2022ApJ...936...12C}. Below, we have provided details of our spectral analysis (empirical and physical synchrotron model). However, we did not explore the physical photospheric models due to the lack of a publicly available robust and validated photospheric model (compatible with \sw{3ML}).

\subsubsection{Empirical spectral modeling} 

For the prompt emission spectral modeling of GRBs, we employed the Multi-Mission Maximum Likelihood framework \citep[\sw{3ML}]{2015arXiv150708343V}. Typically, the GRB spectrum can be adequately described by an empirical \sw{Band} function. Therefore, we initially fitted the spectrum of GRBs of our sample using \sw{Band} function. Subsequently, we explored additional empirical functions such as power-law (PL), \sw{Cutoff power-law} (CPL), and \sw{bkn2pow}, considering model parameters and statistical measures/residuals from spectral fitting with \sw{3ML}. The selection of the best-fit model was determined based on the difference in deviance information criterion (DIC) values obtained from various models. A comprehensive method for empirical spectral modeling is provided in \cite{2023MNRAS.519.3201C}.

\subsubsection{Physical spectral modeling}
\cite{2020NatAs...4..174B} showed that empirical function could be fallacious, and we should use physical spectral modeling to constrain the radiation physics of prompt emission. \cite{2020NatAs...4..174B} further showed that even if the low-energy index of \sw{Band} function exceeds the line of death of the synchrotron model, the spectrum still could be fitted using physical thin shell synchrotron model. Additionally, due to the spectral curvature of empirical functions, the empirical spectral models may lead to incorrect interpretations of the radiation physics of GRBs. So, we have utilized the physical thin shell synchrotron model to accurately interpret the emission mechanism. For the present work, we have applied publicly available \sw{pynchrotron}\footnote{\url{https://github.com/grburgess/pynchrotron}} physical model for the time-integrated and time-resolved spectral fitting of GBM data in \sw{3ML} \citep{2020NatAs...4..174B}. \sw{pynchrotron} model executes the synchrotron emission from a cooling population of electrons in the thin shell case. According to the \sw{pynchrotron} model, the relativistic electrons follow a power-law distribution N($\gamma$) $\propto$ $\gamma^{-p}$ with $\gamma_{inj} \leq$ $\gamma \leq$ $\gamma_{max}$. In this equation, $p$ represents the power-law index of the energy distribution of injected electron, $\gamma_{inj}$  represents the lower limit, and $\gamma_{max}$ represents the upper limit of the relativistic electron spectrum. \sw{pynchrotron} model consists of six model parameters: 1. the power-law index of the energy distribution of injected electron ($p$), 2. The strength of magnetic field  (B), 3. $\gamma_{max}$,  4. $\gamma_{inj}$, 5. Bulk Lorentz factor ($\gamma_{bulk}$) of the relativistic jet, and 6. Lorentz factor for the electron cooling time scale ($\gamma_{cool}$). Although, while the physical spectral modeling of GRBs, we fixed the $\gamma_{inj}$ = 10$^{5}$ (due to degeneracy between B and $\gamma_{inj}$), $\gamma_{max}$ = 10$^{8}$ (slow cooling synchrotron model better fit the prompt spectrum). Additionally, we have also fixed the $\gamma_{bulk}$ for GRBs utilizing the prompt emission correlation between $\gamma_{bulk}$ and isotropic gamma-ray energy.  

\subsection{Search for potential host galaxies using DOT}

The expected polarization fraction from different radiation models depends on the jet viewing geometry, and this can be further verified by investigating the $\Gamma \theta_{\rm j}$ condition, where $\Gamma$ represents the bulk Lorentz factor and $\theta_{\rm j}$ denotes the jet opening angle (see section \ref{Jet composition and emission mechanisms of GRBs} for more information). $\Gamma$ and $\theta_{\rm j}$ could be calculated using the Liang relation (the correlation between isotropic gamma-ray energy and $\Gamma$, \citealt{2010ApJ...725.2209L}) and the jet breaks observed in the afterglow light curve, respectively. However, both of these parameters depend on the redshift. Therefore, redshift is a very important parameter to verify the $\Gamma \theta_{\rm j}$ condition and predict the possible radiation mechanism based on the observed value of polarization fraction. We observed that only two GRBs (GRB 160623A and GRB 160703A) in our sample have redshift constraints. No redshift measurements were found in the literature for the remaining GRBs. To determine their photometric redshift, we attempted to locate the associated host galaxies of the bursts with sub-arcsecond localization in our sample (GRB 160703A and GRB 160821A) using the 3.6m Devasthal Optical Telescope (DOT, \citealt{2023arXiv230715585G}). We conducted observations of GRB 160703A using TANSPEC (in i-filter, \citealt{2022PASP..134h5002S}) on 2022-11-11, with a total exposure time of 5700 seconds. Similarly, observations of GRB 160821A were carried out using a 4K $\times$ 4K IMAGER (in R-filter, \citealt{2018BSRSL..87...42P, 2023JAI....1240009P}) on 2022-12-20, with a total exposure time of 5100 seconds (see Figure \ref{Redshift_GRB160623A} of the appendix). The methods for the optical data reduction of host images taken using TANSPEC and IMAGER are presented in \cite{2022JApA...43...82G, 2023arXiv231216265G}. However, despite our efforts, we were unable to detect any associated host galaxies of these bursts within the best available error circles. Our observations yielded limiting magnitudes of $\sim$ 23 mag for GRB 160703A and 23.6 mag for GRB 160821A, respectively. This suggests that the host galaxies of these GRBs may be intrinsically faint or highly obscured, reflecting the diverse nature of GRB host environments.

\section{Results}
\label{results}

Utilizing the comprehensive analysis outlined above, we proceed to present the detailed spectro-polarimetric results of all five bright GRBs in the subsequent section.

\subsection{Prompt uniform light curves and time-integrated spectra}

The prompt light curve profiles of \fermi detected (GRB 160325A, GRB 160623A, GRB 160802A, and GRB 160821A) and \swift detected (GRB 160703A) GRBs in our sample are presented in Figure \ref{Prompt_Light_Curves} of the appendix. The light curves of GRB 160325A (depicted in red) and GRB 160802A (in green) exhibit similar temporal profiles, characterized by two distinct episodes: a prominent pulse followed by a softer pulse, with a quiescent temporal gap in between. In contrast, GRB 160623A (highlighted in blue) showcases a primary pulse succeeded by weaker emission. Notably, \fermi could not detect the main emission of GRB 160623A due to Earth occultation during the burst's main emission, with the \fermi trigger occurring approximately 50 seconds post-burst \citep{2016GCN.19555....1M}. The light curve of GRB 160821A (depicted in pink) illustrates a faint initial emission followed by a very brighter emission. Meanwhile, GRB 160703A presents multiple overlapping profiles (in grey).

We employed the Bayesian block method on the CZTI Compton light curves to determine the time intervals for the time-integrated spectral analysis of GRBs in our sample. These selected time segments were also utilized for time-integrated polarization measurements, as detailed in section 2.2 of \citealt{2022ApJ...936...12C}. The time-integrated \fermi spectra of GRB 160325A and GRB 160623A were optimally fitted using the \sw{Bkn2pow} function. Conversely, the time-integrated \fermi spectra of GRB 160802A and GRB 160821A exhibited the best fits with the \sw{Band + Blackbody} function. For the \swift BAT-detected GRB 160703A, the time-integrated spectrum was most effectively described by the \sw{Cutoff power-law} function, considering the limitation of energy coverage of BAT. Detailed information regarding the best fit time-integrated spectral parameters for all five GRBs in our sample can be found in Table \ref{tab:TAS} of the appendix.

\subsection{Comparison with \fermi GRBs}

We analyzed the spectral (obtained using time-integrated analysis) and temporal parameters of GRBs in our sample and compared them with a larger sample of \fermi GBM detected GRBs (see Figure \ref{TAS_FERMI_dist}). Such comparison provides valuable insights into the spectral properties and diversity of these cosmic sources. The distribution of the low energy photon index is useful for characterizing the power-law behavior of the photon spectrum at lower energies and identifying the emission mechanisms. The distribution of $\alpha_{\rm pt}$ reveals that a significant number of bursts deviates from the synchrotron emission mechanism. The distribution of \Ep value is crucial and indicates the energy at which the GRB spectrum reaches its maximum intensity. We noted all the bursts in our sample have a harder peak energy than the mean peak value obtained for \fermi GBM detected GRBs. The distribution of high-energy photon indices signifies the steepness of the spectral slope in the high-energy regime. The high energy spectral index ($\beta_{\rm pt}$) values (calculated using the time-integrated spectral measurement) for GRB 160325A, GRB 160623A, and GRB 160802A are steeper than the mean value obtained for \fermi GBM detected GRBs. On the other hand, GRB 160821A has a shallower $\beta_{\rm pt}$ value. Further, we studied the distribution of \tninty duration using \fermi GBM data, and the distribution indicates that all the bursts in our sample belong to the long GRBs class.

\subsubsection{Energy-fluence distribution}

We compared the energy fluence value of GRBs in our sample with \fermi GBM and \swift BAT detected GRBs. Our analysis indicates that GRBs in our sample are significantly brighter than the mean value of observed fluence values (see Figure \ref{TAS_FERMI_dist}). We also represented this result using the distribution of \tninty as a function of energy fluence values for the bursts observed by \fermi GBM (see inset plot in Figure \ref{TAS_FERMI_dist}). High fluence bursts are useful for polarization measurements.

\begin{figure}[!ht]
\centering
\includegraphics[scale=0.31]{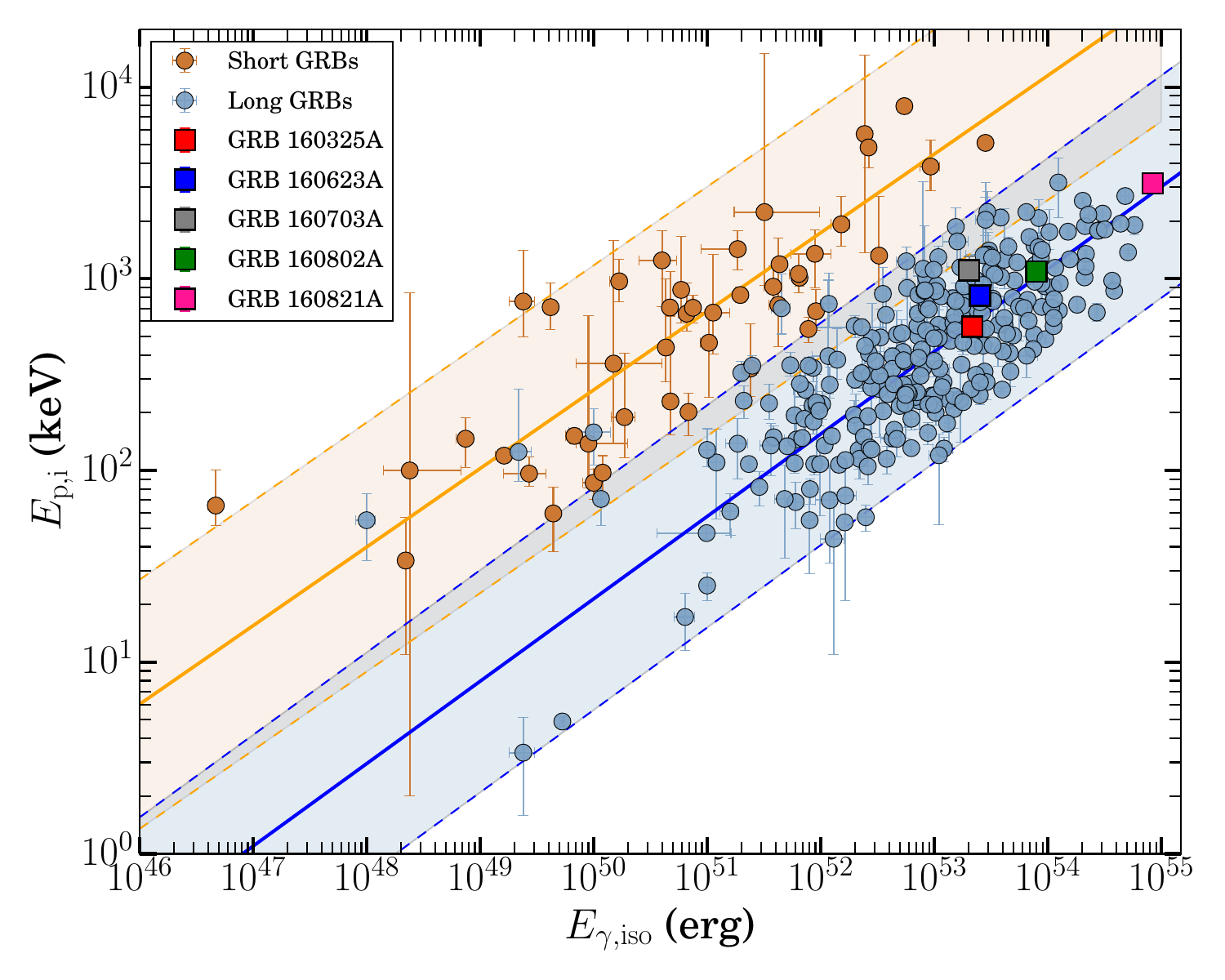}
\includegraphics[scale=0.31]{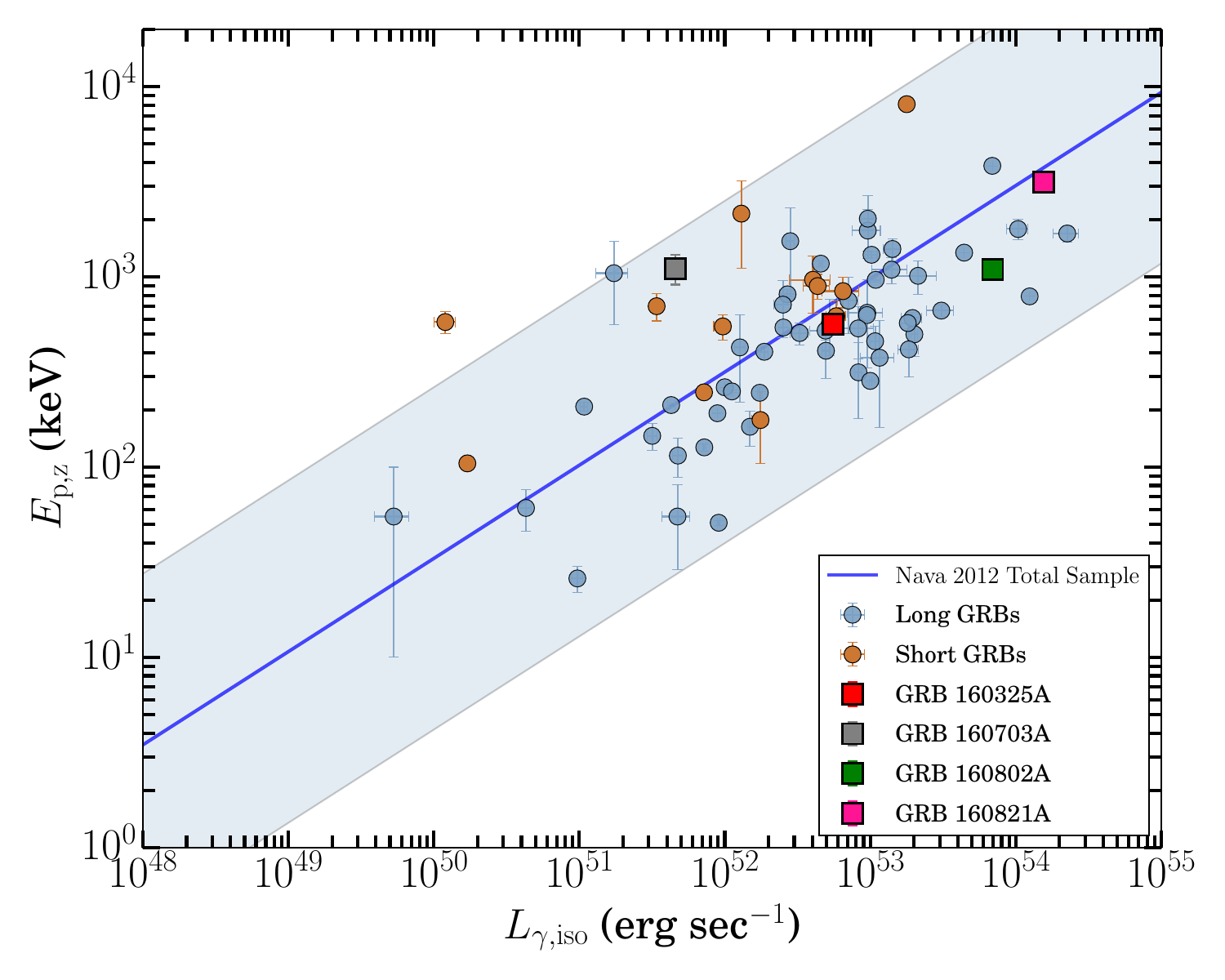}
\caption{{Prompt emission correlation of GRBs:} {Top}: The location of five bright GRBs in Amati correlation. The well-studied long and short bursts extracted from \protect\cite{2020MNRAS.492.1919M} are represented by blue and orange circles, respectively, with solid blue and orange lines depicting the linear fits for these groups. The parallel shaded areas illustrate the 3$\sigma$ variation. {Bottom:} The location of five GRBs in Yonetoku correlation. The well-studied long and short bursts, as studied in \protect\cite{2012MNRAS.421.1256N}, are shown with blue and orange circles. The parallel shaded areas indicate the 3$\sigma$ scatter. The colored squares illustrate the location of the GRBs of our sample. In our analysis, we included all five GRBs in the Amati and Yonetoku relations. However, for the GRBs without measured redshifts, we assumed a  redshift value of 2, mean of redshift distribution for long GRBs \citep{2022JApA...43...82G}.}
\label{Prompt_correlation}
\end{figure}

\subsubsection{Spectral-Hardness plot}

The classification of GRBs primarily relies on the prompt emission properties, such as the duration of the \tninty and the hardness ratio. We studied the spectral hardness distribution for the GRBs in our sample. The peak energy of a GRB's spectrum is related to its duration. Studies have shown that GRBs with longer durations tend to have lower peak energies (soft), while shorter-duration GRBs tend to have higher peak energies (hard). We compiled the \tninty duration and \Ep values of all the GRBs detected by the \fermi GBM instrument from the GBM burst catalog. We noted that all the GRBs in our sample are consistent with the typical characteristics of long GRBs (see Figure \ref{TAS_FERMI_dist}).

\subsubsection{Amati and Yonetoku correlation}

Several global correlations can be observed in the prompt properties of GRBs, and these correlations play a crucial role in characterizing GRBs \citep{2020MNRAS.492.1919M}. We studied the Amati correlation for the GRBs in our sample \citep{2006MNRAS.372..233A}. It is a well-known empirical relationship and relates the isotropic equivalent energy ($E_{\rm \gamma, iso}$) and the spectral peak energy of GRB prompt emission spectra in the rest frame. Amati correlation has important implications for the physics of the prompt emission process, the emission mechanism, or the properties of the GRB progenitor systems. For GRB 160623A, we obtained $E_{\gamma, \rm iso}$ and peak energy values using \kw observations \citep{2017ApJ...850..161T} as the main emission was not detected using \fermi GBM. We noted that all the GRBs in our sample are consistent with the Amati correlation of the long GRBs (see Figure \ref{Prompt_correlation}). The physical explanation for the Amati correlation in the literature remains a subject of debate and lacks consensus. Nevertheless, certain studies suggest that the Amati correlation may be attributed to the viewing angle effect within the context of synchrotron emission \citep{2004ApJ...606L..33Y, 2004ApJ...614L..13E, 2005ApJ...629L..13L}.

We also studied the Yonetoku correlation for our sample \citep{2010PASJ...62.1495Y}. The Yonetoku correlation relates two observables of GRBs: $E_{\rm \gamma, iso}$ and the peak luminosity ($L_{\rm \gamma, iso}$) of the prompt gamma-ray emission. This correlation indicates that GRBs with higher isotropic equivalent energies tend to have higher peak luminosities. The correlation provides constraints and insights into the nature of GRB progenitors, emission processes, and the energy release mechanisms associated with these powerful cosmic explosions. This correlation could potentially be explained by the photospheric dissipation model, taking into account that subphotospheric dissipation occurs at a considerable distance from the central engine \citep{2005ApJ...628..847R}. We noted that all the GRBs in our sample are consistent with the Yonetoku correlation (see Figure \ref{Prompt_correlation}). Furthermore, the photospheric model has been useful in explaining both the Amati and Yonetoku relations. Recent studies have shown that these correlations can be naturally accounted for by considering the effects of the viewing angle relative to the jet axis. When the photospheric emission is viewed at different angles, the observed spectral properties and the inferred energetics can vary significantly. This variation can lead to the observed Amati and Yonetoku relations \citep{2019NatCo..10.1504I, 2022Univ....8..310P, 2024ApJ...961..243I, 2022ApJ...926..104P}.

\subsection{Time-resolved spectral measurements}

The GRBs spectrum shows strong evolution within the burst; therefore, the derived time-integrated spectral parameters may not provide intrinsic spectral behavior and can be artifacts due to strong spectral evolution. Thus, time-resolved spectral measurements are needed to verify the underlying radiation mechanisms of GRBs. We studied the time-resolved spectral analysis of those bursts (GRB 160325A, GRB 160802A, and GRB 160821A) for which \fermi GBM observations were available. \fermi GBM wide spectral coverage is crucial for detailed spectral analysis.

We selected the temporal bins for time-resolved spectral analysis using the Bayesian Block method. After selecting bins, we calculated the significance of individual bins and only selected those with a signification greater than 10. Further, we fitted all these bins with \sw{Band} and \sw{CPL} models and calculated the difference of DIC values to identify the best-fit model for individual bins. The comparison between DIC values of \sw{Band} and \sw{CPL} models for all three GRBs are plotted with red squares in Figure \ref{DIC} of the appendix. The DIC comparison indicates that individual bins of GRB 160325A, GRB 160802A, and GRB 160821A are preferred \sw{Band} function over \sw{CPL} model (no bins have $\Delta$ DIC $\leq$ -10). For some of the bins, \sw{CPL} model has $\Delta$ DIC value in between zero and -10, indicating that \sw{Band} and \sw{CPL} both models have an equivalent fit. After selecting the best-fit function between \sw{Band} and \sw{CPL} models, we added the additional \sw{Blackbody} (BB) function. We again selected the best-fit model between \sw{Band} or \sw{CPL} with \sw{Band+BB} or \sw{CPL+BB} using the difference of DIC values obtained for the two models. A detailed selection method for the different empirical functions is present in \cite{2023MNRAS.519.3201C}. Furthermore, we have also compared the fits between the best fit empirical and physical models. However, there are some time bins for which the physical synchrotron parameters are not very well constrained (due to the low signification).  

\begin{figure*}
\centering
\includegraphics[scale=0.28]{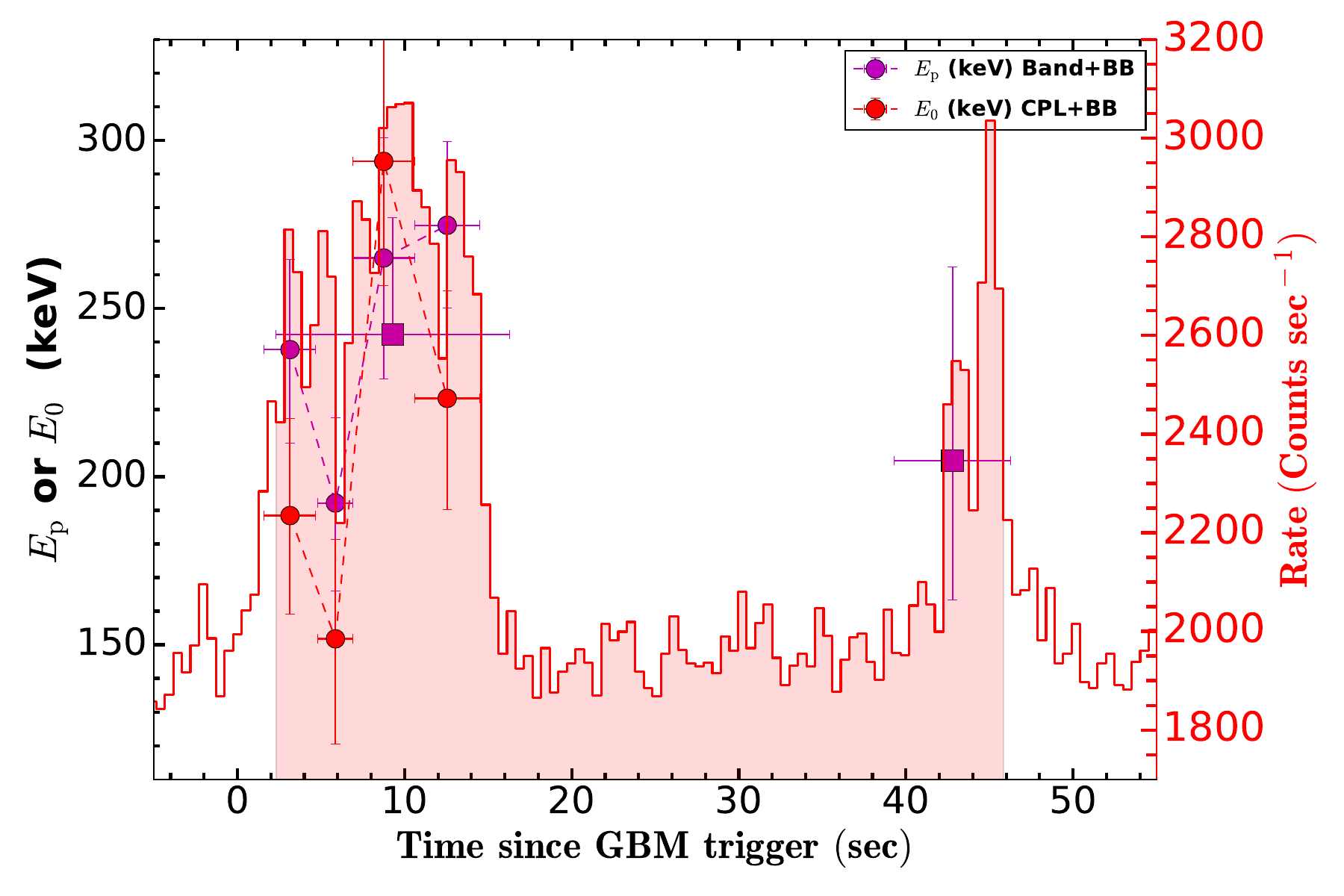}
\includegraphics[scale=0.28]{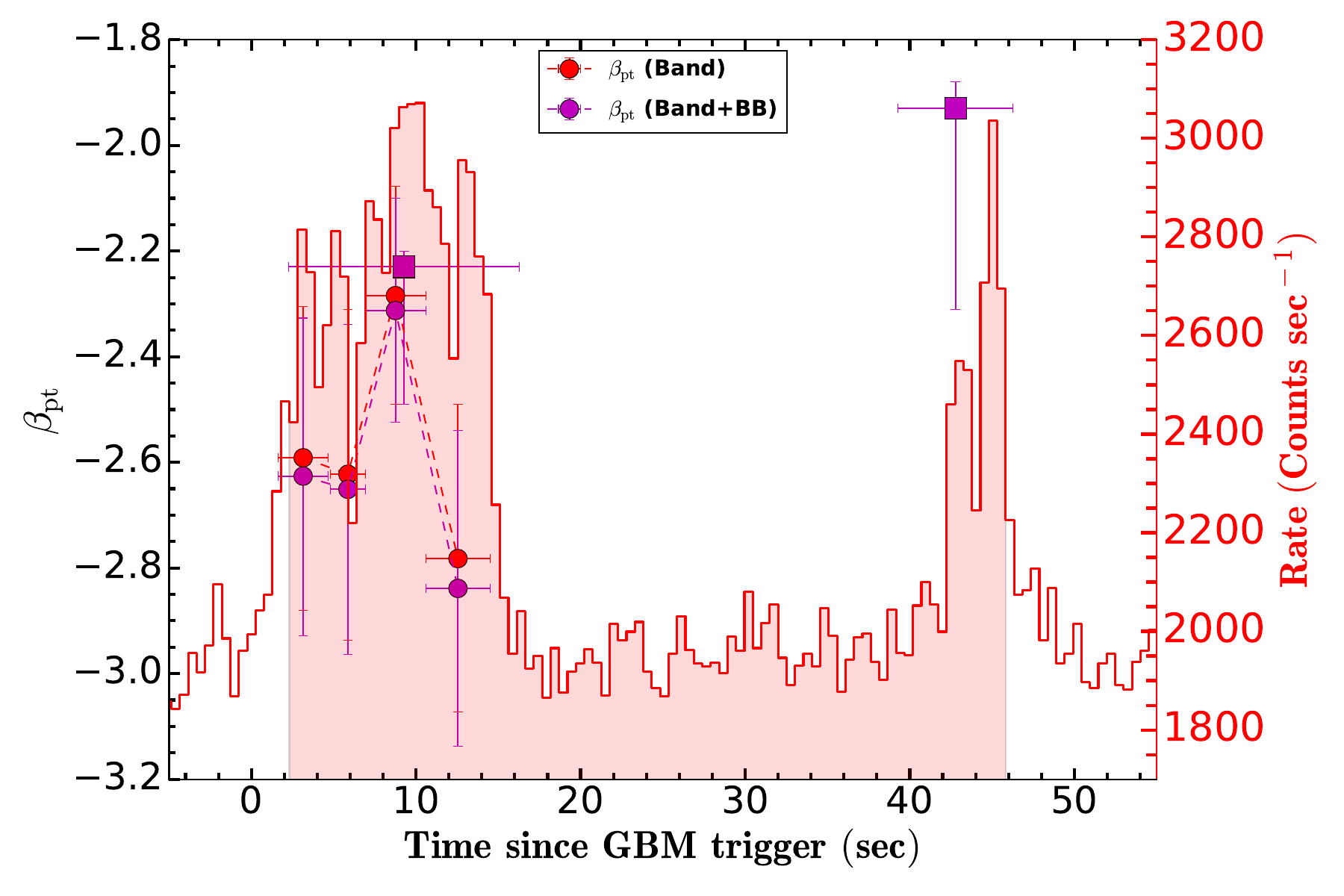}
\includegraphics[scale=0.28]{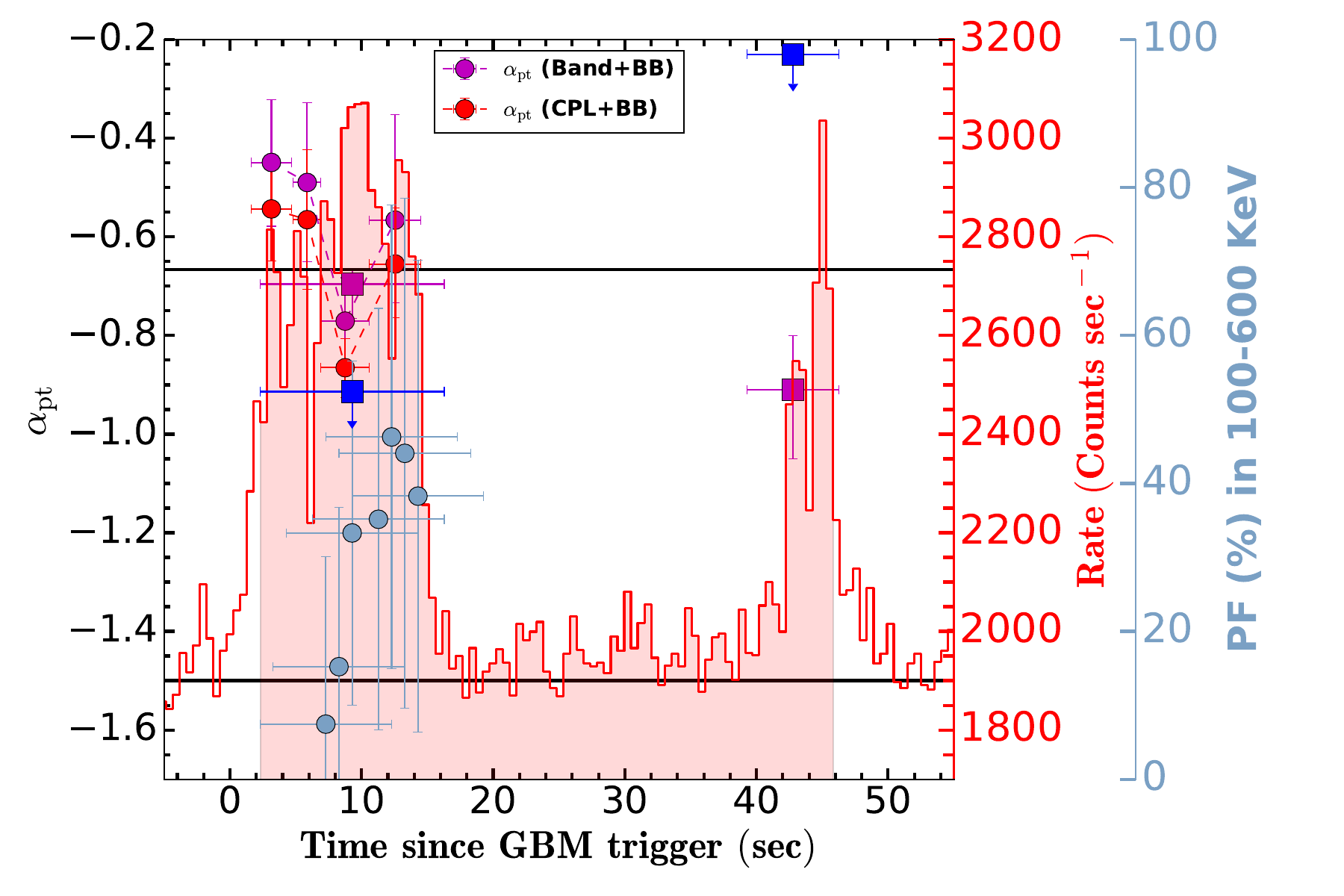}
\includegraphics[scale=0.28]{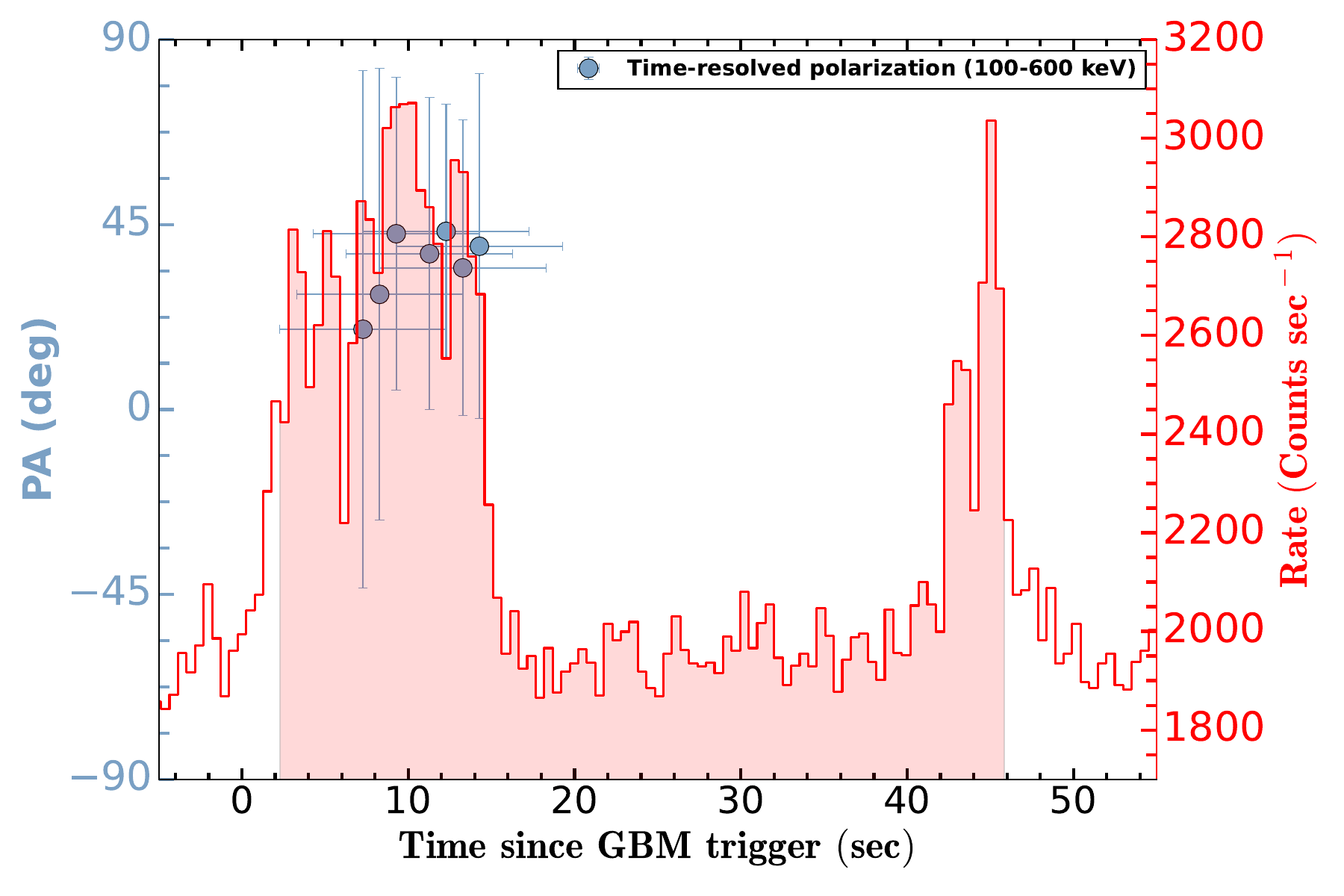}
\includegraphics[scale=0.28]{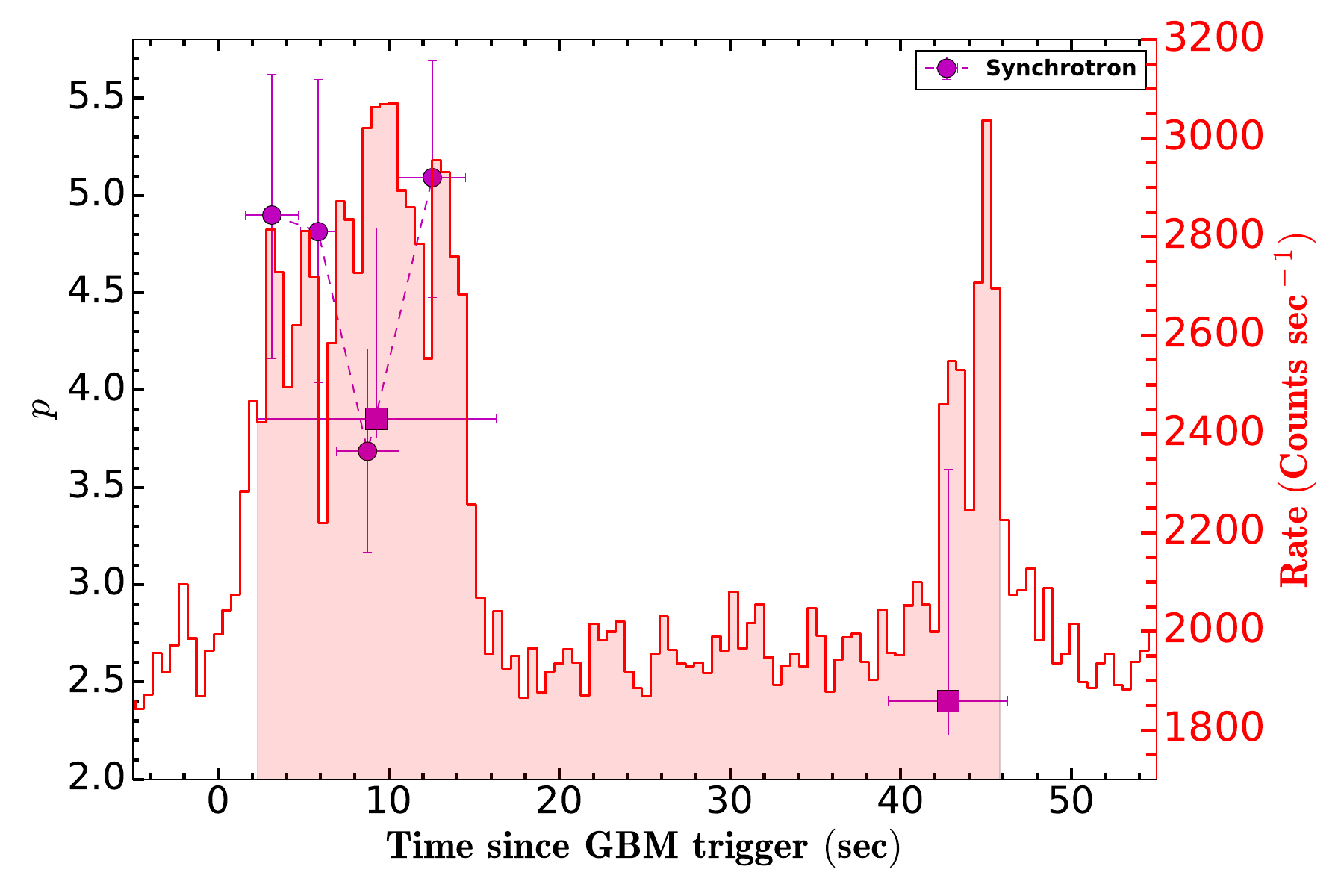}
\includegraphics[scale=0.28]{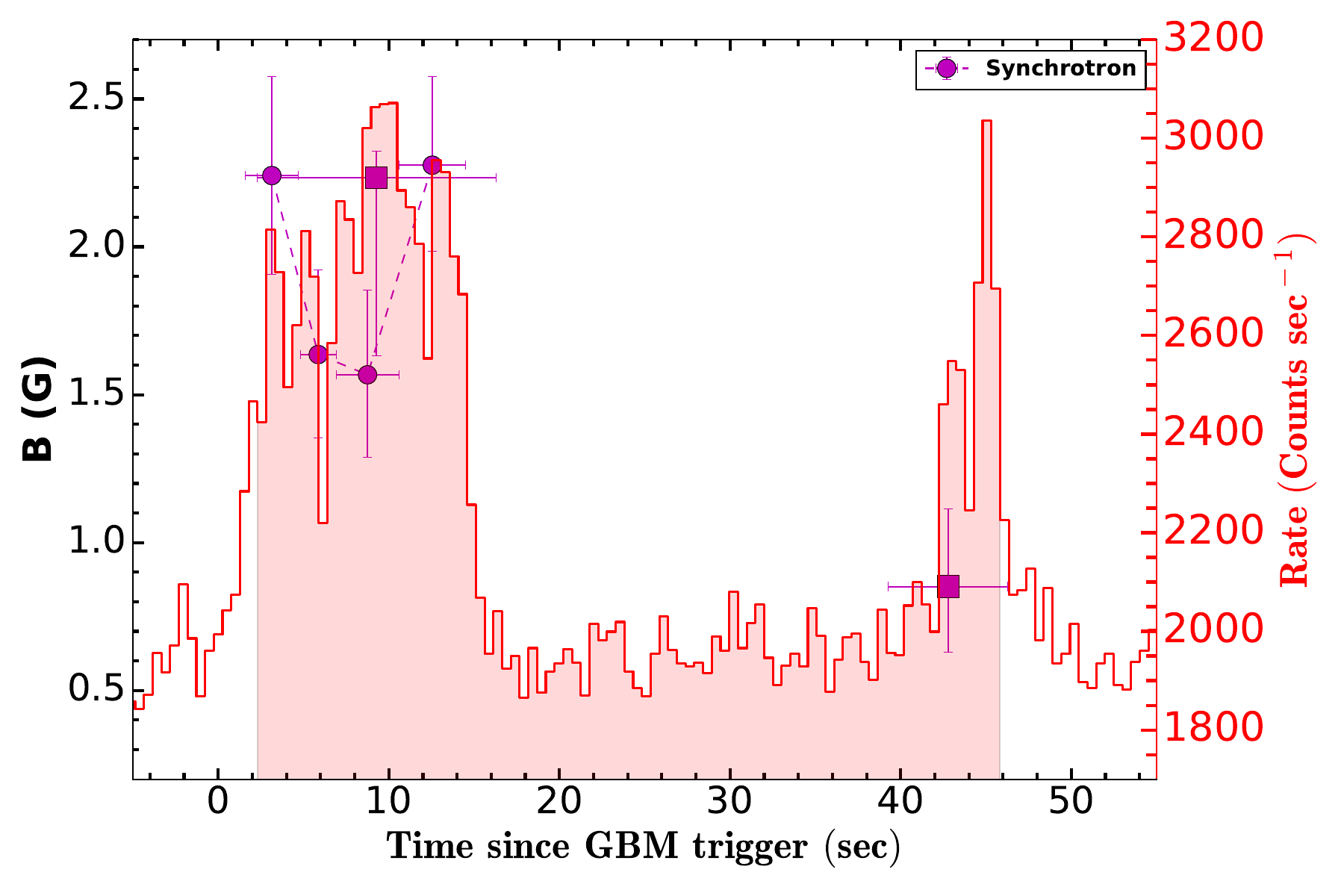}
\caption{{Time-resolved spectro-polarimetric characteristic of GRB 160325A.} {Top-left:} Temporal evolution of peak energy or cutoff energy obtained using empirical spectral fitting. {Top-right:} Temporal evolution of high energy photon index. {Middle-left:} Temporal evolution of low energy photon index. The black solid lines correspond to the theoretically predicted values of the low energy photon index from thin-shell synchrotron emission models. The pulsed-wise time-resolved polarization fraction is shown using blue squares. The right side y-scale (light blue) represents the evolution of polarization fraction over time obtained using time-resolved polarization analysis (sliding mode). {Middle-right:} Temporal evolution of polarization angle over time obtained using time-resolved polarization analysis (sliding mode). {Bottom-left:} Temporal evolution of the power-law index of the energy distribution of injected electron obtained using physical spectral fitting. {Bottom-right:} Temporal evolution of the strength of the magnetic field. Squares show the results for pulse-wise time-resolved spectro-polarimetric analysis.}
\label{TRS_GRB160325A}
\end{figure*}

\begin{figure*}
\centering
\includegraphics[scale=0.28]{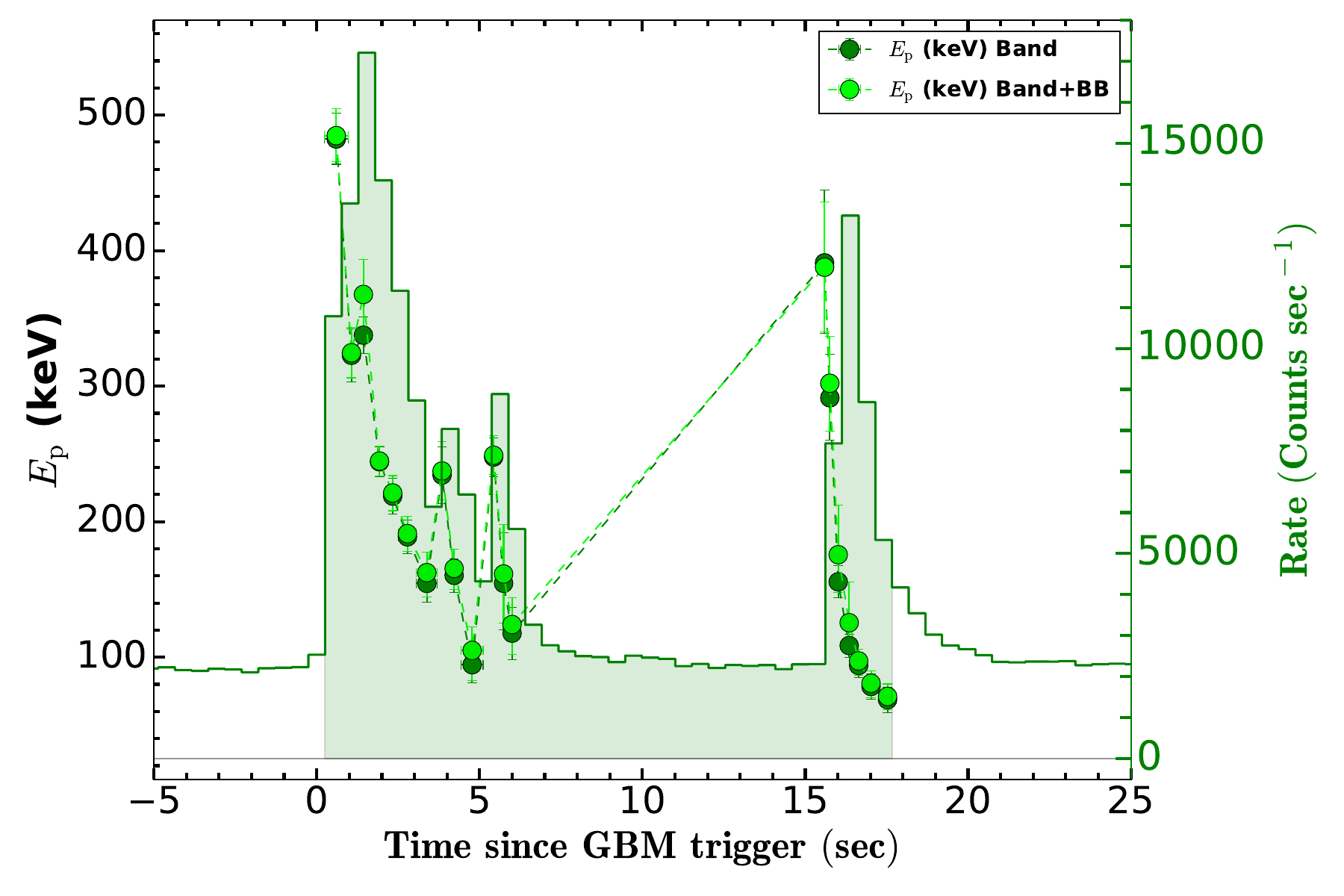}
\includegraphics[scale=0.28]{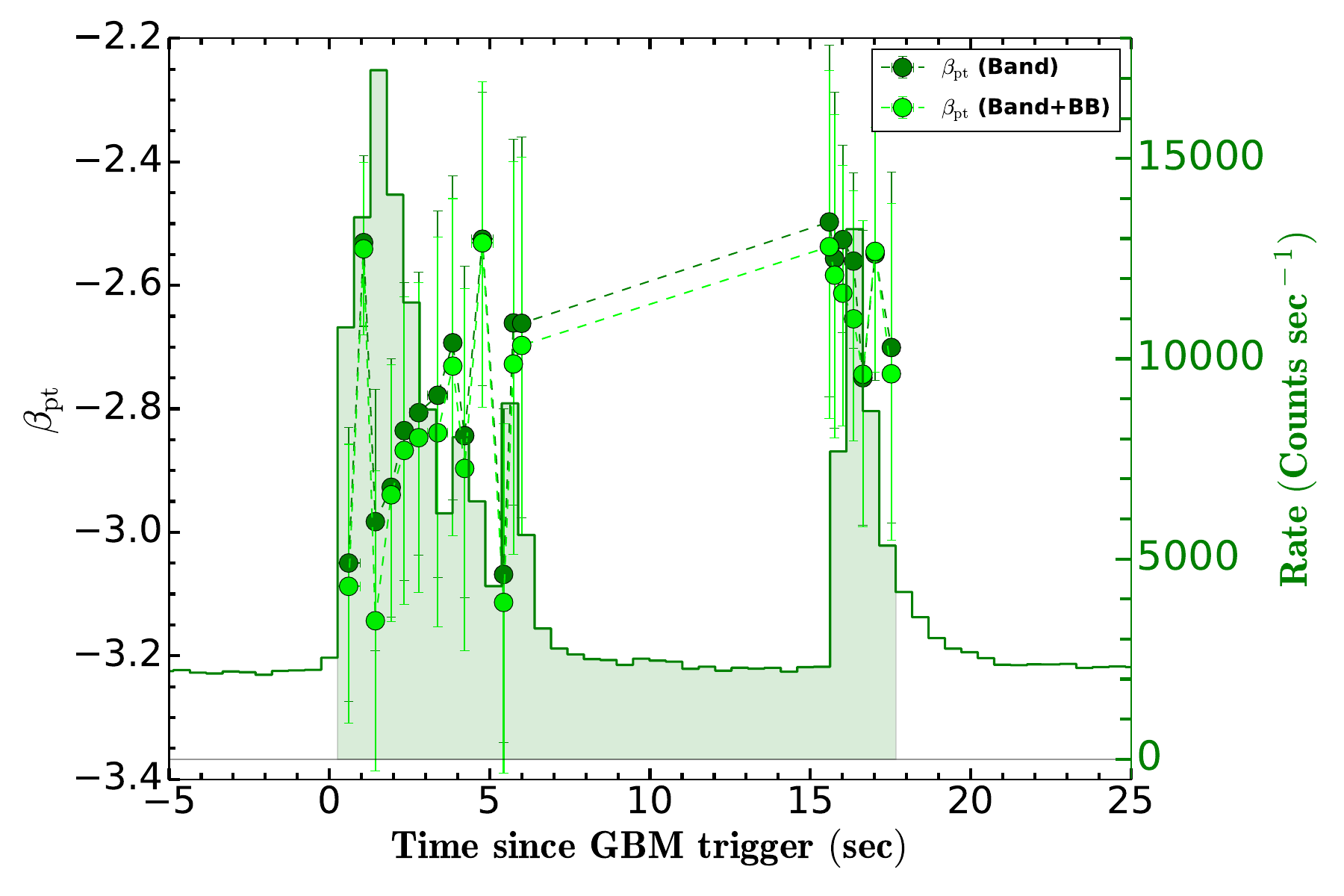}
\includegraphics[scale=0.28]{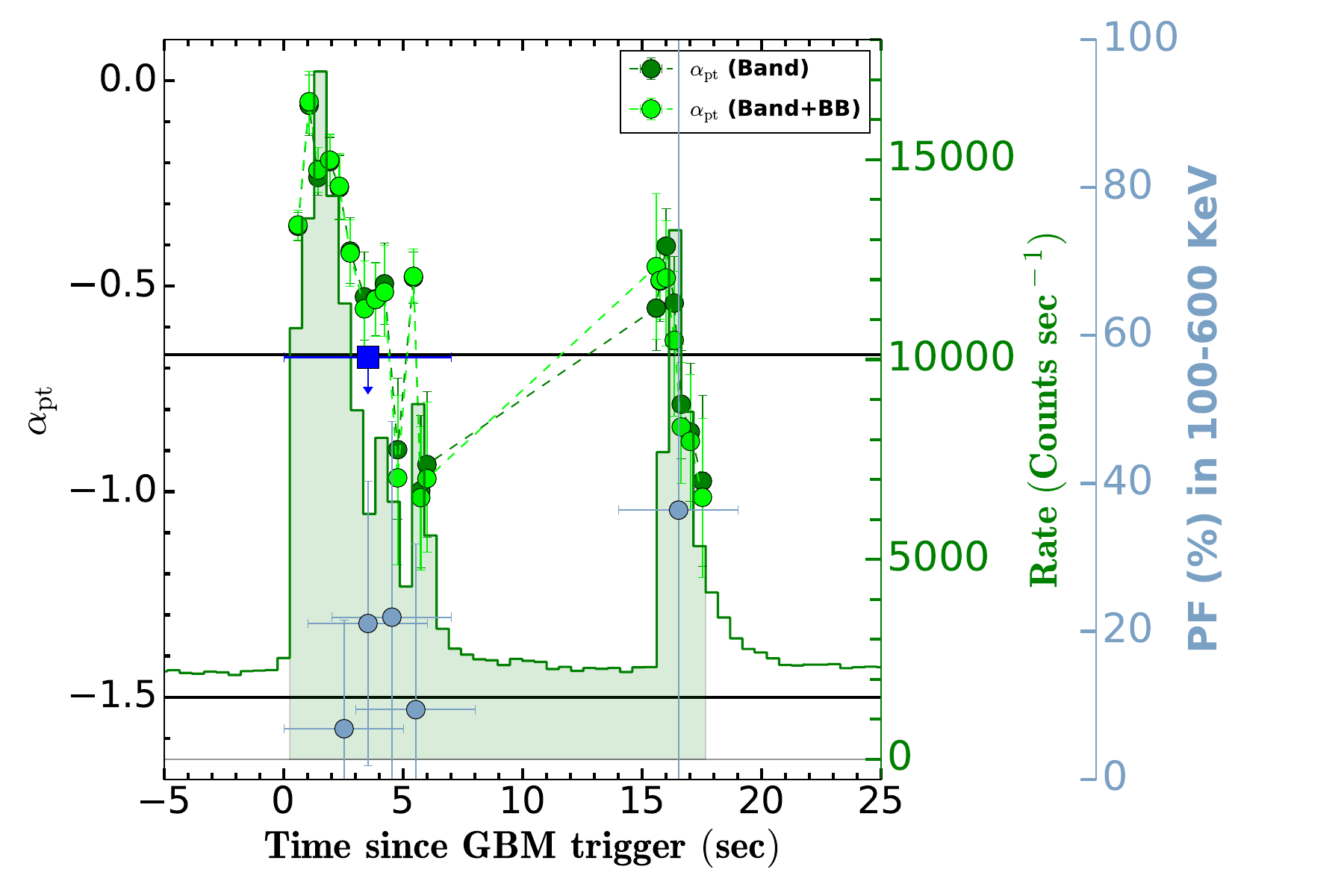}
\includegraphics[scale=0.28]{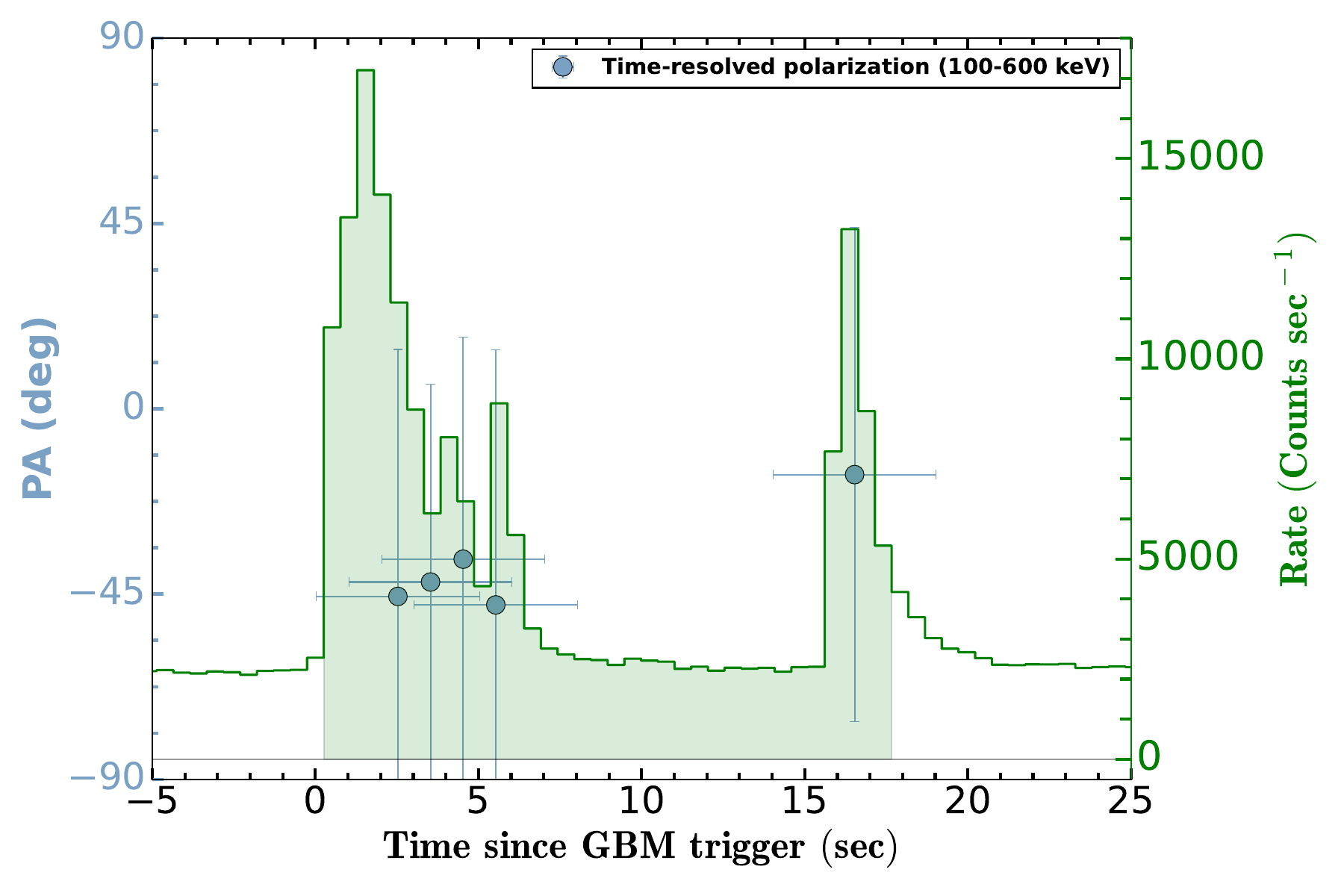}
\includegraphics[scale=0.28]{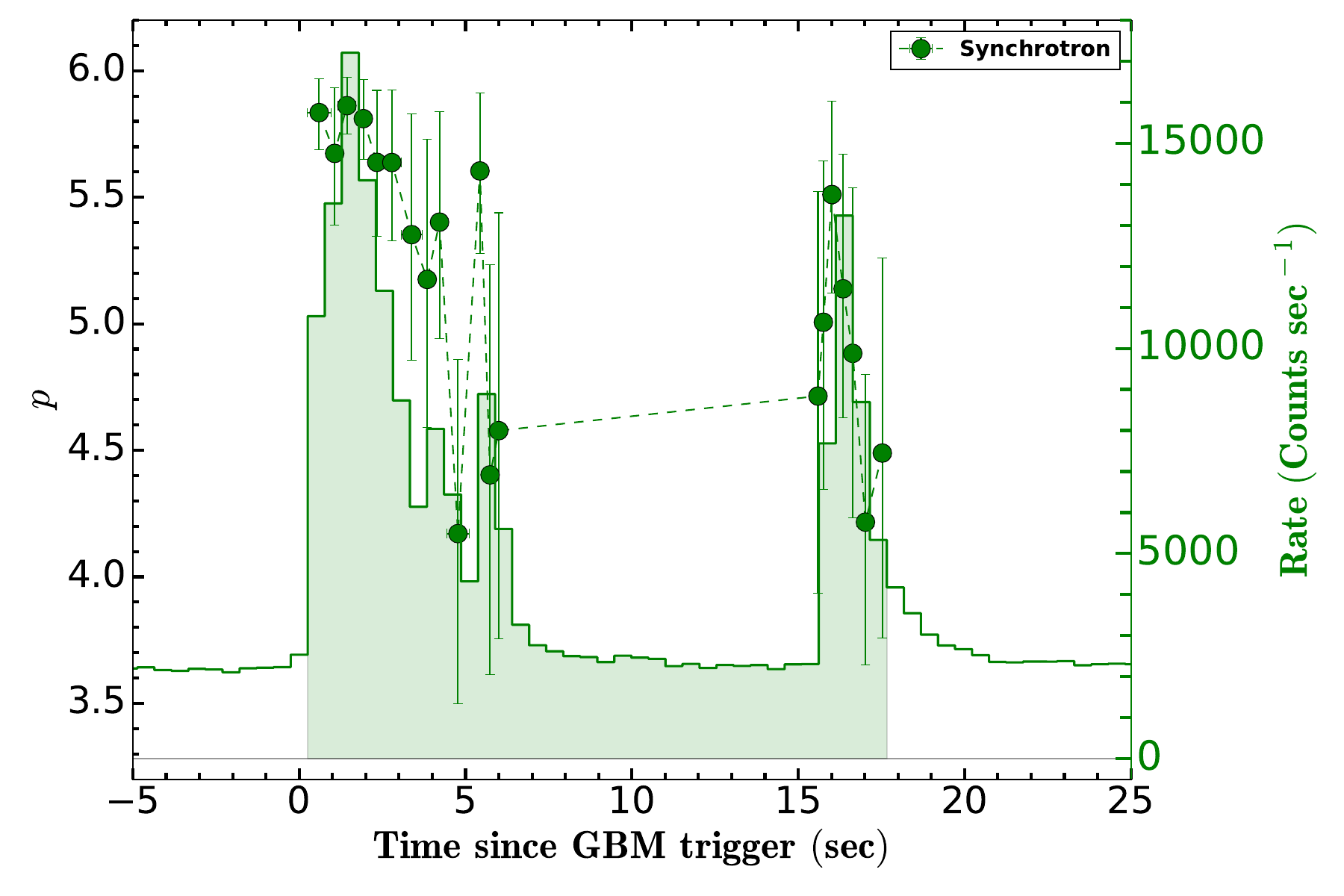}
\includegraphics[scale=0.28]{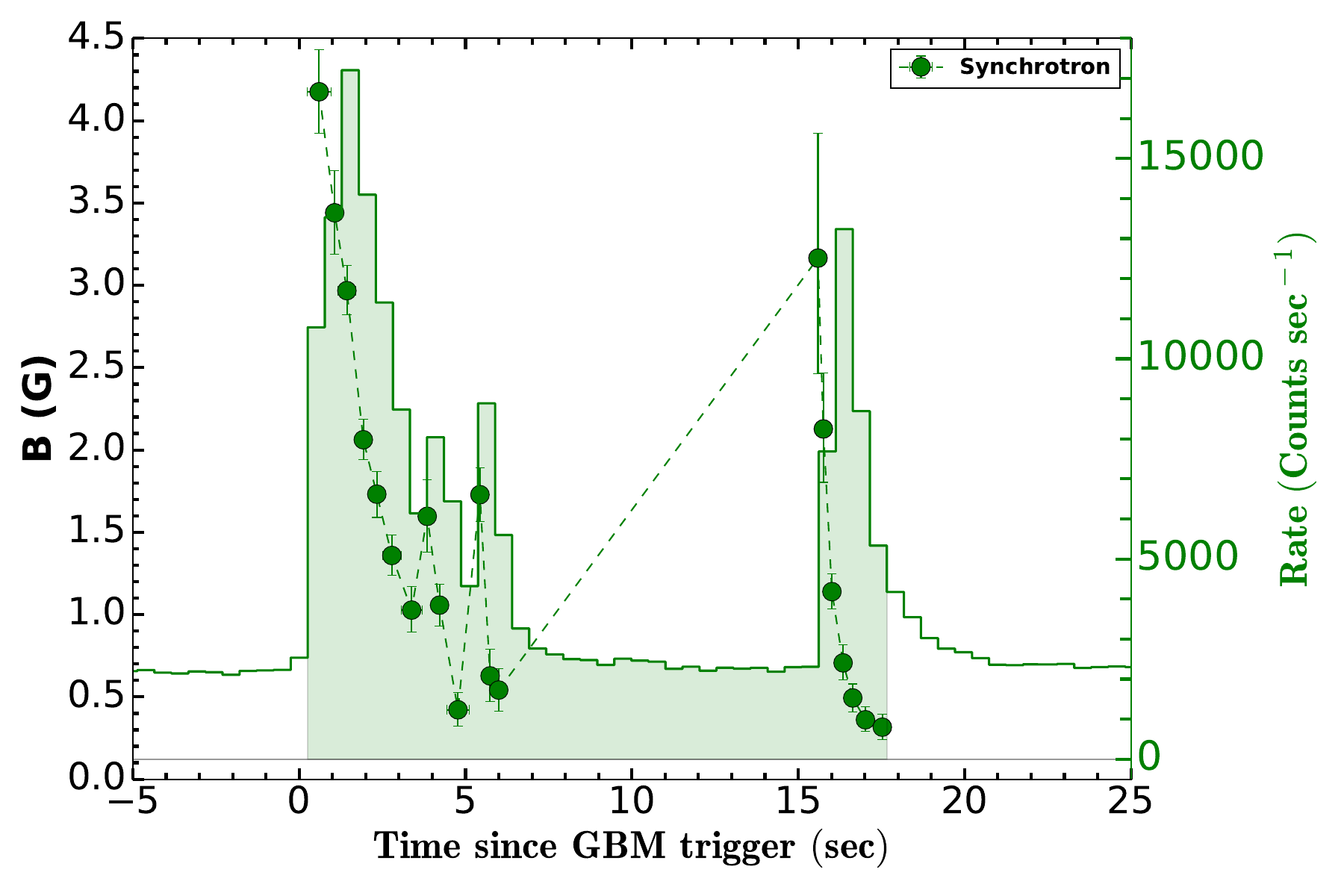}
\caption{{Time-resolved spectro-polarimetric characteristic of GRB 160802A.} {Top-left:} Temporal evolution of peak energy obtained using empirical spectral fitting of GRB 160802A. {Top-right:} Temporal evolution of high energy photon index. {Middle-left:} Temporal evolution of low energy photon index. The pulsed-wise time-resolved polarization fraction is shown using blue squares. The right side y-scale (light blue) represents the evolution of polarization fraction over time obtained using time-resolved polarization analysis (sliding mode). {Middle-right:} Temporal evolution of polarization angle over time obtained using time-resolved polarization analysis (sliding mode). {Bottom-left:} Temporal evolution of the power-law index of the energy distribution of injected electron obtained using physical spectral fitting. {Bottom-right:} Temporal evolution of the strength of the magnetic field.}
\label{TRS_GRB160802A}
\end{figure*}

\begin{figure*}[!ht]
\centering
\includegraphics[scale=0.28]{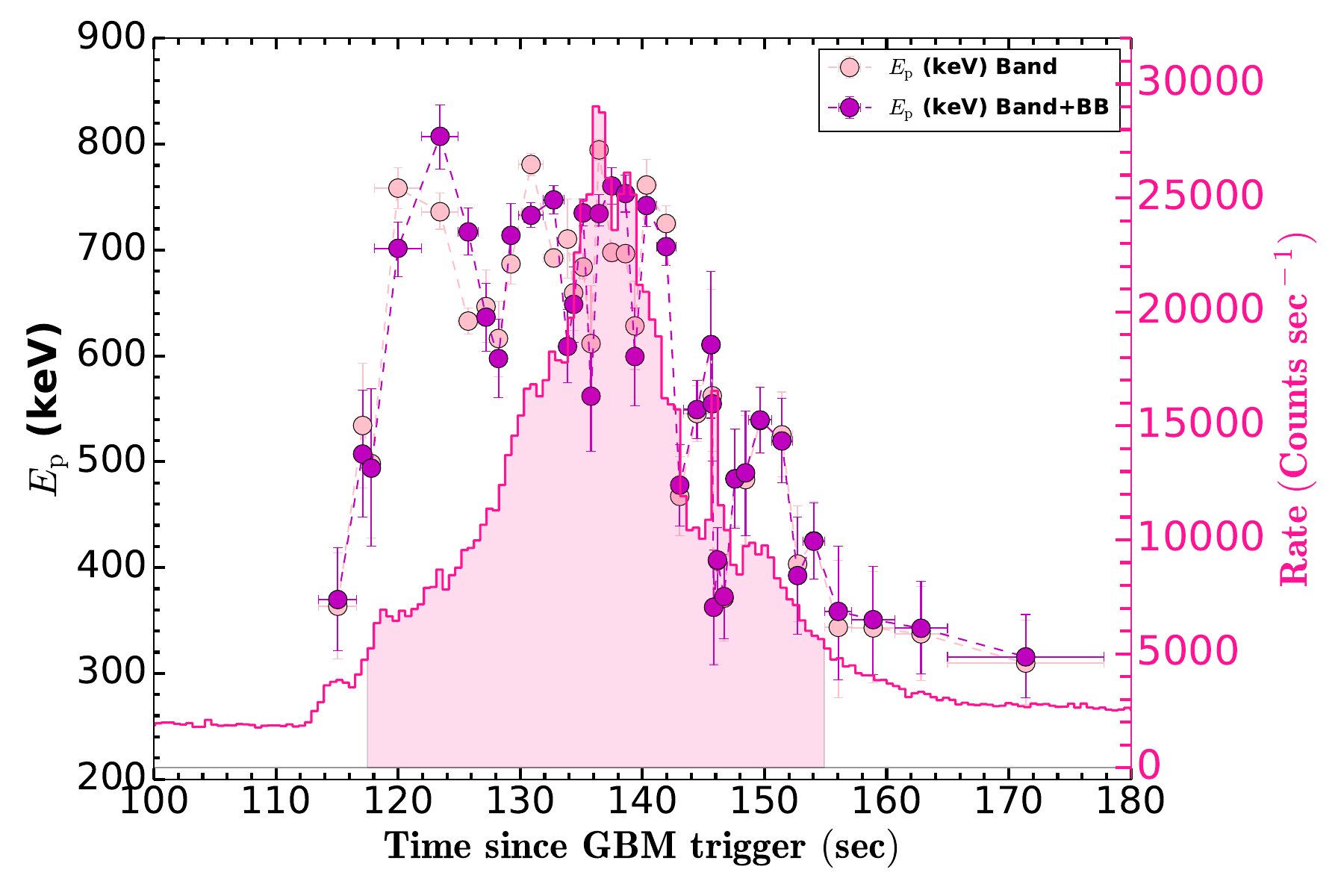}
\includegraphics[scale=0.28]{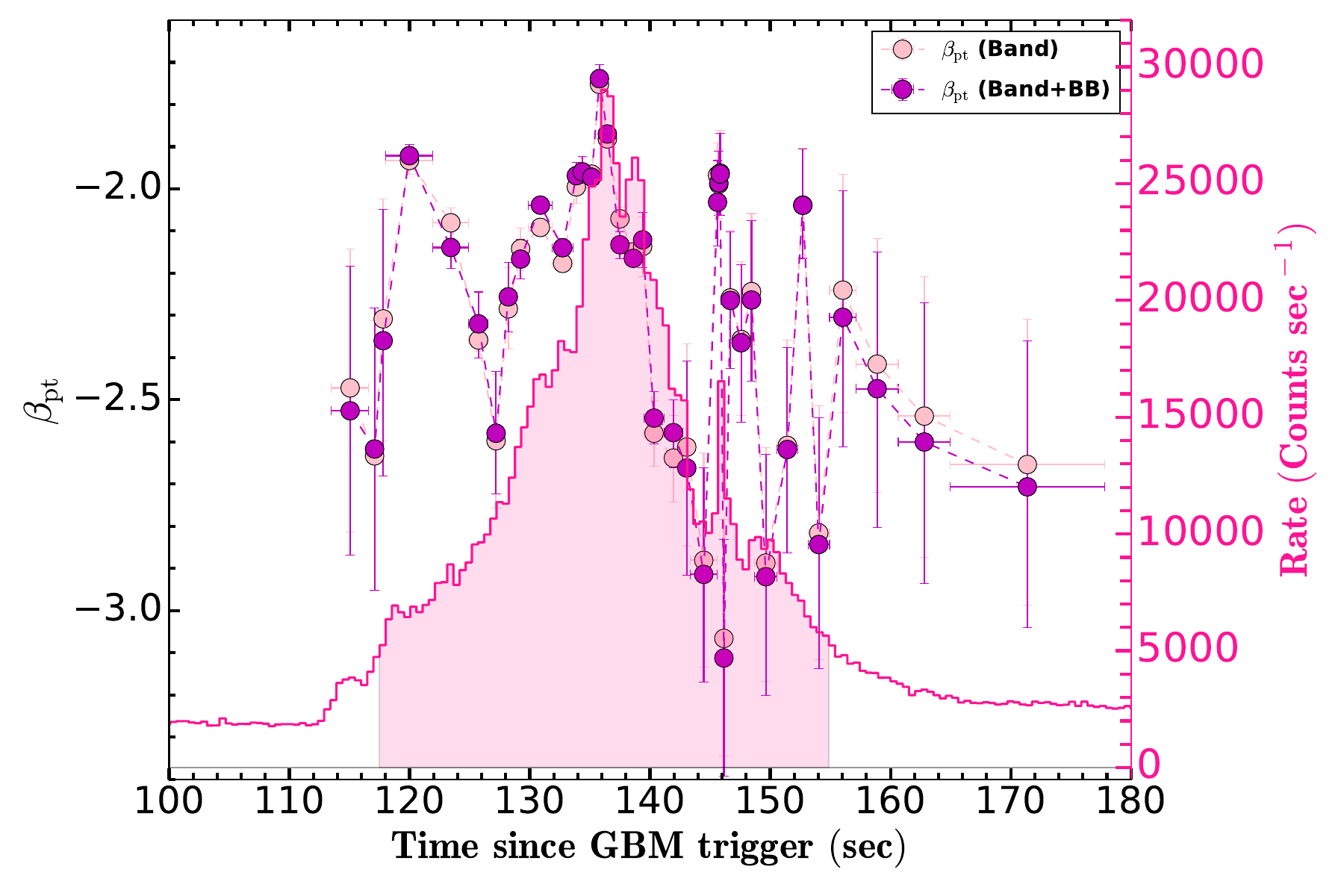}
\includegraphics[scale=0.28]{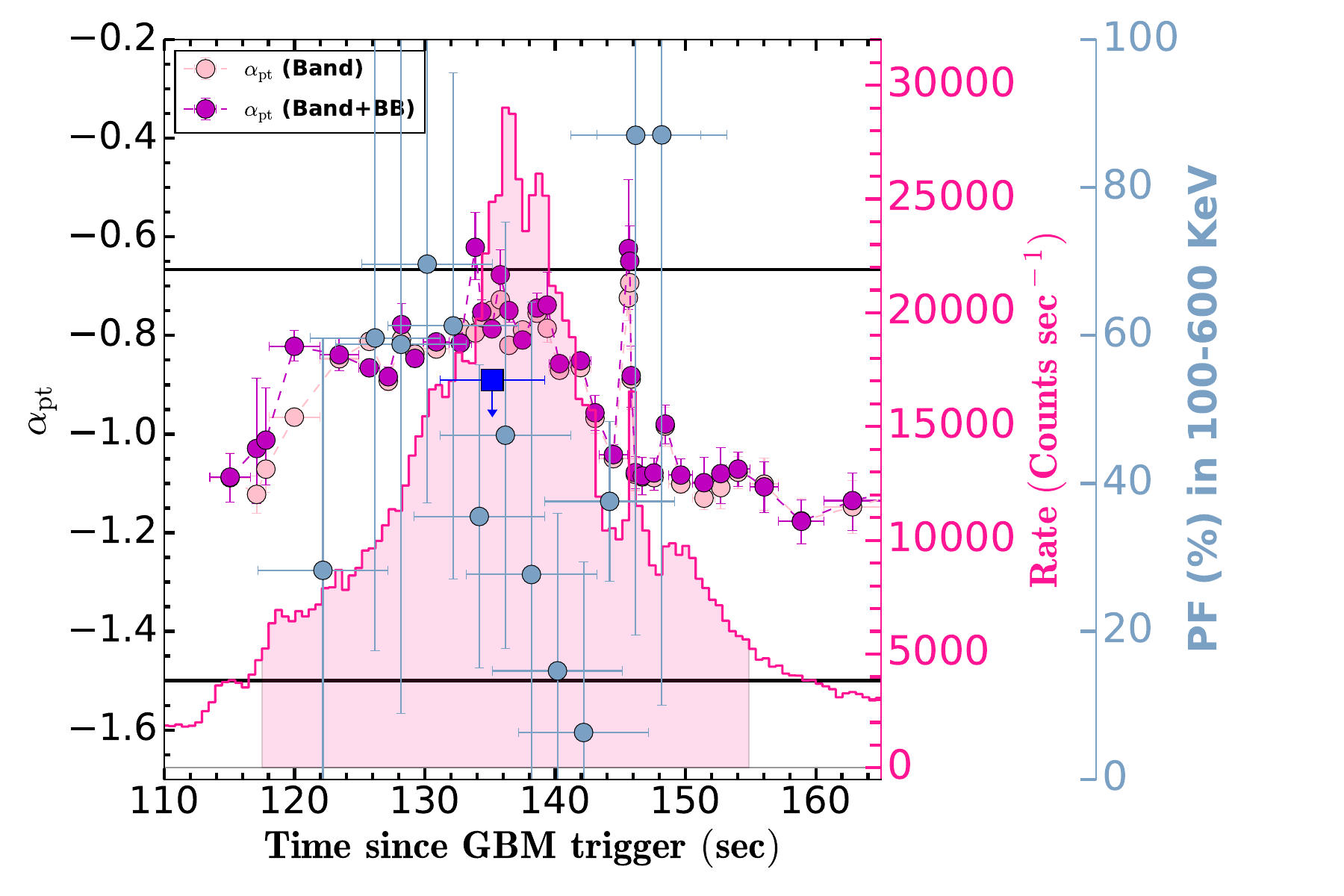}
\includegraphics[scale=0.28]{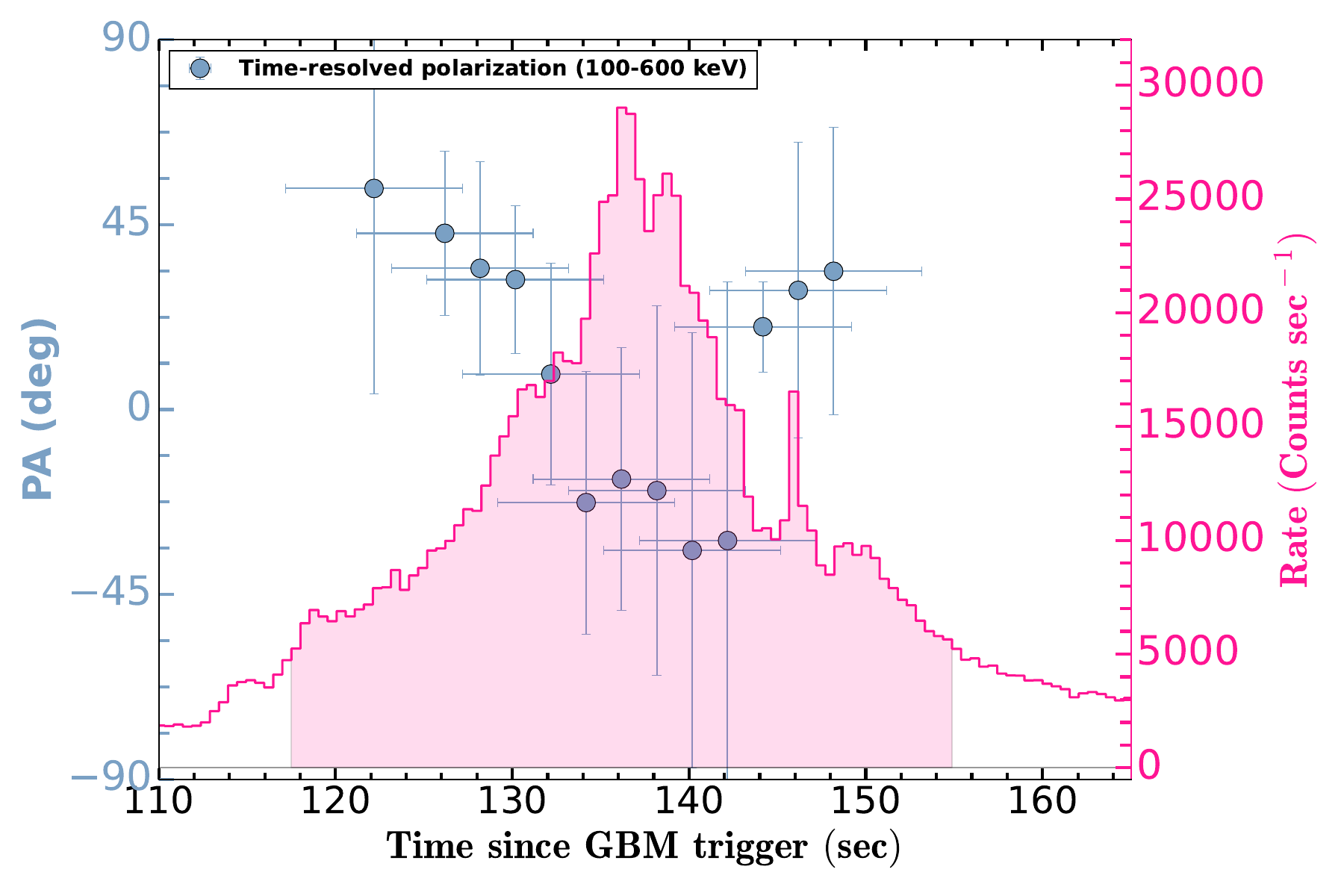}
\includegraphics[scale=0.28]{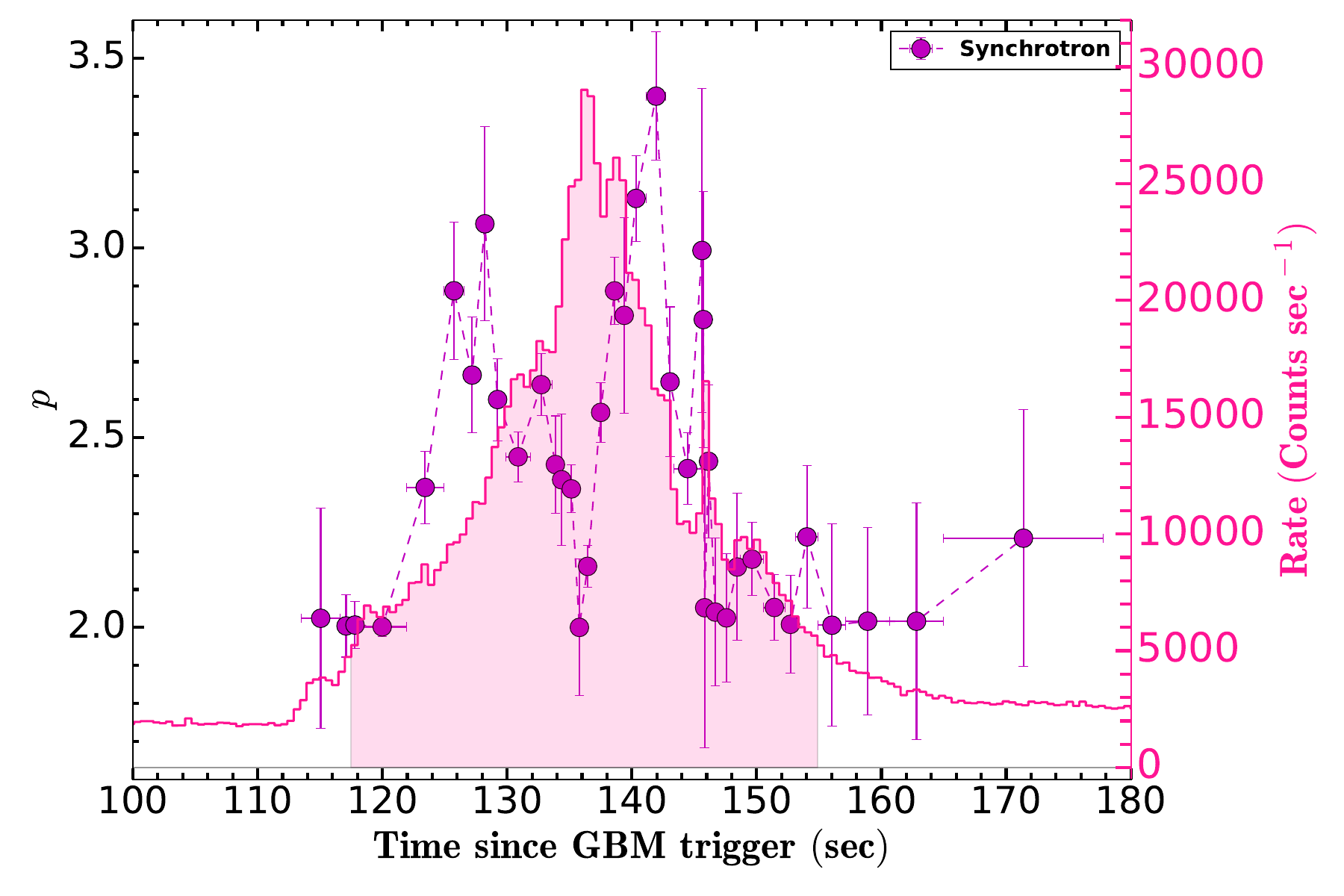}
\includegraphics[scale=0.28]{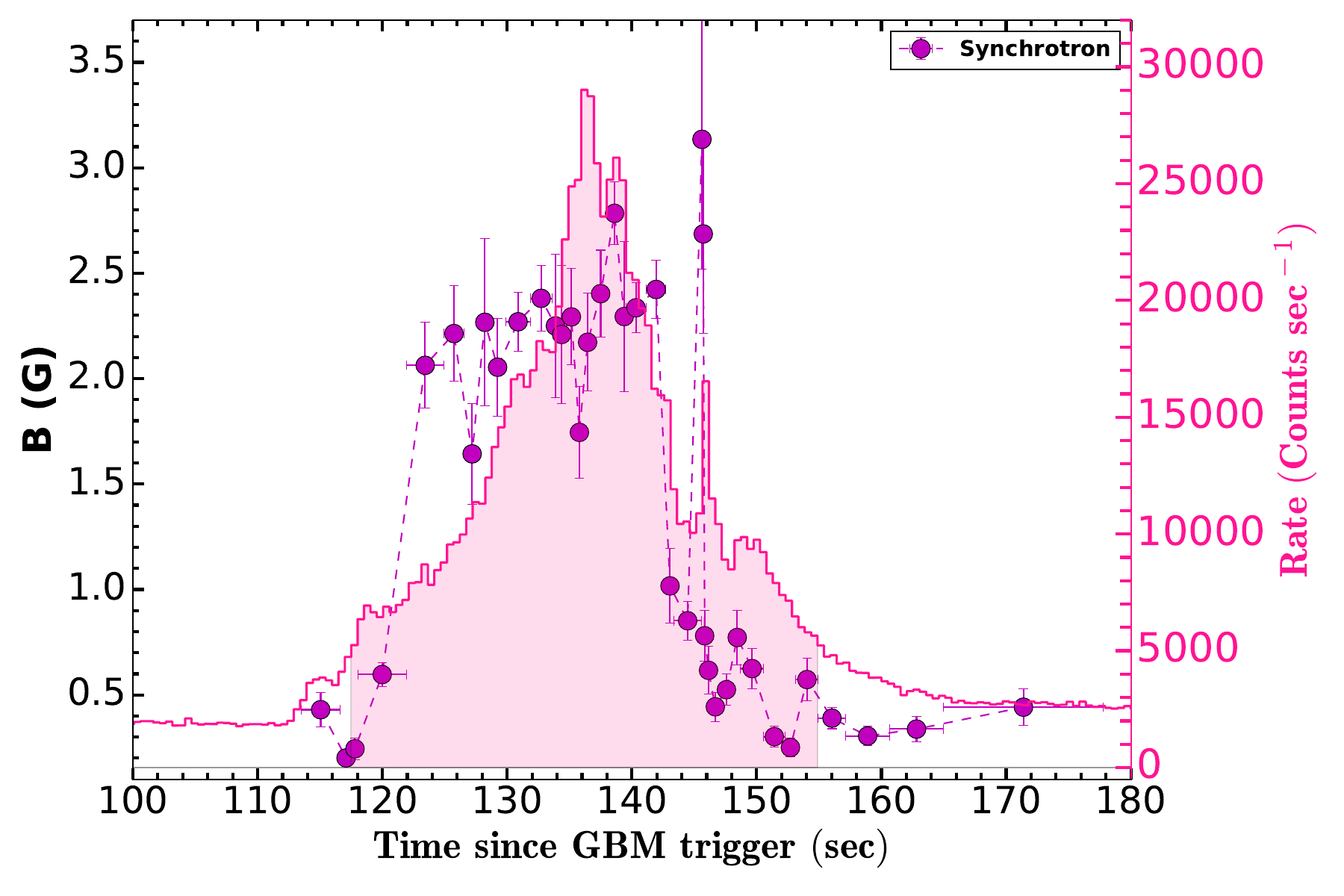}
\caption{{Time-resolved spectro-polarimetric characteristic of GRB 160821A.} {Top-left:} Temporal evolution of peak energy obtained using empirical spectral fitting of GRB 160821A. {Top-right:} Temporal evolution of high energy photon index. {Middle-left:} Temporal evolution of low energy photon index. The pulsed-wise time-resolved polarization fraction is shown using blue squares. The right side y-scale (light blue) represents the evolution of polarization fraction over time obtained using time-resolved polarization analysis (sliding mode). {Middle-right:} Temporal evolution of polarization angle over time obtained using time-resolved polarization analysis (sliding mode). {Bottom-left:} Temporal evolution of the power-law index of the energy distribution of injected electron obtained using physical spectral fitting. {Bottom-right:} Temporal evolution of the strength of the magnetic field.}
\label{TRS_GRB160821A}
\end{figure*}

We used the derived spectral parameters using time-resolved spectral analysis to study their evolution and correlation among them. The spectral evolution of empirical parameters \Ep, low and high energy photon indices for GRB 160325A, GRB 160802A, and GRB 160821A is presented in Figures \ref{TRS_GRB160325A}, \ref{TRS_GRB160802A}, and \ref{TRS_GRB160821A}, respectively. The evolution of physical parameters (electron spectral index and magnetic field strength) obtained using synchrotron modeling is also shown in these figures. We observed that \Ep evolution of all three GRBs has an intensity tracking behavior. Additionally, $\alpha_{\rm pt}$ evolution for GRB 160802A and GRB 160821A have the same tracking behavior, supporting a double-tracking nature. The correlation among different empirical and physical spectral parameters of time-resolved spectral parameters was also studied. The correlation results between different model parameters are listed in the appendix in Table \ref{tab:correlation}. Our correlation analysis indicates that the peak energy of the burst (obtained using empirical fitting) is strongly correlated with flux evolution for all three GRBs. We also observed that $\alpha_{\rm pt}$ is strongly correlated with flux evolution for GRB 160802A and GRB 160821A. However, it is anti-correlated for GRB 1603025A (correlation analysis for GRB 160325A is not statistically significant due to less number of available bins). The physical parameters B and $p$ calculated using synchrotron modeling are found to be correlated with each other for GRB 160802A and GRB 160821A. Moreover, the physical parameters B and $p$ strongly correlate with empirical parameters \Ep, $\alpha_{\rm pt}$, and flux for GRB 160802A and GRB 160821A.

\subsection{Time-resolved polarization measurements}

Previous studies on the polarization of a few GRBs, such as GRB 100826A, GRB 160821A, and GRB 170114A, have suggested that their polarization properties could exhibit temporal evolution \citep{2011ApJ...743L..30Y, 2019ApJ...882L..10S, 2019A&A...627A.105B}. However, it's important to note that these GRBs were observed using different instruments and analyzed through distinct pipelines. The observed hints regarding the evolution in polarization properties of GRB 100826A, GRB 160821A, and GRB 170114A were obtained using the GAP, \AstroSat/CZTI, and POLAR instruments, respectively. These findings imply that the polarization properties of GRBs may undergo intrinsic changes over time, potentially resulting in null or low polarization fractions in time-integrated polarization measurements. 

\begin{figure}[!ht]
\centering
\includegraphics[scale=0.28]{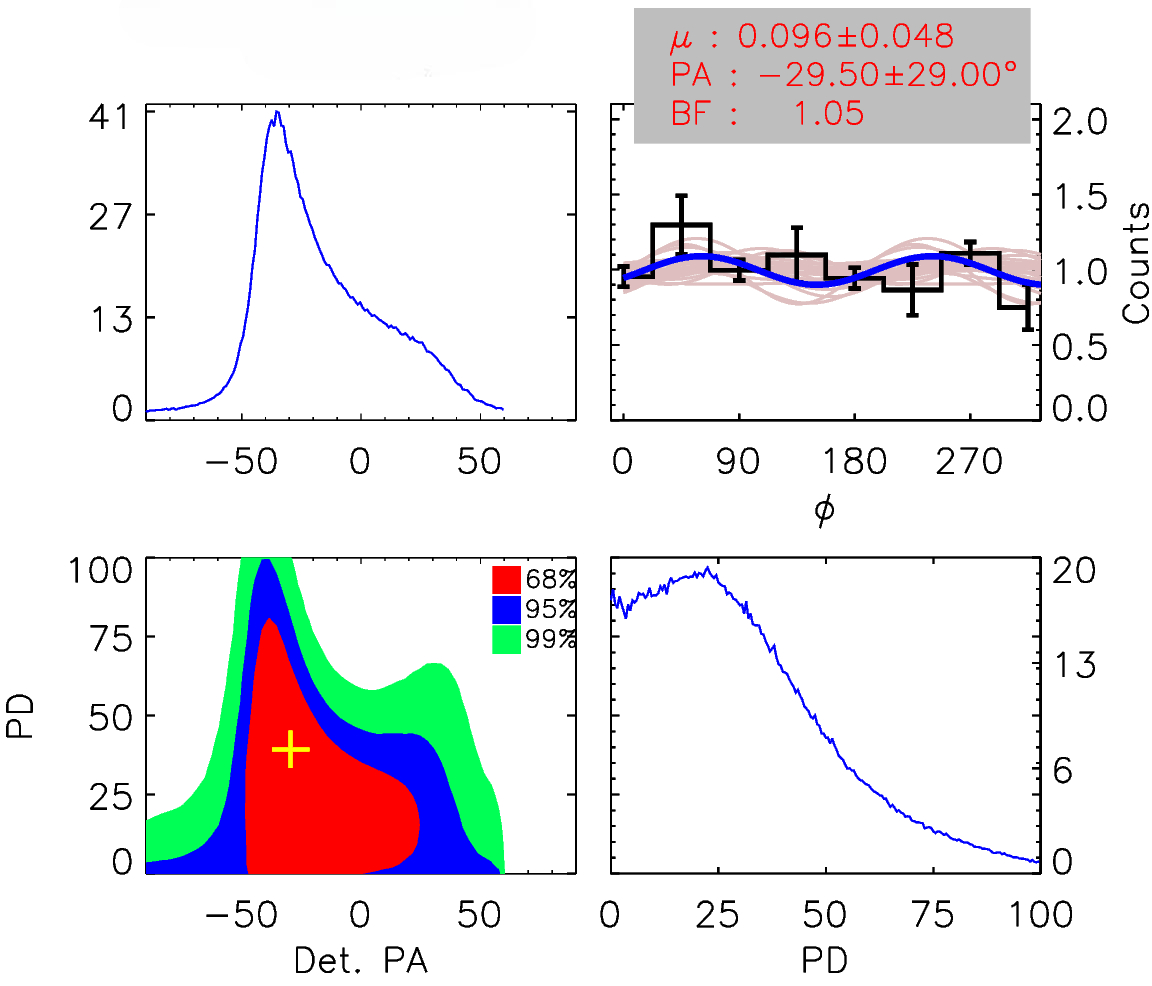}
\caption{An example of the posterior probability distribution (polarization angle in the top left and polarization degree in the bottom right) obtained using polarization analysis (using MCMC) of GRB 160623A (during the peak duration).In the top right panel, the modulation curve and the sinusoidal fit are illustrated by a solid blue line, accompanied by 100 random Markov Chain Monte Carlo (MCMC) iterations. In the bottom left panel, the confidence area for the polarization angle and degree is represented by red, blue, and green contours, corresponding to confidence levels of 68\%, 95\%, and 99\%, respectively.}
\label{Time_resolved_GRB160623A}
\end{figure}

In our recent five-year catalog paper \citep{2022ApJ...936...12C}, we highlighted a notable observation: approximately 75\% of GRBs exhibit low or null polarization fractions in our time-integrated polarization analysis. However, to ascertain whether these bursts are intrinsically unpolarized or if their polarization properties undergo changes within the bursts, leading to null or low polarization, a detailed time-resolved polarization analysis is imperative. In the present work, we studied a detailed time-resolved polarization analysis of five GRBs detected in the first year of operation of \AstroSat. We applied two distinct binning techniques to the GRB light curves and subsequently conducted polarization measurements. In case the GRB light curve has more than one pulse (for example, GRB 160325A and GRB 160802A), we selected individual pulses for time-resolved polarization measurement. Conversely, for GRBs exhibiting a single pulse, namely GRB 160623A, GRB 160703A, and GRB 160821A, we selected the peak duration of the burst. The results of our time-resolved polarization measurement are tabulated in Table \ref{time-resolved_table}. Additionally, we present an illustrative example of the posterior probability distribution obtained through polarization analysis of GRB 160623A (during the peak duration) in Figure \ref{Time_resolved_GRB160623A}.

\begin{table*}[!ht]
\caption{The calculated values of time-resolved polarization fraction (pulsed or peak-wise time bins) of all the five bursts in 100-600 \keV.}
\label{time-resolved_table}
\begin{scriptsize}
\begin{center}
\begin{tabular}{|c|c|c|c|c|}
\hline
 \bf GRB name& Time interval (sec) & No. of Compton events  & \bf PF (\%) & BF \\
\hline
GRB 160325A &2.28-16.28 & 556 & $<$ 52.42 & 0.75\\
GRB 160325A &39.28-46.28& 144& $<$ 98.04  & 2.78\\ \hline
GRB 160623A & 5.16-10.16 & 1089 & $<$ 58.86 & 1.05\\ \hline
GRB 160703A  & 0.22-9.22 & 172& unconstrained & 0.80\\ \hline
GRB 160802A &0.03-7.03 &1234 & $<$ 57.06  & 0.90\\
GRB 160802A &14.03-19.03 & 273 & unconstrained  & 0.79\\ \hline
GRB 160821A & 131.18-139.18 & 1246& $<$ 53.98 & 1.87\\ 
\hline
\end{tabular}
\end{center}
\end{scriptsize}
\end{table*}

\begin{figure*}[!ht]
\centering
\includegraphics[scale=0.35]{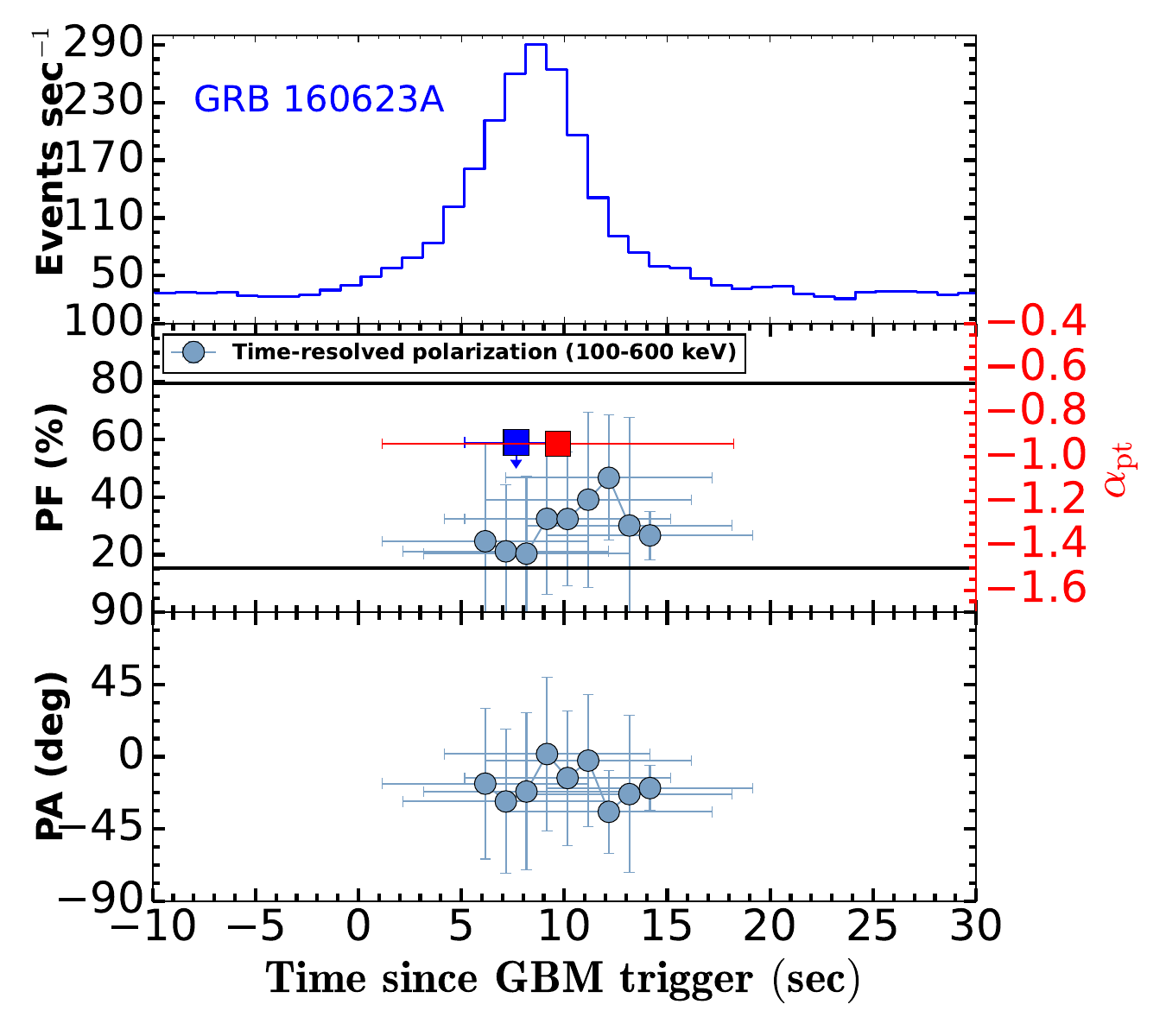}
\includegraphics[scale=0.35]{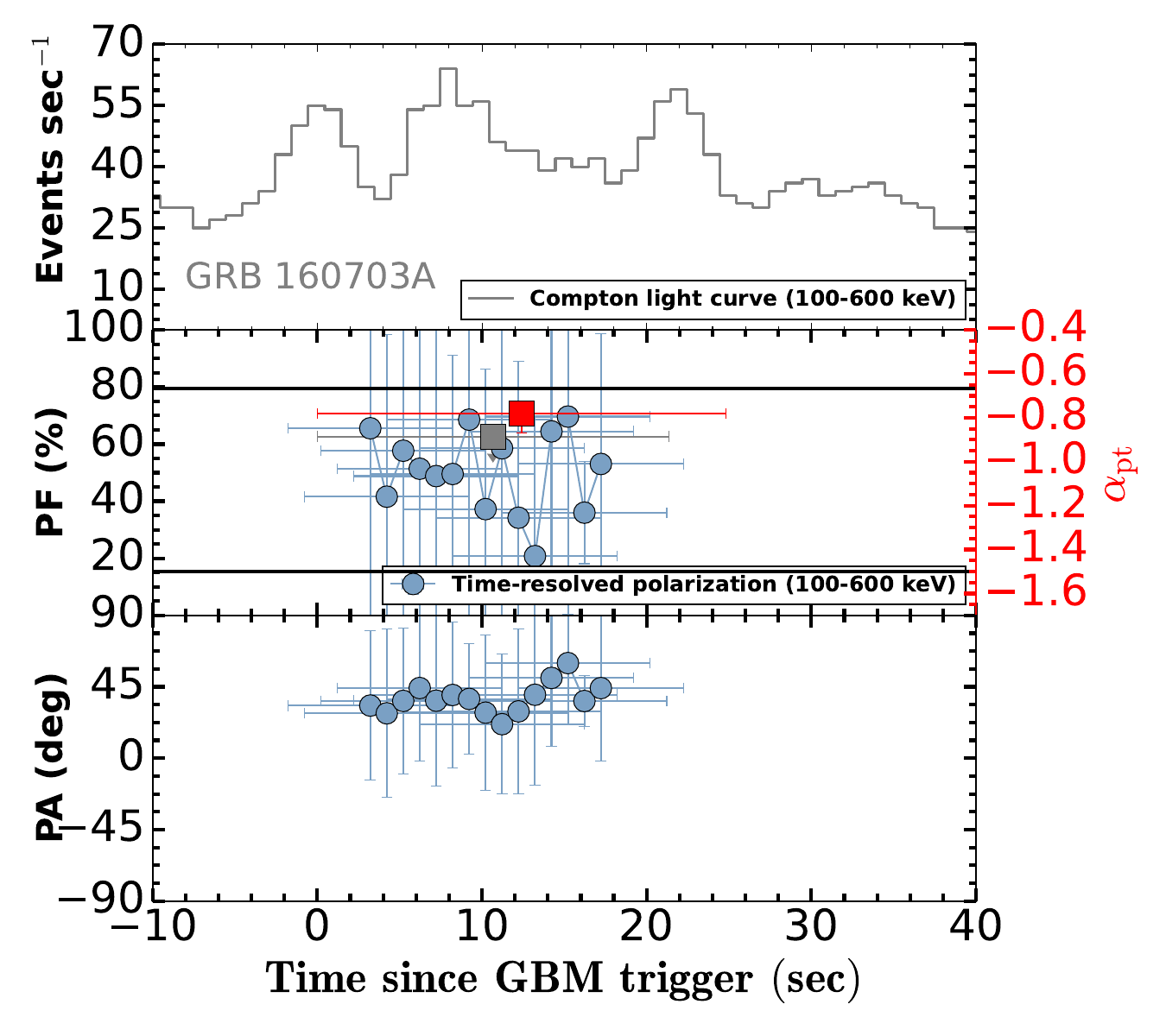}
\caption{Time-resolved polarization analysis of GRB 160623A (left) and GRB 160703A (right). {Top panels:} 1-sec bin size Compton light curves obtained using CZTI data. {Middle panels:} The evolution of PF over time (time-sliding mode). The PF obtained during the peak or averaged analysis is shown using blue (GRB 160623A) and grey (GRB 160703A) squares, respectively. The right side Y-scales (red) show the values of $\alpha_{\rm}$ of GRB 160623A and GRB 160703A, respectively. The black solid lines correspond to the theoretically predicted values of the low energy photon index from thin-shell synchrotron emission models. {Bottom panels:} The evolution of PA over time (time-sliding mode).}
\label{Time_resolved}
\end{figure*}

Further, we also selected the time bins using the sliding mode temporal binning method (since the GRB light curves exhibit rapid or irregular variations) with a bin width of 10 sec (for GRB 160325A, GRB 160623A, GRB 160703A, and GRB 160821A) or 5 sec (for GRB 160802A) for the time-resolved polarization measurements. We initially divided the light curve into smaller time intervals of the bin width from 0-10 sec or 0-5 sec and slid these average intervals across the entire duration of the burst with increasing order of 1 sec (GRB 160325A, GRB 160623A, GRB 160703A, and GRB 160802A) or 2 sec (GRB 160821A). Using the sliding mode binning, we calculated the average values of polarization parameters within each bin. The polarization results obtained using the temporal sliding binning along with pulsed/peak-wise binning algorithms are displayed in Figures \ref{TRS_GRB160325A}, \ref{TRS_GRB160802A}, \ref{TRS_GRB160821A}, and \ref{Time_resolved}, respectively. The time-resolved (pulsed-wise) analysis of the first pulses of GRB 160325A and GRB 160802A constrains the higher PF values (see Table \ref{time-resolved_table}), although sliding mode analysis of the same pulses indicates lower PF values. We noted that the polarization angles of GRB 160325A, GRB 160623A, GRB 160703A, and GRB 160802A obtained for different burst intervals remain within their respective error bars. This suggests that there is no substantial change in the polarization properties as these bursts evolve. However, we noted that the polarization angles of GRB 160821A changed twice within the burst, consistent with our previous results reported in 100-300 \keV with the previous polarization pipeline \citep{2019ApJ...882L..10S}. Our time-resolved polarization analysis gives a hint that the polarization properties of GRB 160821A depend on the temporal window of the burst. 

\subsection{Energy-resolved polarization measurements}

\begin{figure*}
\centering
\includegraphics[scale=0.32]{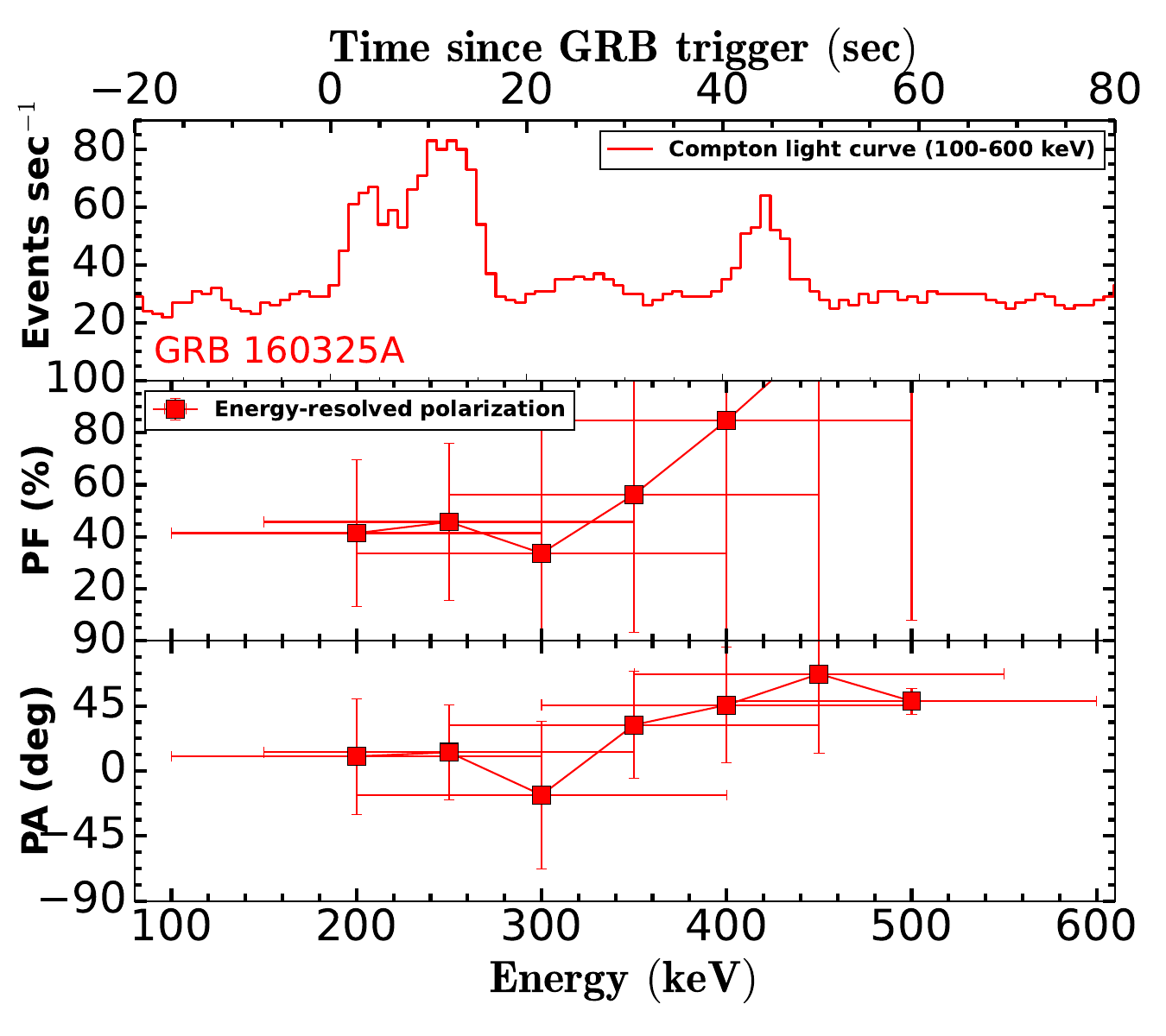}
\includegraphics[scale=0.32]{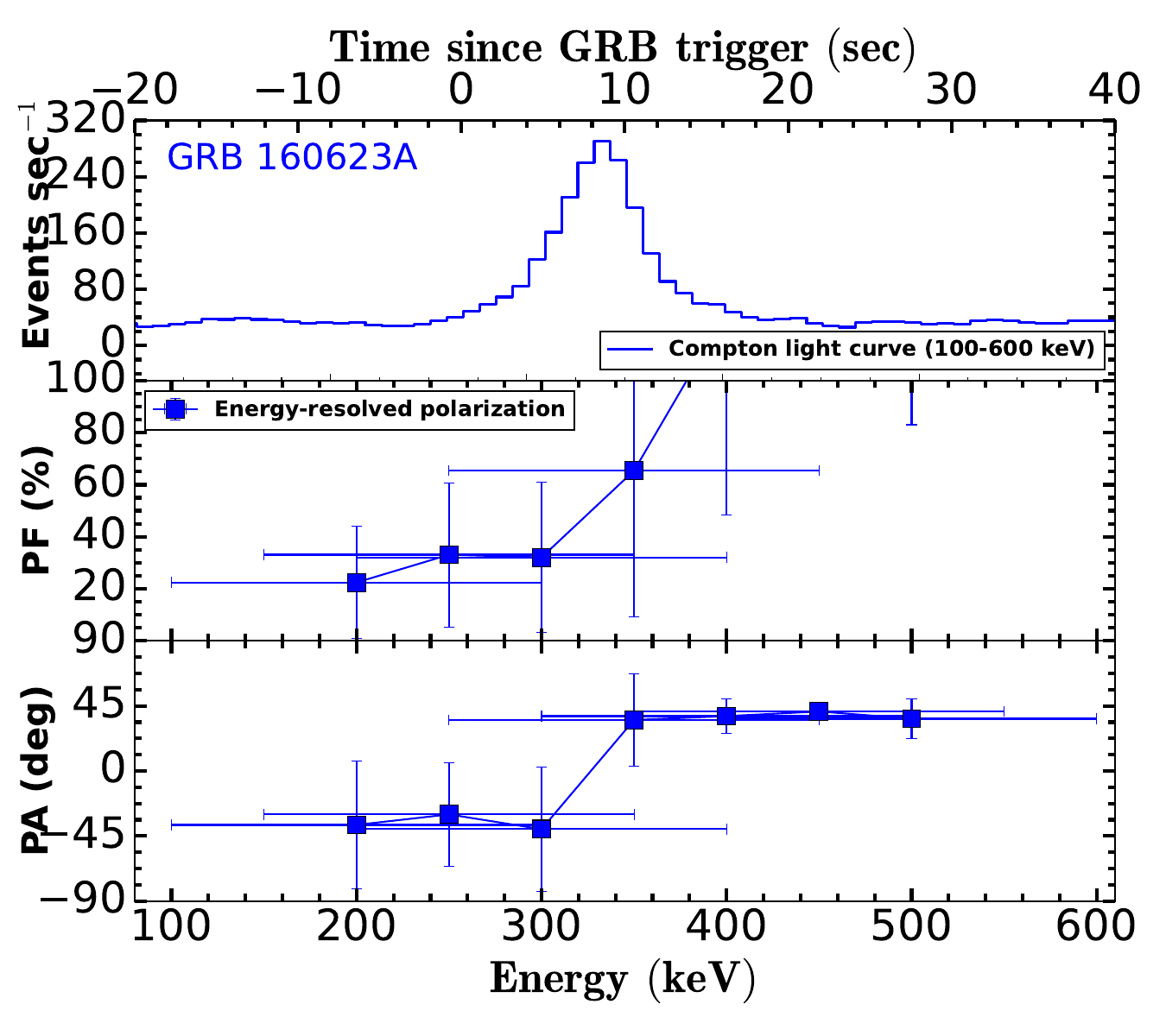}
\includegraphics[scale=0.32]{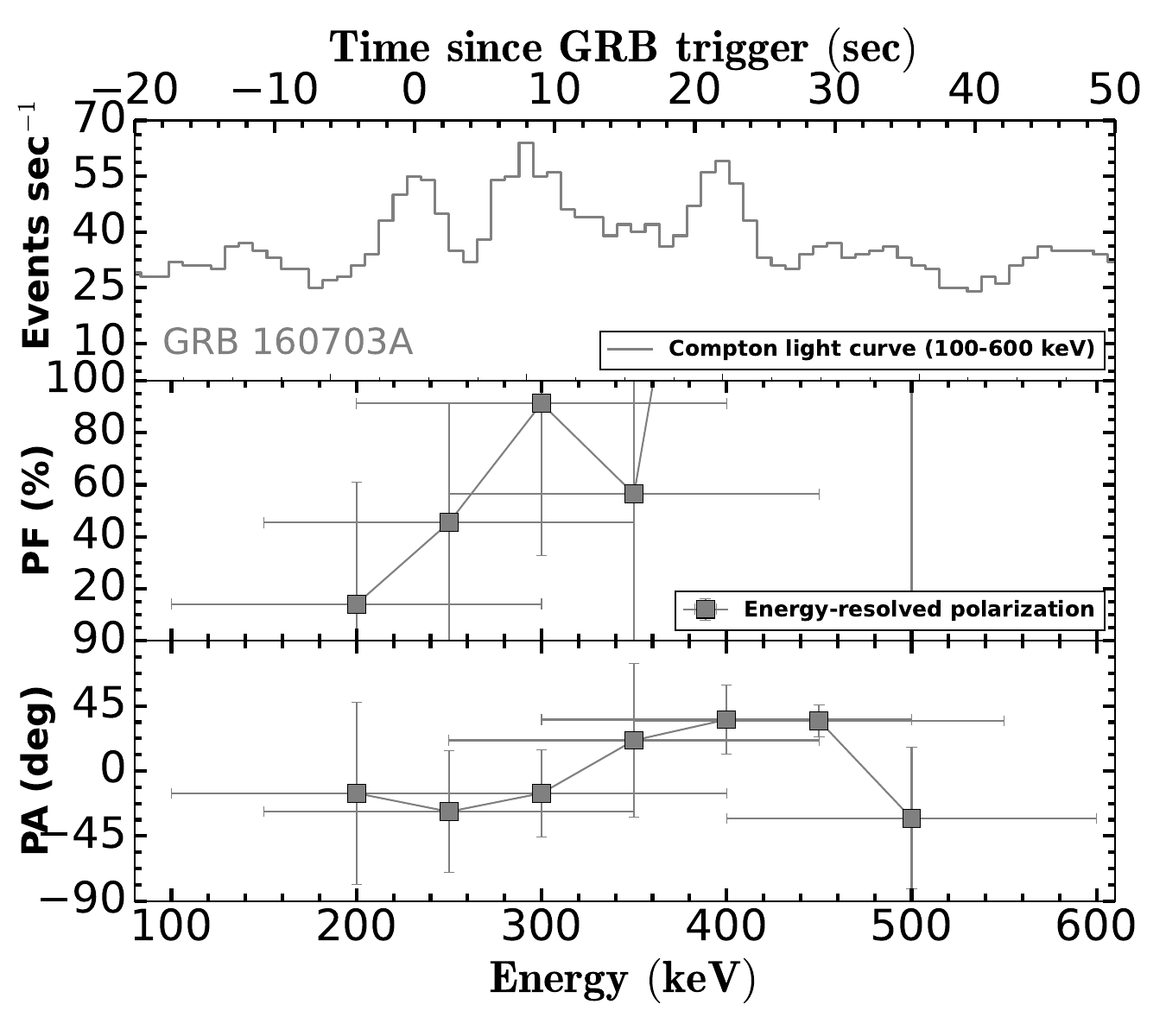}
\includegraphics[scale=0.32]{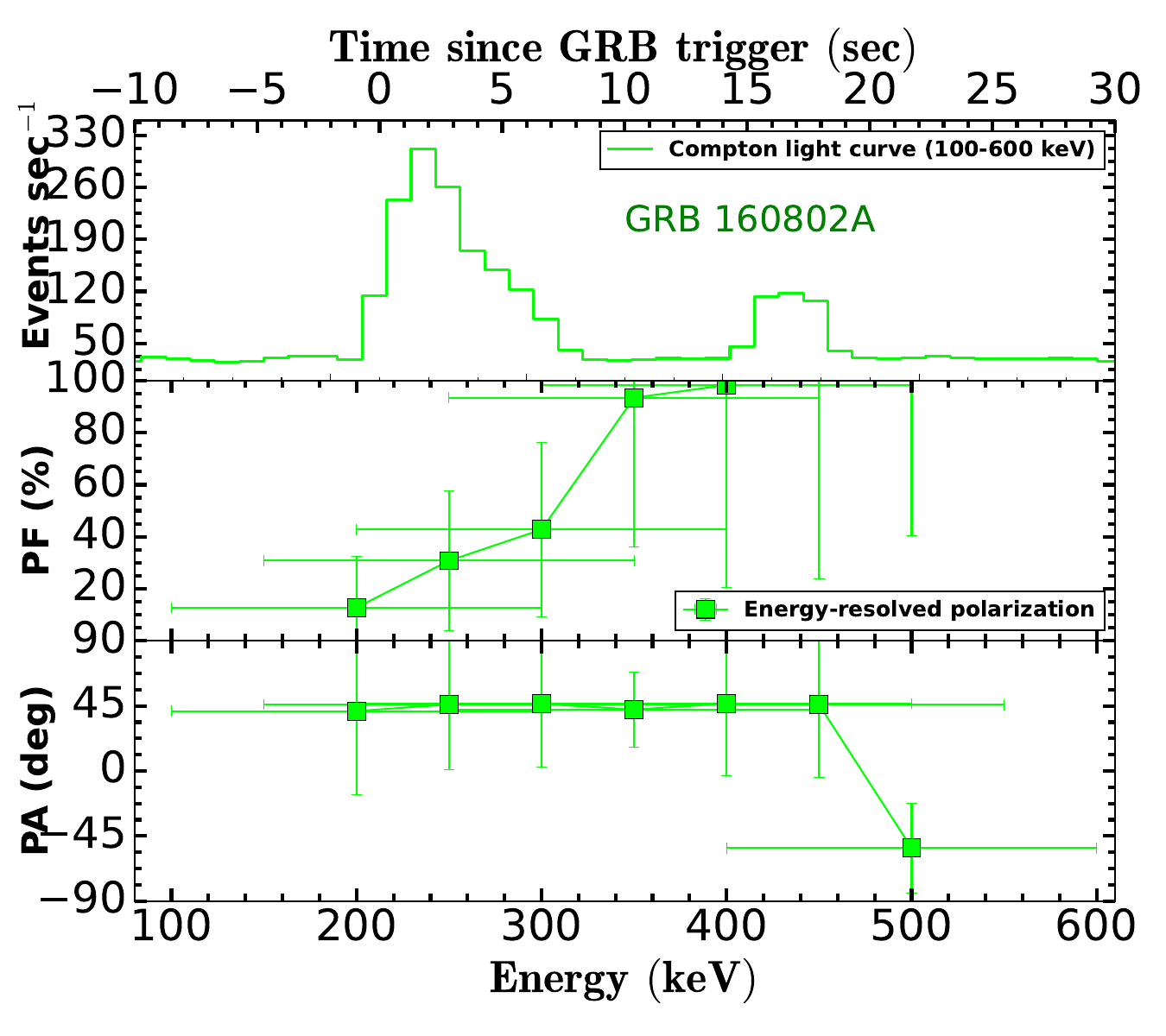}
\includegraphics[scale=0.32]{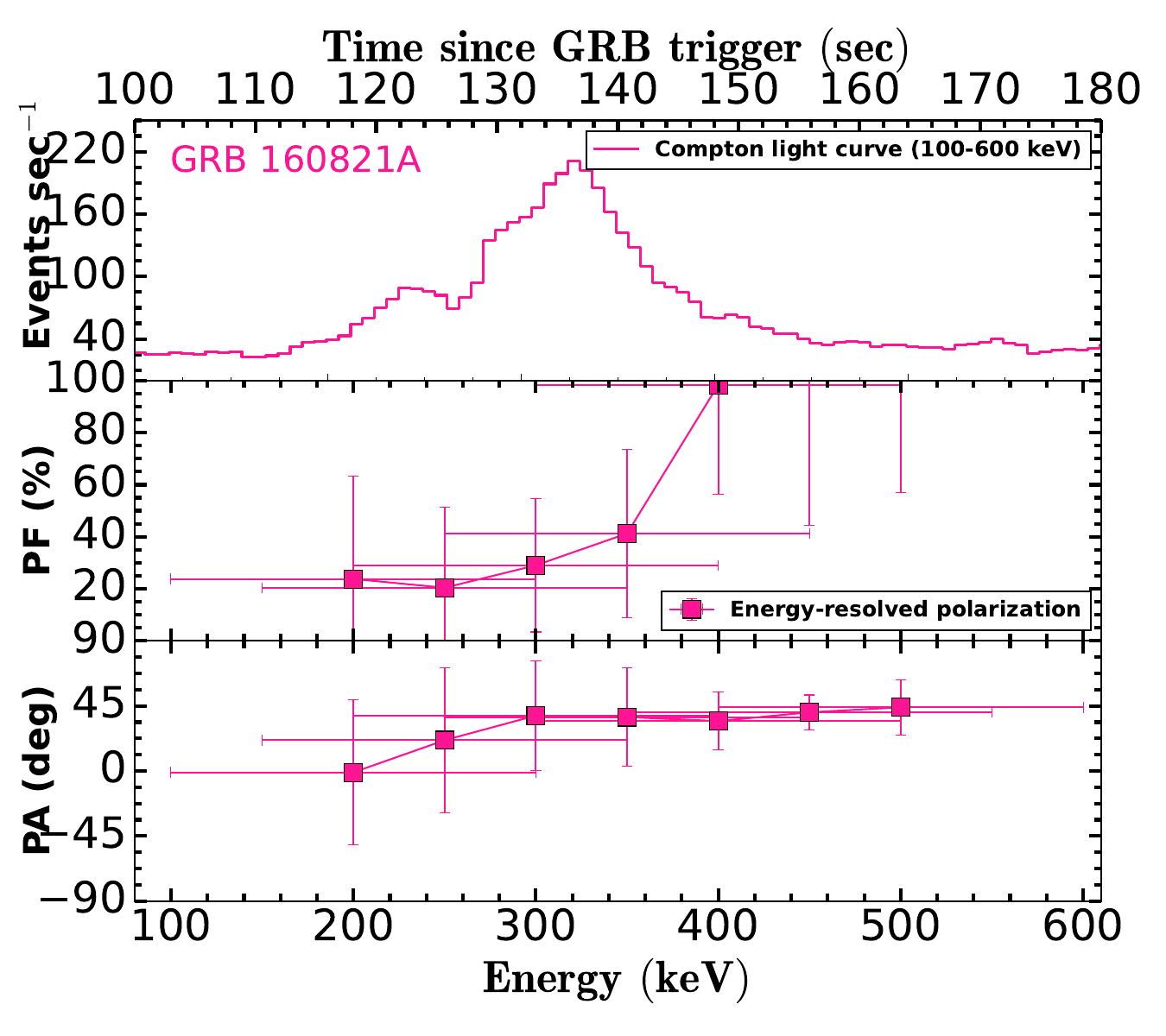}
\caption{{Energy-resolved polarization measurements:} {Top:} Compton light curves of the GRBs with 1-sec bin size obtained using CZTI data. {Middle and bottom:} The evolution of polarization faction and polarization angle with energy. The energy binning has been carried out based on the sliding mode algorithm.}
\label{Energy_resolved}
\end{figure*}

In addition to conducting time-resolved polarization measurements, we also performed an energy-resolved polarization analysis. A comparison of the polarization fraction obtained using the \AstroSat CZTI and POLAR missions catalog revealed that \AstroSat CZTI detected approximately 20 \% higher polarization compared to POLAR measurements \citep{2022ApJ...936...12C}. We suggested that the discrepancy between the observed time-integrated and energy-integrated polarization fractions of prompt emission by \AstroSat CZTI and POLAR missions could be attributed to the fact that both instruments report the polarization fractions values in different energy channels (CZTI values in 100-600 \keV, and POLAR values in 50-500 \keV). 

In this work, we carried out energy-resolved polarization measurements to investigate the energy-dependent behavior of polarized radiation (polarization degree and angle) during the prompt phase of GRBs. We have employed two methods for selecting energy bins of individual bursts. Initially, we selected the bins based on observed peak energy calculated from the time-averaged spectral analysis. We created two bins: one ranging from 100-\Ep ~\keV and the other from \Ep-600 \keV. In cases where the observed peak energy exceeded 600 \keV (the maximum allowed energy range for the polarization measurements using CZTI), we selected the following bins: 100-300 \keV and 300-600 \keV, considering that the mean value of peak energy for long GRBs is approximately 200-300 \keV (refer to Figure \ref{TAS_FERMI_dist}). The calculated values of the energy-resolved polarization fraction of all five bursts are listed in Table \ref{energy-resolved_table}.

Further, we also selected the energy bins using the sliding mode spectral binning method (since the GRB spectra exhibit rapid variations) with a bin width of 50 \keV for the energy-resolved polarization of all five GRBs in our sample. We initially divided the spectrum into smaller energy intervals of the bin width from 100-300 \keV and slid these average energy intervals across the total energy range of the CZTI with an increasing order of 50 \keV. Using the sliding mode binning, we calculated the average values of polarization parameters within each spectral bin. The polarization results obtained using the energy sliding binning algorithm are shown in Figure \ref{Energy_resolved}. We noted that the polarization angles of GRB 160325A, GRB 160703A, and GRB 160802A, GRB 160821A obtained for different energy segments remain mostly consistent (no substantial change in the polarization angles); however, we noted that the polarization angles of GRB 160623A changed with energy. 
Additionally, we noted that the polarization fraction values have increasing trends with energy, although the analysis might be limited due to fewer Compton counts in later energy bins. The energy-resolved polarization analysis gives a hint that polarization measurements depend on the energy channels of the detectors. 

\begin{table*}
\caption{The calculated values of energy-resolved polarization fraction (100-\Ep or 300 \keV and \Ep or 300 -600 \keV) of all the five bursts in our sample.}
\label{energy-resolved_table}
\begin{scriptsize}
\begin{center}
\begin{tabular}{|c|c|c|c|c|}
\hline
 \bf GRB name& Energy range (\keV) & No. of Compton events  & \bf PF (\%) & BF \\
\hline
GRB 160325A &100-187 & 391 & $<$ 70.54 & 1.41\\
GRB 160325A &187-600 & 380& $<$ 33.28  & 0.80\\ \hline
GRB 160623A & 100-300 & 1428 & $<$ 24.42 & 0.74\\ 
GRB 160623A & 300-600 & 277 & unconstrained & 2.82\\ \hline
GRB 160703A  & 100-351 & 376& $<$ 18.64 & 0.73\\ 
GRB 160703A  & 351-600 & 51 & unconstrained & 1.02\\ \hline
GRB 160802A &100-363 &1360 & $<$ 27.92  & 0.70\\
GRB 160802A &363-600 & 152 & $<$ 69.97  & 0.69\\ \hline
GRB 160821A & 100-300 & 2387& $<$ 20.03 & 0.85\\ 
GRB 160821A & 300-600 & 468 & unconstrained & 2.27\\ 
\hline
\end{tabular}
\end{center}
\end{scriptsize}
\end{table*}

\section{Discussion}
\label{discussion}

Based on the above data analysis and results, we present the key discussion on the spectro-polarimetric properties of individual GRBs in this section.

\subsection{Jet composition and emission mechanisms of the sample}
\label{Jet composition and emission mechanisms of GRBs}

\begin{table*}[!ht]
\caption{The calculated values of the Lorentz factor and jet opening angle of all the five bursts in our sample. $\Gamma \theta_{\rm j}$ $>>$ 1 suggests that the jet from all the GRBs in our sample is observed from an on-axis view.}
\label{gamma_thetaj}
\begin{center}
\begin{tabular}{|c|c|c|c|c|}
\hline
 \bf GRB name& $E_{\rm \gamma, iso}$ $\times ~10^{52}$ (erg) & $\Gamma$ & $\theta_{\rm j}$ (degree) & $\Gamma \theta_{\rm j}$ \\
\hline
GRB 160325A & 21.63 & 392.48 & $>$ 1.32 & $\sim$ 9.05\\
GRB 160623A & 25.3 & 408.17 & $>$ 13 &  $\sim$ 92.61\\ 
GRB 160703A & 20.27 &386.18 & $>$ 4.68&  $\sim$ 31.52\\ 
GRB 160802A &79.44 & 543.36  & $>$ 2.1 &  $\sim$ 19.91\\
GRB 160821A & 844.03&980.98 & $>$ 2.1&  $\sim$ 35.95\\ 
\hline
\end{tabular}
\end{center}
\end{table*}

The main objective of this study is to investigate the possible jet composition and emission mechanisms of GRBs through time-resolved and energy-resolved spectro-polarimetric analysis. Different radiation models in GRBs are associated with different polarization fraction values. However, it is important to note that the observed polarization fraction values also depend on the viewing geometry of the bursts. To assess the viewing geometry of individual bursts, we employed the $\Gamma \theta_{\rm j}$ condition. By applying this condition, we sought to gain insights into the viewing perspective of the bursts and their implications on their polarization properties. When viewing the jet from an on-axis perspective, the value of $\Gamma \theta_{\rm j}$ is significantly greater than 1. Conversely, for off-axis observations, $\Gamma \theta_{\rm j}$ is expected to be much smaller than 1. In the case of a narrow jetted view, the $\Gamma \theta_{\rm j}$ value is expected to be approximately 1. The value of $\Gamma$ of the fireball can be derived from prompt emission as well as afterglow features of GRBs \citep{2010ApJ...725.2209L, 2018A&A...609A.112G}. In this work, we constrain the value of bulk Lorentz factor using well-studied Liang correlation\footnote{$\Gamma_{0}$ $\approx$ 182 $\times$ $E_{\gamma, \rm iso, 52}^{0.25 \pm 0.03}$}, the strong correlation between bulk Lorentz factor and isotropic gamma-ray energy of the fireball \citep{2010ApJ...725.2209L}. The derived values of bulk Lorentz factor are tabulated in Table \ref{gamma_thetaj}. For GRB 160623A, we obtained $E_{\gamma, \rm iso}$ value using \kw observations \citep{2017ApJ...850..161T} as the main emission was not detected using \fermi GBM. Additionally, we derive the jet opening angle (lower limits) using the X-ray afterglow light curves observed using \swift XRT and equation 4 of \cite{2001ApJ...562L..55F}. The $\theta_{\rm j}$ value depends on micro-physical afterglow parameters (medium number density ($n_{0}$) and electrons thermal energy fraction ($\epsilon_{e}$)). We assume typical values of $n_{0}$ = 1 and $\epsilon_{e}$ = 0.2 to constrain $\theta_{\rm j}$ values \citep{2022JApA...43...11G}. However, detailed afterglow modeling and a good data set will be needed to constrain these parameters better \citep{2022MNRAS.511.1694G}. For GRB 160623A, we obtained the $\theta_{\rm j}$ value from \cite{2020ApJ...891L..15C}. However, in the case of GRB 160802A and GRB 160821A, no \swift XRT observations are available, so we used $\theta_{\rm j}$ = 2.1 degree, which is the mean value of jet opening angle for typical \fermi-detected long GRBs \citep{2021ApJ...908L...2S}. After calculating the bulk Lorentz factor and jet opening angle values for individual bursts, we determine the viewing geometry ($\Gamma \theta_{\rm j}$). The calculated values of $\Gamma \theta_{\rm j}$ are tabulated in Table \ref{gamma_thetaj}. We noted that the calculated for all five bursts in our sample have $\Gamma \theta_{\rm j}$ $>>$ 1, suggesting that the jet from these GRBs is observed from an on-axis perspective. Further, we utilized $\Gamma \theta_{\rm j}$ condition and our spectro-polarimetric results for each of the five GRBs in our sample to investigate GRBs' possible jet composition and emission mechanisms.

\subsubsection{GRB 160325A}

We studied the spectro-polarimetric properties (analysis of spectral properties using \fermi as well as the polarization of emitted radiation using \AstroSat) of GRB 160325A for both pulses (the light curve of this burst exhibits two separate emission episodes with a quiescent period in between). The $\alpha_{\rm pt}$ values seem harder during the first episode, and we observed a low polarization fraction (using time-resolved polarization measurements) during this episode. Conversely, $\alpha_{\rm pt}$ value becomes softer during the second emission episode, and we observed a hint of high polarization fraction (an upper limit of 98 \%). The observed spectro-polarimetric properties during the first episode suggest that the emission during this episode originated from a thick shell photosphere with localized dissipation occurring below it. In contrast, the emission during the second episode is dominated by thin-shell synchrotron emission. Furthermore, our time-resolved polarization measurements of GRB 160325A indicate the transition of a baryonic-dominated jet composition during the first episode to a subdominant Poynting flux jet composition during the second episode. Our results (with an updated polarization analysis pipeline) are consistent with our previous spectro-polarimetric analysis of the bursts reported in 100-300 \keV \citep{2020MNRAS.493.5218S}.

\subsubsection{GRB 160623A}

The prompt light curve of GRB 160623A obtained using \kw exhibits a broad emission episode (main), followed by a weaker emission episode \citep{2016GCN.19554....1F}. However, the main emission episode of GRB 160623A was occluded for \fermi mission. Therefore, we utilized the time-integrated spectral analysis results reported by us using \kw observations to constrain the radiation mechanism of GRB 160623A \citep{2022ApJ...936...12C}. We noted that the observed value of $\alpha_{\rm pt}$ using the time-integrated \kw spectrum lies within the synchrotron slow and fast cooling prediction. Additionally, the time-integrated and time-resolved (however, it is important to note that within a 2-sigma confidence interval, these polarization measurements are consistent with low polarization) polarization analysis using CZTI data gives a hint for the high degree of polarization, supporting the synchrotron emission in an ordered magnetic field (see Figure \ref{Time_resolved}). The possibility of no significant polarization cannot be entirely ruled out based on the current measurements. Our spectro-polarimetric analysis of GRB 160623A suggests a Poynting flux jet composition throughout the burst's emission.

\subsubsection{GRB 160703A}

The light curve of GRB 160703A, as observed by \kw, displays multiple overlapping emission pulses \citep{2016GCN.19649....1F}, consistent with \swift BAT light curve (see Figure \ref{Prompt_Light_Curves} of the appendix). However, since this GRB was not detected by the \fermi mission, we were unable to perform a detailed time-resolved spectral analysis of this burst. To investigate the radiation mechanism of GRB 160703A, we relied on the time-integrated spectral analysis results previously reported by us using \kw observations \citep{2022ApJ...936...12C}. The low energy photon index obtained from the time-integrated \kw spectrum is consistent with the synchrotron emission model. Furthermore, our $\alpha_{\rm pt}$ value calculated using the time-integrated \swift BAT spectral analysis is also consistent with the synchrotron emission model (see Table \ref{tab:TAS} of the appendix). Similar to the previous case, both the time-integrated and time-resolved (however, it is important to note that the observed polarization is also consistent with low polarization within 2-sigma confidence interval) polarization analysis using \AstroSat CZTI data provide indications of a hint for the high degree of polarization, supporting the presence of synchrotron emission in an ordered magnetic field (see Figure \ref{Time_resolved}). Our spectro-polarimetric analysis of GRB 160703A suggests a Poynting flux jet composition throughout the emission of the burst.

\subsubsection{GRB 160802A}

The light curve of GRB 160802A displays two distinct emission episodes separated by a quiescent period (see Figure \ref{Prompt_Light_Curves} of the appendix). A detailed spectro-polarimetric analysis was conducted for both episodes, revealing a notable similarity in spectral behavior to GRB 160325A. The spectral analysis of GRB 160802A indicates that the low energy photon index remains (hard) above the synchrotron emission ``line of death" for most of the temporal bins in the first episode (see Figure \ref{TRS_GRB160802A}). Time-resolved polarization measurements (sliding mode analysis) during this episode constrain the polarization fraction to low values. Since the jet of this burst was observed on-axis (see section \ref{Jet composition and emission mechanisms of GRBs}), our spectro-polarimetric analysis of the first episode is consistent with the photospheric emission model. Such hard values of $\alpha_{\rm pt}$ and low polarization fraction can be explained using a Baryonic dominated jet with subphotospheric dissipation. In contrast, $\alpha_{\rm pt}$ value becomes softer (than the first episode) during the second emission episode. Although we obtained a hint of a high degree of polarization fraction (with respect to the time-resolved measurements during the first episode), we were unable to obtain a more precise measurement due to the low number of Compton counts during this episode. The observed spectro-polarimetric properties during the second episode suggest that it is dominated by thin-shell synchrotron emission. Furthermore, our time-resolved polarization measurements of GRB 160802A suggest a possible transition of a baryonic-dominated jet composition during the first episode to a subdominant Poynting flux jet composition during the second episode. However, the limited number of Compton events during the second episode of GRB 160802A prevents us from making a definitive claim for such a transition.

\subsubsection{GRB 160821A}

The light curve of GRB 160821A observed by \fermi GBM reveals an initial fainter emission followed by a highly intense emission. However, the initial weaker emission was not detected by \AstroSat CZTI. Therefore, this study focuses solely on the spectro-polarimetric analysis of the main emission episode of GRB 160821A. The exceptional brightness of the main emission episode of GRB 160821A helps us to perform a detailed time-resolved spectro-polarimetric analysis of the burst. The observed evolution of $\alpha_{\rm pt}$ lies within the predicted range of the thin shell synchrotron emission model. The high flux suggests that the bursts are observed on-axis, as discussed in Section \ref{Jet composition and emission mechanisms of GRBs}. During the rising and peak phase of the main pulse, we observed the swing in the polarization angle by approximately 90 degrees. Subsequently, from the peak to the decay phase of the pulse, the polarization angle swings back. Our time-resolved polarization analysis indicates that the lower value of the time-integrated polarization fraction reported in \cite{2022ApJ...936...12C} may be attributed to the variation in the polarization angle. The spectro-polarimetric analysis of GRB 160821A provides further support for synchrotron emission occurring within an ordered magnetic field. The results also suggest that the jet composition throughout the burst's emission is dominated by Poynting flux. These results align with our previous spectro-polarimetric analysis of bursts reported in the 100-300 \keV energy range \citep{2019ApJ...882L..10S}.

\section{Summary and Conclusion}
\label{conclusion}

The spectro-polarimetric analysis of GRBs has been investigated for a limited number of GRBs, and most of the studies explored only the time-integrated polarization measurements due to the transient behavior of GRBs, in particular, as well as the challenge of X-ray polarization measurement, in general, \citep{2020MNRAS.491.3343G, 2020A&A...644A.124K, 2022ApJ...936...12C}. In our recent study, we suggested that the majority of bursts in the sample exhibit minimal or no polarization in our time-integrated measurements within the 100-600 \keV energy range, as observed with \AstroSat CZTI \citep{2022ApJ...936...12C}. However, a detailed time-resolved and energy-resolved polarization analysis was needed to identify if the observed low-polarization fraction is intrinsic or due to variation in polarization fraction and polarization angle with time and energy within the burst. In this paper, we investigated the prompt emission temporal, spectral, and polarization properties of five bright bursts observed using the CZTI onboard \AstroSat in its first year of operation. Our study focuses on the application of time-resolved and energy-resolved spectro-polarimetry techniques to obtain detailed polarization information and characterize the emission properties of these GRBs. The primary objective of our study is to delve into the jet compositions of these bright GRBs and constrain the different radiation models of prompt emission. This issue has been a subject of long-standing debate, and prompt emission spectroscopy on its own has been insufficient to resolve these questions independently.

By exploiting the high-angular-resolution CZTI data, we have derived time-resolved polarization profiles for a sample of GRBs. We studied the $\Gamma \theta_{\rm j}$ condition to constrain the jet geometry of these bursts, as observed polarization also depends on the jet geometry. We utilized 10.4m GTC and 3.6m DOT telescopes to contain the redshift/ host search of the bursts, which further helps to verify the $\Gamma \theta_{\rm j}$ condition. Our analysis suggests that the jet emissions from these GRBs were observed on-axis. Furthermore, our comprehensive spectro-polarimetric analysis suggests that GRB 160623A, GRB 160703A, and GRB 160821A have a Poynting flux-dominated jet, and emission could be explained using a thin shell synchrotron emission model in an ordered magnetic field. On the other hand, GRB 160325A and GRB 160802A have the first pulse with a thermal signature followed by non-thermal emission during the second pulse. Our analysis indicates that GRB 160325A and GRB 160802A have a Baryonic dominated jet with mild magnetization. We do not observe any rapid evolution in the polarization angles of GRB 160325A, GRB 160623A, GRB 160703A, and GRB 160802A. However, we observe a rapid change in polarization angle by $\sim$ 90 degrees within the main pulse of very bright GRB 160821A, consistent with our previous results reported in 100-300 \keV \citep{ 2019ApJ...882L..10S}. The profile of GRB 160821A (time-resolved polarization analysis) reveals temporal variations in the angle of polarization, shedding light on the radiation mechanisms and geometry involved in this extreme event. We noted that some authors performed the theoretical simulations and reproduced such large temporal variation in polarization angle under the photospheric emission model. They also discussed the physics and implications of observing such changes \citep{2024ApJ...961..243I, 2020ApJ...896..139P}. However, our analysis reveals a hint of high degree of polarization for GRB 160821A, which contrasts with the predictions of the photospheric emission model.

Additionally, we have studied the polarization properties as a function of energy, suggesting a hint of variations in the polarization degree and angle across different energy bands. We noted that the polarization angles of GRB 160325A, GRB 160703A, and GRB 160802A, GRB 160821A obtained for different energy segments remain mostly consistent; however, the polarization angles of GRB 160623A changed with energy (though large associated error due to the limited number of Compton events). Further, we noted that the polarization fraction values have increasing trends with energy, although the analysis might be limited due to fewer Compton counts in later energy bins. The energy-resolved polarization analysis gives a hint that polarization properties depend on the energy channels of the detectors.

Our results demonstrate the capability of \AstroSat CZTI for detailed time-resolved and energy-resolved spectro-polarimetry of GRBs. The combination of high-angular-resolution imaging, broad energy coverage, and polarization sensitivity provides a unique opportunity to unravel the complex physics governing these explosive phenomena. By studying the polarization of these GRBs, we obtain important insights into the geometry and magnetic field structures associated with these bursts. Our findings suggest that prompt emission polarization analysis, when combined with spectral and temporal data, possesses a distinct capacity to resolve the long-standing debate surrounding the emission mechanisms of GRBs. A comprehensive analysis that delves into both time-resolved and energy-resolved spectro-polarimetry offers greater insight into the emission mechanisms of GRBs compared to a time-averaged spectro-polarimetric analysis \citep{2023arXiv231216265G}.

Our time-resolved and energy-resolved analysis may be somewhat limited due to the relatively low number of Compton events in the finer time/energy bins. We need more observations (extremely bright GRBs with more Compton counts) or more sensitive GRB polarimeters with larger effective areas and refined theoretical models to improve our understanding of the physical processes that drive these energetic and enigmatic events. Additionally, examination of the correlation between spectral parameters and measured polarization parameters for more bright GRBs will provide further constraints on the radiation physics of GRBs. The findings presented in this study pave the way for future investigations and highlight the potential of \AstroSat CZTI for advancing our understanding of GRBs and their role in the Universe. Further, the insights gained from this study have profound implications for our understanding of high-energy astrophysics and the physical processes associated with GRBs. The scientific community is actively engaged in preparing for the next generation of gamma-ray missions, including COSI, eAstroGAM, AMEGO, AMEGO-X, and POLAR 2. Our research contributes valuable insights for these forthcoming missions, particularly through our time-resolved polarization measurements. This information is instrumental for the development and optimization of upcoming GRB polarimeters such as LEAP, POLAR 2 \citep{2020SPIE11444E..2VH}, COSI, and other missions.

\begin{acknowledgments}
We thank the anonymous referee for providing positive and encouraging comments on our manuscript. RG and SBP are very grateful to Prof. A. R. Rao for the excellent suggestions and discussion with the project. RG is also thankful to Dr. Tyler Parsotan for reading the manuscript and fruitful discussion. RG, SBP, DB, and VB acknowledge the financial support of ISRO under AstroSat archival Data utilization program (DS$\_$2B-13013(2)/1/2021-Sec.2). This publication uses data from the \AstroSat mission of the Indian Space Research Organisation (ISRO), archived at the Indian Space Science Data Centre (ISSDC). CZT-Imager is built by a consortium of institutes across India, including the Tata Institute of Fundamental Research (TIFR), Mumbai, the Vikram Sarabhai Space Centre, Thiruvananthapuram, ISRO Satellite Centre (ISAC), Bengaluru, Inter University Centre for Astronomy and Astro-physics, Pune, Physical Research Laboratory, Ahmedabad, Space Application Centre, Ahmedabad. This research also has used data obtained through the HEASARC Online Service, provided by the NASA-GSFC, in support of NASA High Energy Astrophysics Programs. RG was sponsored by the National Aeronautics and Space Administration (NASA) through a contract with ORAU. The views and conclusions contained in this document are those of the authors and should not be interpreted as representing the official policies, either expressed or implied, of the National Aeronautics and Space Administration (NASA) or the U.S. Government. The U.S. Government is authorized to reproduce and distribute reprints for Government purposes notwithstanding any copyright notation herein. AA acknowledges funds and assistance provided by the Council of Scientific \& Industrial Research (CSIR), India, under file no. 09/948(0003)/2020-EMR-I. AA also acknowledges the Yushan Young Fellow Program by the Ministry of Education, Taiwan, for financial support.  This research is based on observations obtained at the 3.6m Devasthal Optical Telescope (DOT), which is a National Facility run and managed by Aryabhatta Research Institute of Observational Sciences (ARIES), an autonomous Institute under the Department of Science and Technology, Government of India.
\end{acknowledgments}

\vspace{5mm}
\facilities{AstroSat(CZTI), Fermi(GBM), Swift(BAT), 10.4m GTC, 3.6m DOT}

\software{astropy version 5.1 \citep{2013A&A...558A..33A, 2018AJ....156..123A}, 3ML version 2.4.0  \citep{2015arXiv150708343V}, Fermi GBM Data Tools version 1.1.1.\citep{GbmDataTools}, gtburst version 03-00-00p0, numpy version 1.21.6, matplotlib version 3.3.2
          }

\bibliography{CZTI_bib}{}
\bibliographystyle{aasjournal}

\appendix
\restartappendixnumbering

\section{Figures}

\begin{figure*}[!ht]
\centering
\includegraphics[scale=0.55]{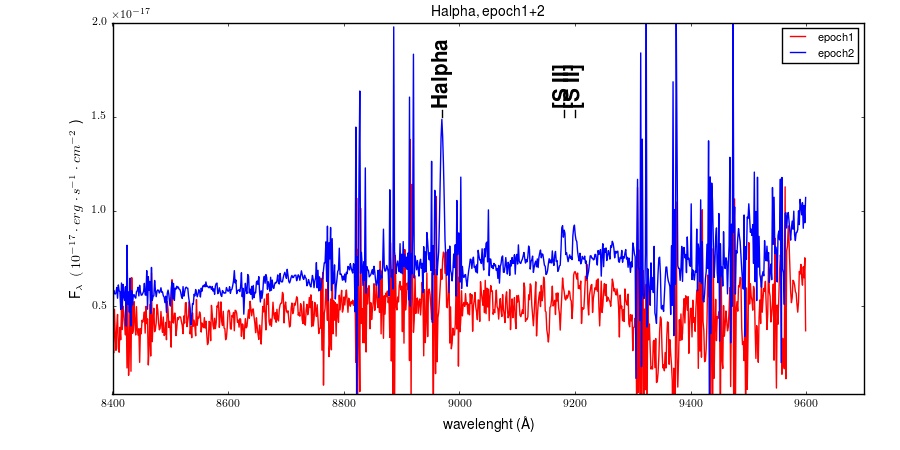}
\includegraphics[scale=0.28]{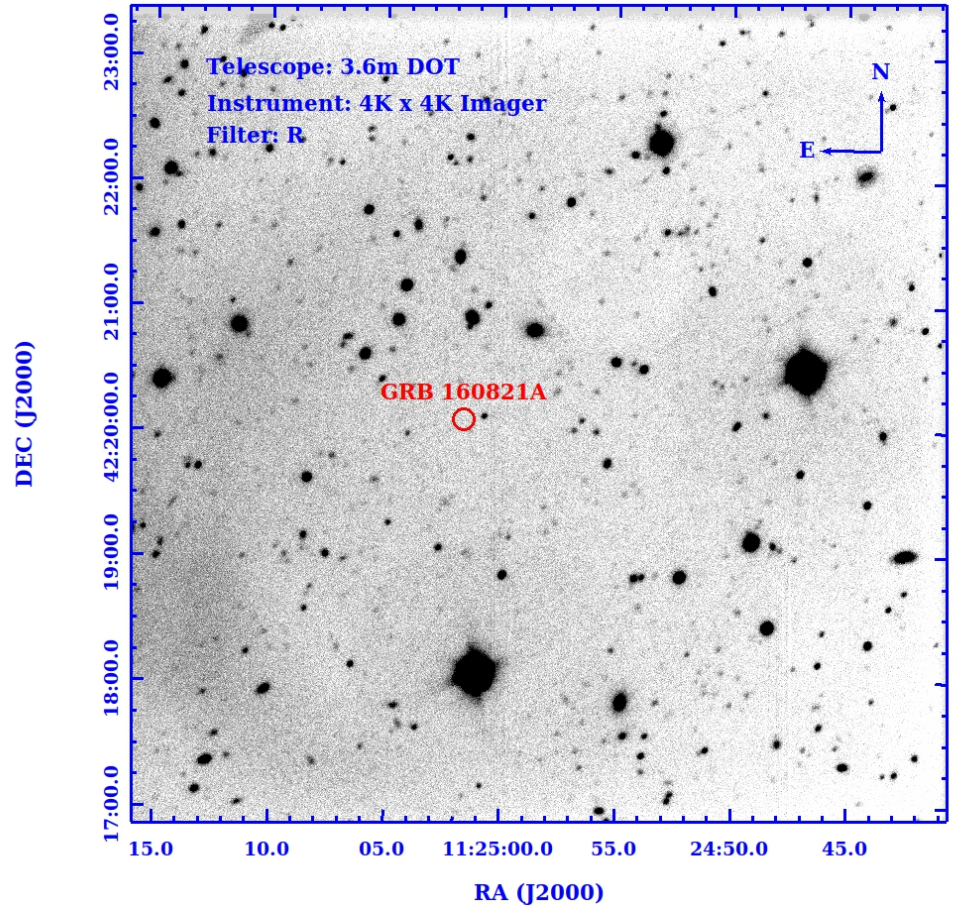}
\caption{{\bf Our efforts to constrain the redshift of the GRBs using larger optical telescopes.} {Top:} Redshift measurement ($z$ = 0.367) of GRB 160623A using 10.4m GTC observations. Our analysis revealed emission lines of H-alpha and [SII] at a common redshift of $z$ = 0.367. {Bottom:} R-filter stacked image of the field of GRB 160821A taken using 3.6m DOT/4K$\times$4K IMAGER \protect\citep{2018BSRSL..87...42P}. The red circle marks the associated uncertainties in the position of the burst.}
\label{Redshift_GRB160623A}
\end{figure*}

\begin{figure*}[!ht]
\centering
\includegraphics[scale=0.28]{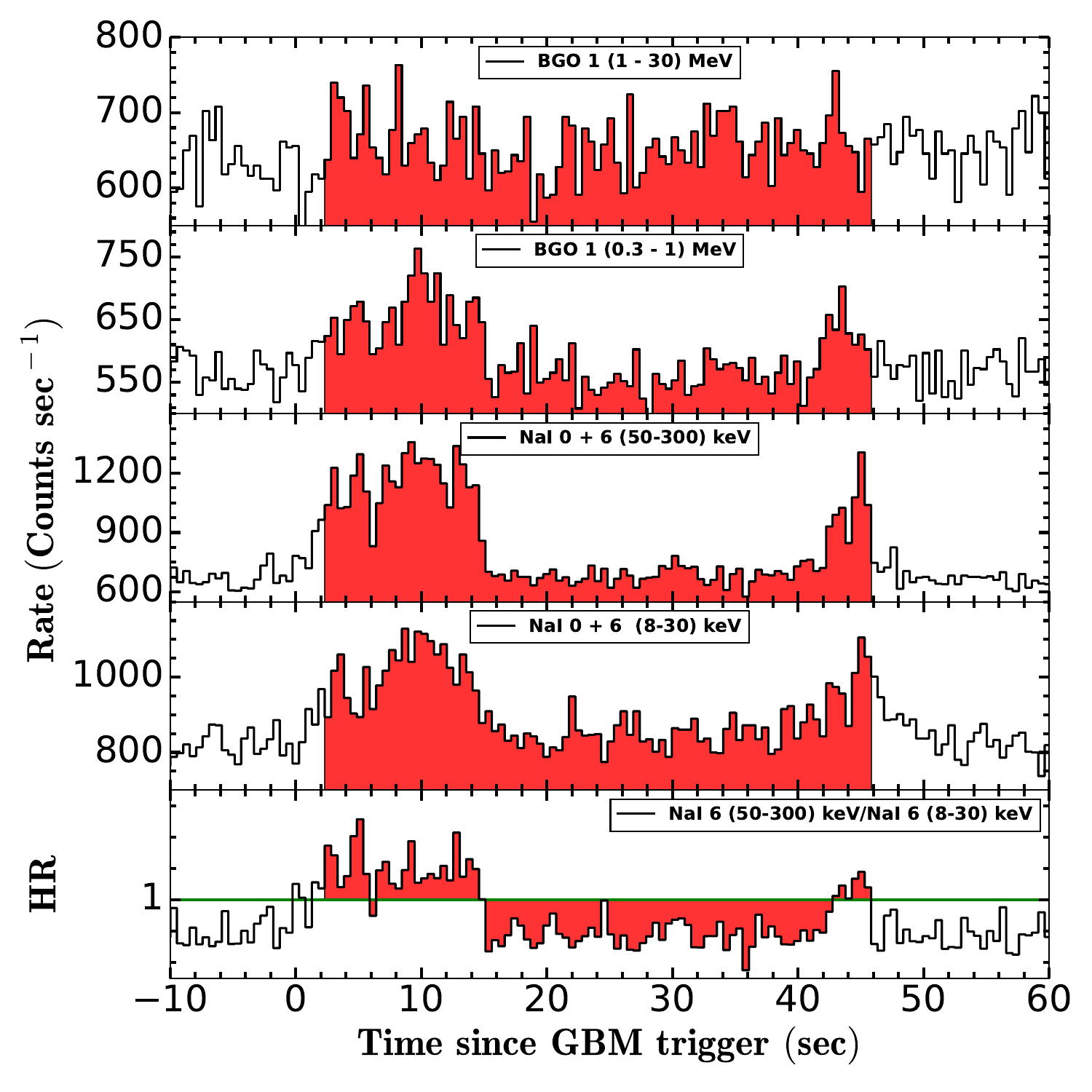}
\includegraphics[scale=0.28]{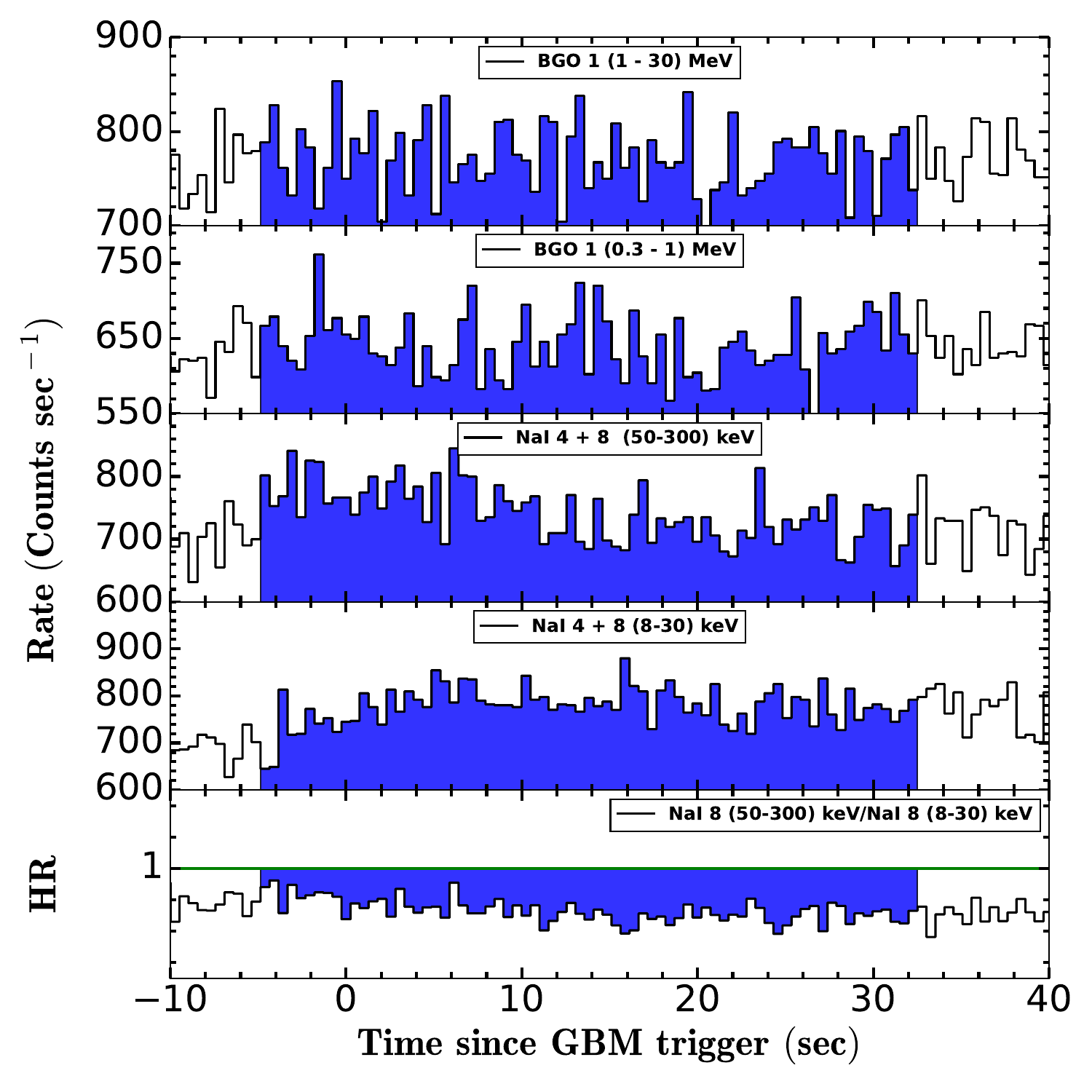}
\includegraphics[scale=0.28]{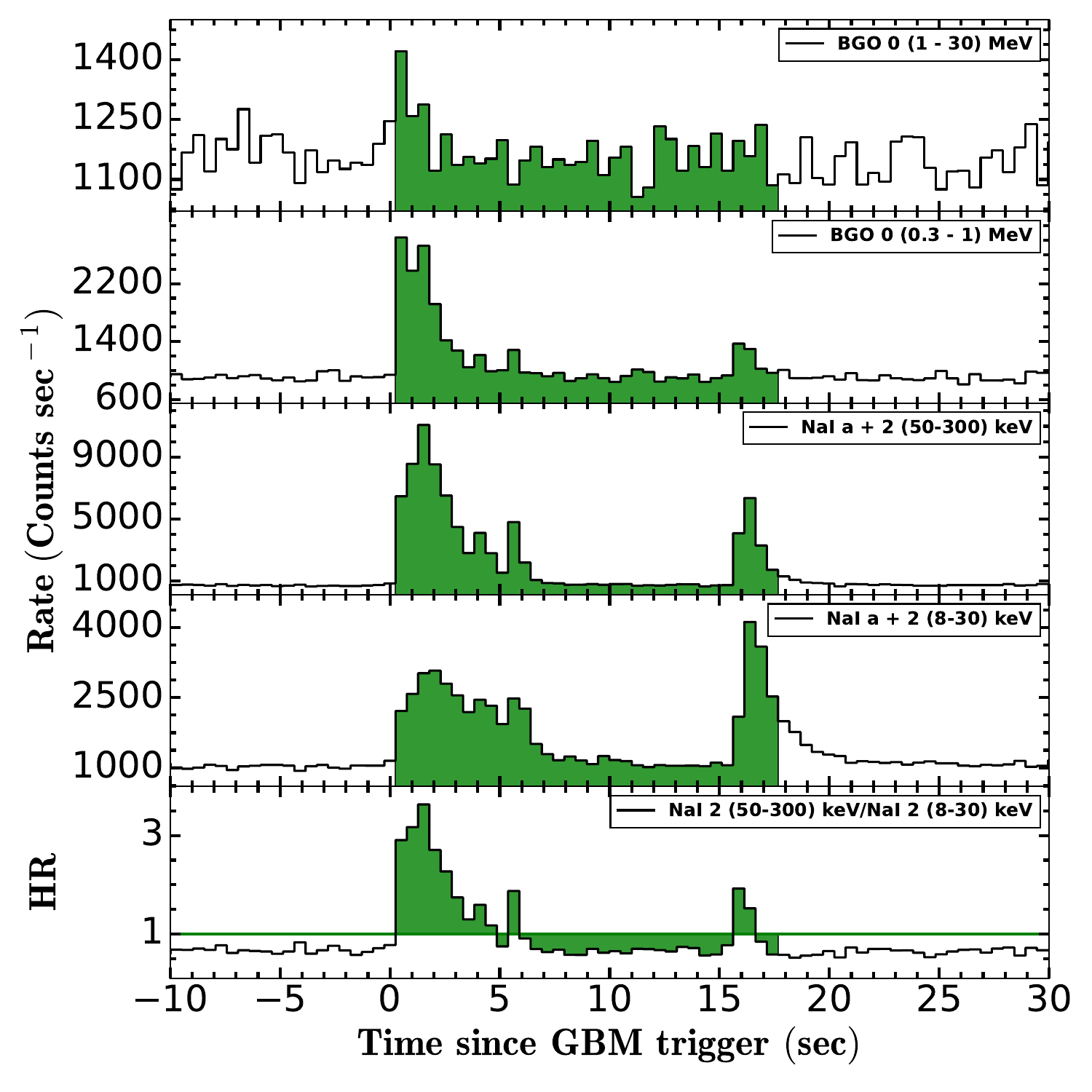}
\includegraphics[scale=0.28]{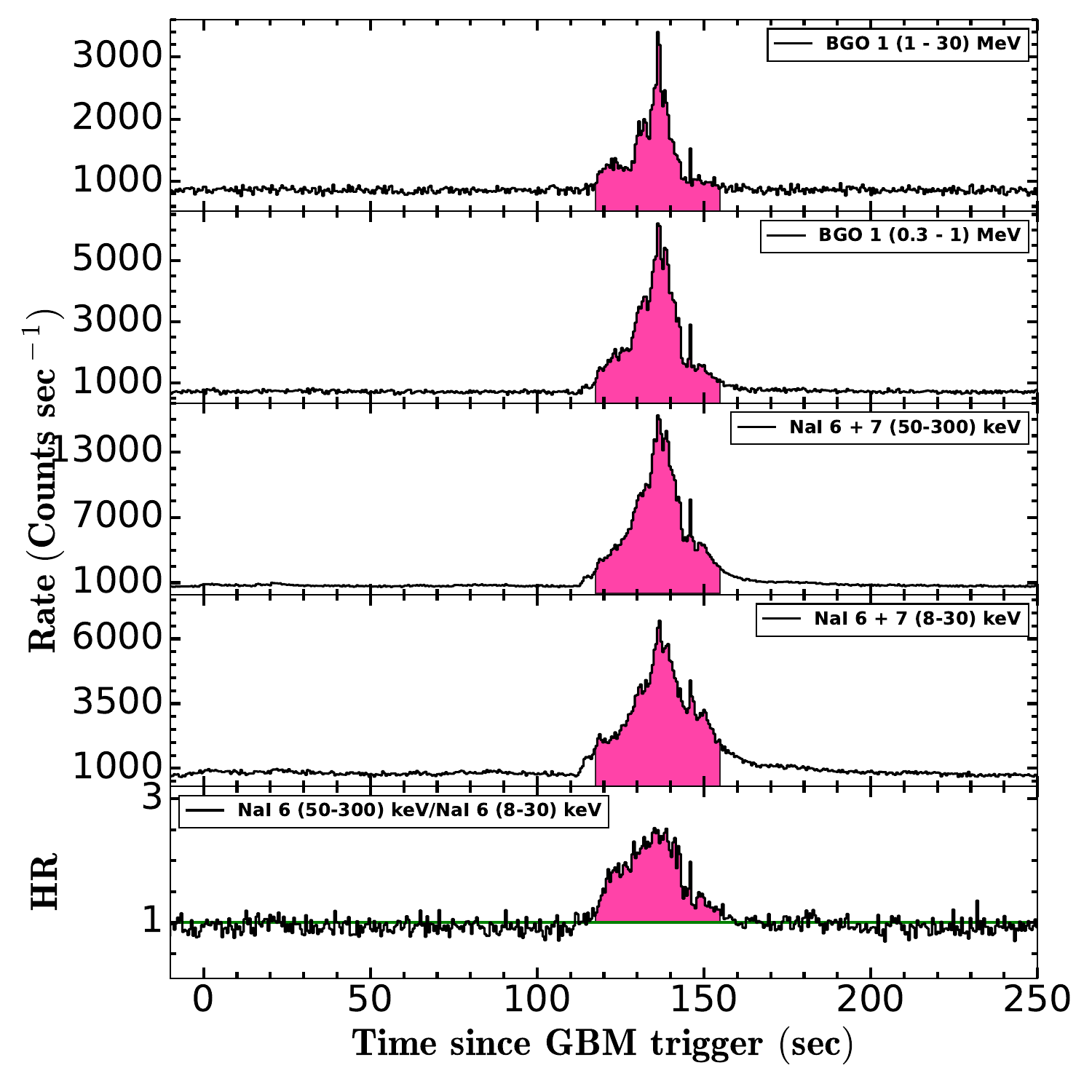}
\includegraphics[scale=0.32]{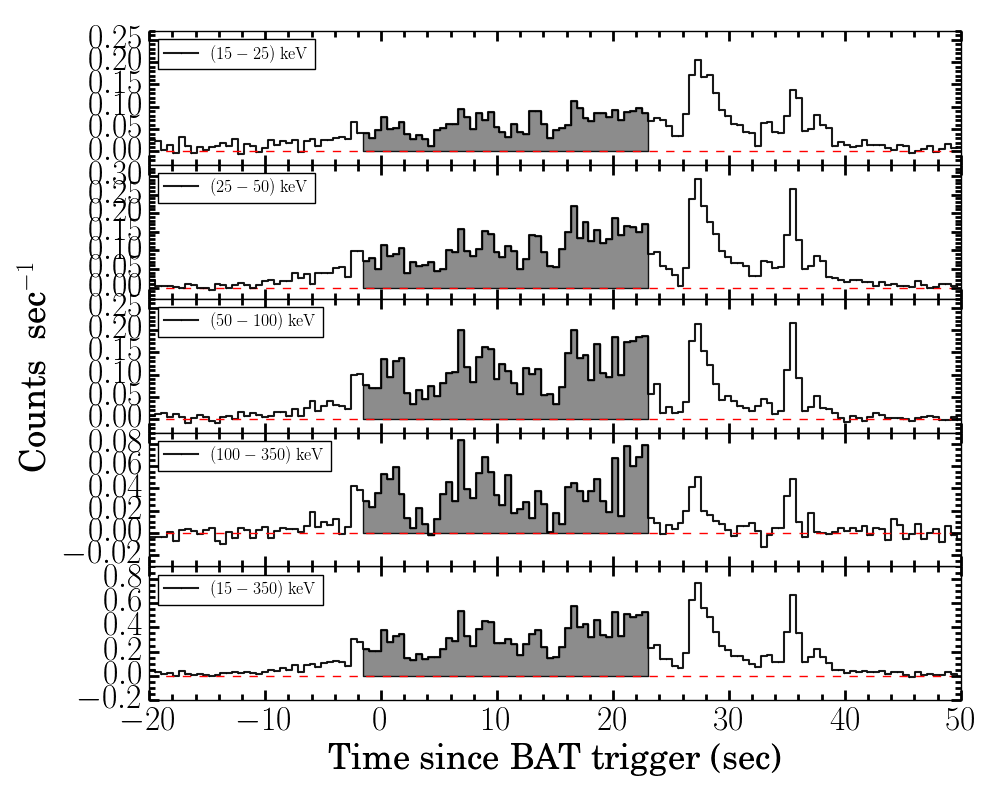}
\caption{Multi-channels light curve profiles of \fermi detected (GRB 160325A (top-left), GRB 160623A (top-right), GRB 160802A (middle-left), and GRB 160821A (middle-right)) and \swift detected (GRB 160703A, bottom) GRBs in our sample. The HR (50-300 \keV/8-30 \keV) evolution for \fermi GRBs is shown in the bottom sub-panels of each GRBs. The shaded colored regions correspond to the time interval used for the time-integrated spectro-polarimetric analysis.}
\label{Prompt_Light_Curves}
\end{figure*}

\begin{figure}
\centering
\includegraphics[scale=0.32]{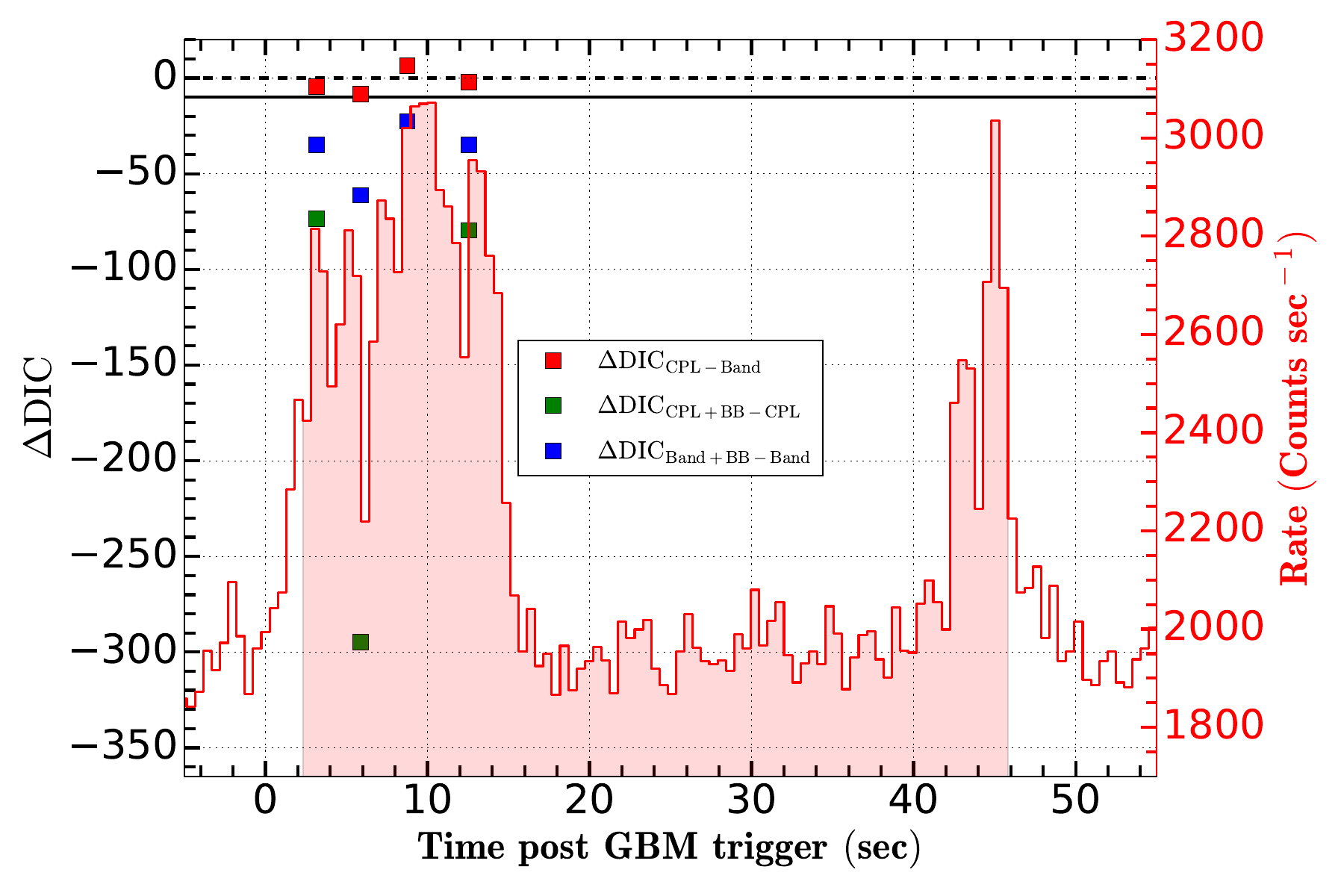}
\includegraphics[scale=0.32]{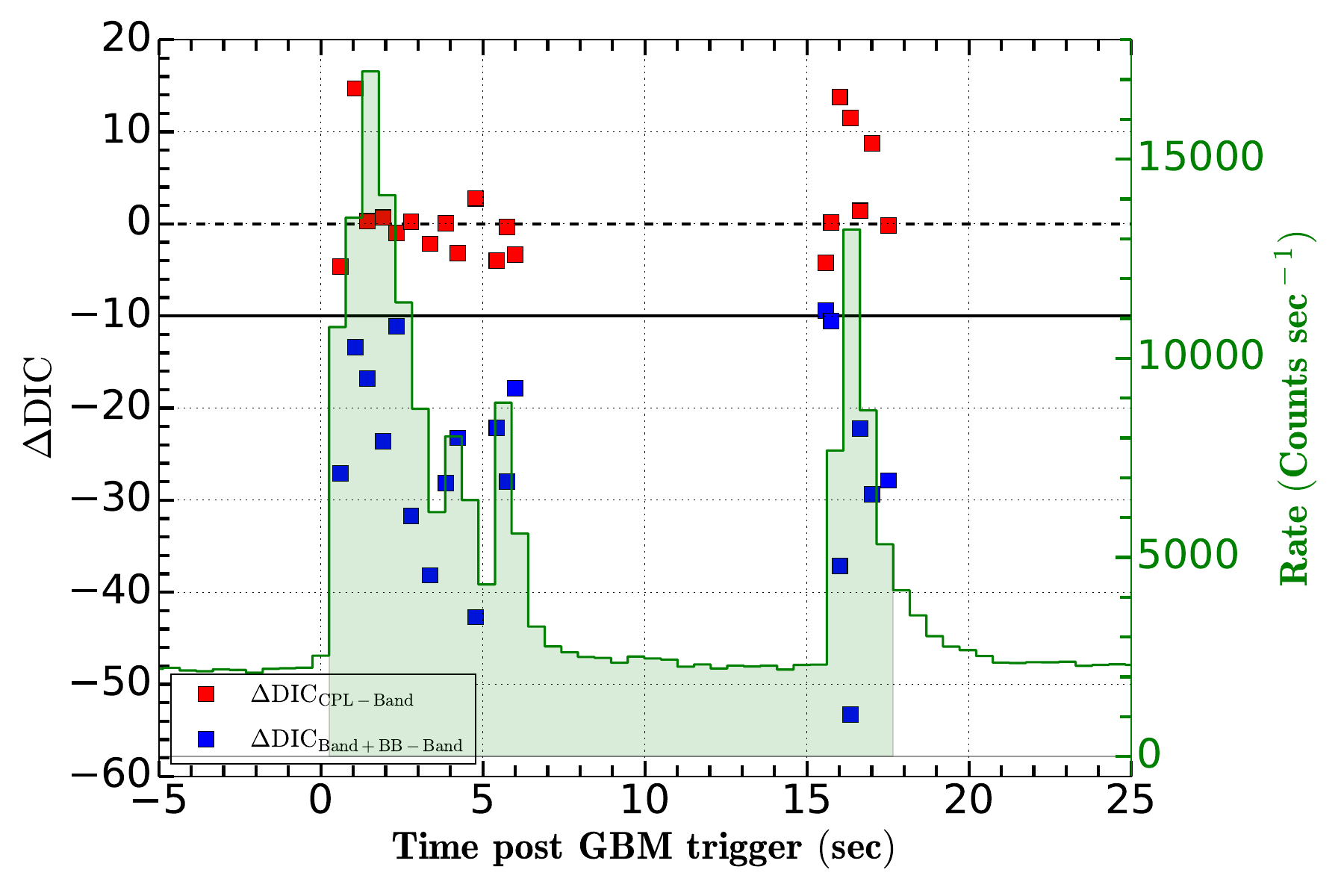}
\includegraphics[scale=0.32]{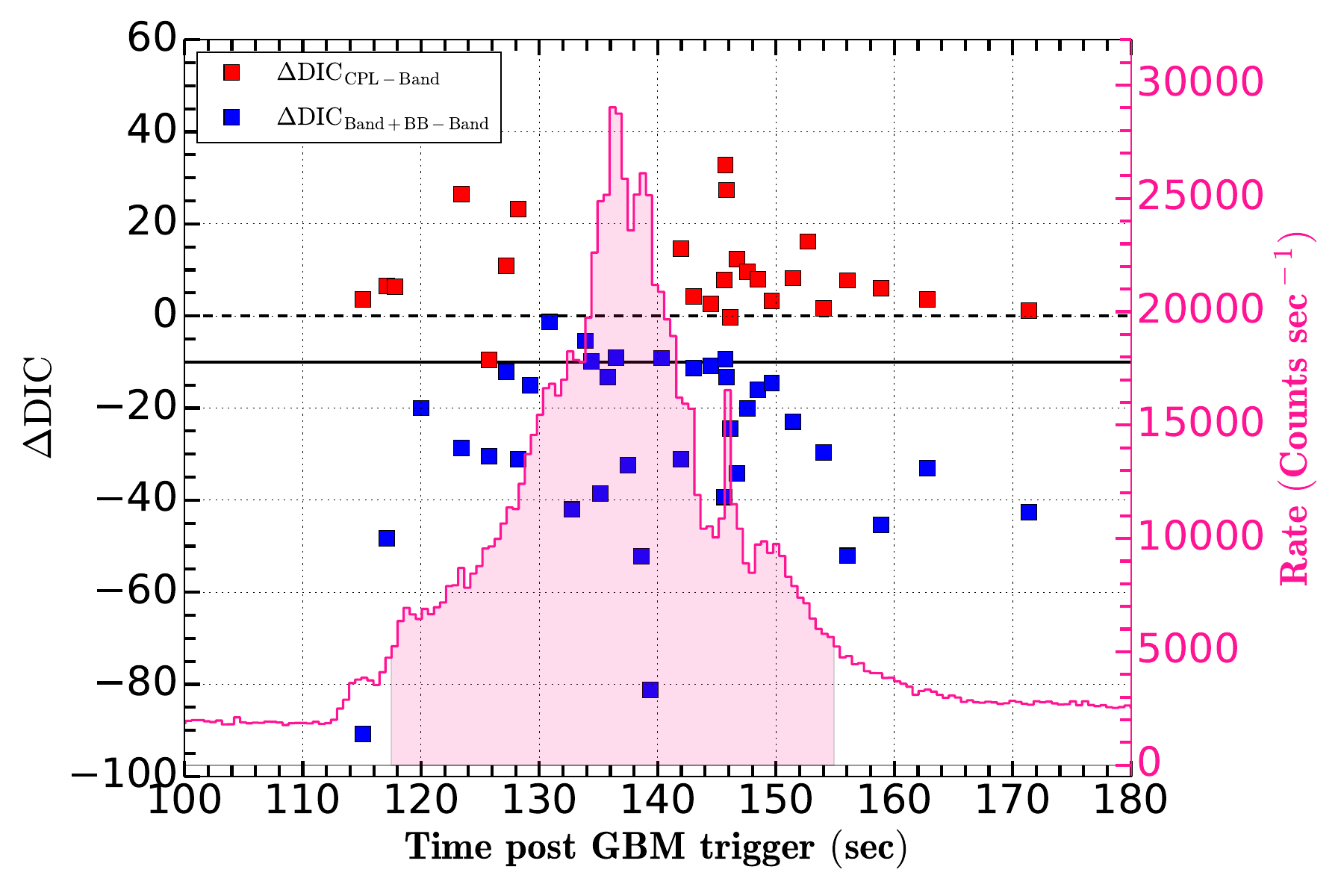}
\caption{The selection of the best-fit model using the difference of deviance information criterion values obtained from different models. The horizontal black dashed and solid lines demonstrate $\Delta$ DIC corresponding to zero and -10, respectively. The top, middle, and bottom panels illustrate the difference in DIC values for GRB 160325A, GRB 160802A, and GRB 160821A, respectively.}
\label{DIC}
\end{figure}

\section{Tables}

\begin{table*}[!ht]
\caption{Empirical and physical spectral fitting of the time-averaged spectrum of GRBs of our sample. Time-integrated flux has been calculated from 10 \keV to 40 MeV energy range. For the \swift detected GRB 160703A, the time-integrated flux has been calculated from 15 \keV to 150 \keV energy range.}
\label{tab:TAS}
\setlength{\tabcolsep}{2pt}
\flushleft
\begin{scriptsize}
\setlength\tabcolsep{1.0pt} 
\begin{tabular}{|c|c|c|c|c|c|c|c|} \hline 
\textbf{Spectral model} & \multicolumn{5}{c|}{\textbf{Spectral parameters}} & \multicolumn{2}{c|}{\textbf{Statistics}} \\ \hline
\hline
   \multicolumn{8}{|c|}{\textbf{GRB 160325A}} \\

\hline\hline

 & $\bf \alpha_{\rm \bf pt}$ & \bf \Ep /$\bf E_{\it c}$ (\keV) & $\bf \beta_{\rm \bf pt}$ & $\rm \bf k{\it \bf T}_{\rm \bf BB}$ (\keV) & Flux (erg/cm$^{2}$/sec)& \textbf{DIC} & \textbf{$\rm \bf \Delta~ {DIC}$}\\ \hline
\sw{Band}&$-0.79_{-0.07}^{+0.07}$&$228.54_{-21.69}^{+21.76}$&$-2.60_{-0.28}^{+0.27}$&-&5.28 $\times 10^{-07}$&5300.69&-\\
\sw{Band+BB}&$-0.78_{-0.07}^{+0.07}$&$227.57_{-21.34}^{+21.04}$&$-2.53_{-0.24}^{+0.24}$&$11.84_{-7.56}^{+7.44}$& 5.52 $\times 10^{-07}$ &5289.52& -11.17\\
\sw{CPL}&$-0.83_{-0.06}^{+0.06}$&$215.12_{-26.87}^{+27.30}$&-&-&4.05 $\times 10^{-07}$ &5303.23&- \\
\sw{CPL+BB}&$-0.83_{-0.06}^{+0.06}$&$216.69_{-26.25}^{+26.33}$&-&$12.07_{-7.74}^{+7.66}$&4.07 $\times 10^{-07}$ &5291.95&-11.28\\\hline
\textbf{\sw{bkn2pow}} & \bf $\it \bf \alpha_{1,2,3}$= 0.80$\bf^{+0.17}_{-0.18}$, 1.24$\bf^{+0.07}_{-0.07}$, 2.51$\bf^{+0.20}_{-0.19}$ & \multicolumn{3}{c|}{$\bf E_{\rm \bf b1,b2}$= \bf 39.64$^{\bf+10.04}_{-9.23}$, 187.81$\bf^{+21.53}_{-20.98}$} &\bf 5.26 $\times 10^{-07}$  &\bf 5250.34 &\bf -- \\
\hline

\end{tabular}
\end{scriptsize}
\end{table*}

\begin{table*}[!ht]
\caption{The time-resolved spectral analysis of GRB 160325A was conducted using empirical models, namely the \sw{Cutoff power-law} and \sw{CPL + Blackbody}. The flux values (in erg $\rm cm^{-2}$ $\rm s^{-1}$) reported in this study were calculated within the energy range of 8 \keV to 40 MeV.}
\label{TRS:cpl}
\setlength{\tabcolsep}{4pt}
\flushleft
\begin{scriptsize}
\setlength\tabcolsep{4.0pt} 
\begin{center}
\begin{tabular}{|c|c|ccc|cccc|ccc|} \hline
 & &\multicolumn{3}{c|}{\bf CPL}&\multicolumn{4}{c|}{\bf CPL+BB}& {\bf CPL} & {\bf CPL+BB} &\\
T$_{\rm start}$ & T$_{\rm stop}$ &\boldmath $\Gamma$ &\boldmath E$_{\bf c}$  & \bf Flux $\times 10^{-06}$  & {\boldmath $\Gamma$} & {\boldmath E$_{\bf c}$} &\bf kT & \bf Flux $\times 10^{-06}$&\bf DIC & \bf DIC & \bf $\Delta$DIC\\
\bf (s)&  \bf (s) & & (\boldmath$\keV$)& &  & (\boldmath$\keV$) & (\boldmath$\keV$) & & & &  \\[1ex] \hline
1.605 & 4.684 & $-0.55_{-0.10}^{0.11}$ & $186.13_{-28.91}^{28.38}$ & 0.92 & $-0.54_{-0.11}^{0.10}$ & $188.44_{-29.25}^{28.93}$ & $11.92_{-7.85}^{8.27}$ & 0.92 & 2008.58 & 1934.9 & -73.68 \\
4.817 & 6.911 & $-0.56_{-0.14}^{0.14}$ & $146.97_{-28.41}^{28.42}$ & 0.75 & $-0.56_{-0.14}^{0.14}$ & $151.78_{-31.42}^{29.47}$ & $11.97_{-8.24}^{8.40}$ & 0.75 & 1610.44 & 1315.75 & -294.69 \\
6.911 & 10.594 & $-0.88_{-0.05}^{0.05}$ & $292.80_{-36.06}^{34.77}$ & 1.5 & $-0.87_{-0.06}^{0.06}$ & $293.82_{-37.04}^{37.76}$ & $12.20_{-7.53}^{7.13}$ & 1.48 & 2289.79 & -14.76 & -2304.55 \\
10.594 & 14.505 & $-0.71_{-0.07}^{0.07}$ & $234.40_{-28.46}^{28.28}$ & 1.19 & $-0.65_{-0.11}^{0.11}$ & $223.32_{-33.18}^{31.92}$ & $9.85_{-5.31}^{5.97}$ & 1.13 & 2361.18 & 2281.56 & -79.62 \\
44.13 & 45.471 & $-0.67_{-0.12}^{0.12}$ & $162.67_{-30.15}^{29.99}$ & 0.98 & $-0.67_{-0.13}^{0.13}$ & $164.65_{-31.46}^{31.41}$ & $11.41_{-7.86}^{8.27}$ & 0.97 & 1130.91 & 772.52 & -358.39 \\
\hline
\end{tabular}
\end{center}
\end{scriptsize}
\end{table*}

\begin{table*}[!ht]
\caption{The time-resolved spectral analysis of GRB 160325A was conducted using empirical models, namely the \sw{Band} and \sw{Band + Blackbody}. The flux values (in erg $\rm cm^{-2}$ $\rm s^{-1}$) reported in this study were calculated within the energy range of 8 \keV to 40 MeV.}
\label{TRS:band}
\setlength{\tabcolsep}{2pt}
\flushleft
\begin{scriptsize}
\setlength\tabcolsep{2.0pt} 
\begin{center}
\begin{tabular}{|c|c|cccc|ccccc|ccc|}\hline
T$_{\rm start}$ & T$_{\rm stop}$ & \boldmath $\it \alpha_{\rm pt}$ & \boldmath $\it \beta_{\rm pt}$ & \boldmath \Ep & \bf Flux &\boldmath $\it \alpha_{\rm pt}$ & \boldmath $\it \beta_{\rm pt}$ & \boldmath \Ep &   kT  & \bf Flux & \bf DIC$_{\rm Band}$ & \bf DIC$_{\rm Band+BB}$ &  \bf $\Delta$DIC\\  \hline
(s) &(s) &  & & \boldmath (\keV) & $\times 10^{-06}$& &  & \boldmath (\keV) &  (\keV)  & $\times 10^{-06}$&  &  &  \\ \hline
1.605 & 4.684 & $-0.46_{-0.12}^{0.12}$ & $-2.59_{-0.29}^{0.29}$ & $235.52_{-26.46}^{26.64}$ & 1.26 & $-0.45_{-0.13}^{0.13}$ & $-2.63_{-0.30}^{0.30}$ & $237.82_{-27.80}^{26.83}$ & $11.01_{-7.17}^{7.56}$ & 1.24 & 2013.13 & 1978.2 & -34.93 \\
4.817 & 6.911 & $-0.48_{-0.16}^{0.16}$ & $-2.62_{-0.31}^{0.31}$ & $186.38_{-23.78}^{24.09}$ & 1.01 & $-0.49_{-0.16}^{0.16}$ & $-2.65_{-0.31}^{0.31}$ & $192.18_{-26.16}^{25.30}$ & $11.95_{-8.02}^{8.68}$ & 1.0 & 1618.76 & 1557.52 & -61.24 \\
6.911 & 10.594 & $-0.79_{-0.08}^{0.08}$ & $-2.28_{-0.21}^{0.21}$ & $260.82_{-34.79}^{35.33}$ & 2.38 & $-0.77_{-0.09}^{0.09}$ & $-2.31_{-0.21}^{0.21}$ & $265.08_{-35.89}^{35.73}$ & $11.02_{-6.60}^{6.72}$ & 2.3 & 2283.19 & 2260.57 & -22.62 \\
10.594 & 14.505 & $-0.69_{-0.08}^{0.07}$ & $-2.78_{-0.29}^{0.29}$ & $279.91_{-24.44}^{24.56}$ & 1.46 & $-0.57_{-0.17}^{0.21}$ & $-2.84_{-0.30}^{0.30}$ & $274.81_{-24.57}^{24.86}$ & $9.55_{-4.63}^{5.27}$ & 1.42 & 2363.46 & 2328.57 & -34.89 \\
44.13 & 45.471 & $-0.63_{-0.13}^{0.13}$ & $-2.81_{-0.31}^{0.32}$ & $198.19_{-23.51}^{23.17}$ & 1.22 & $-0.62_{-0.14}^{0.14}$ & $-2.85_{-0.32}^{0.33}$ & $200.19_{-24.16}^{23.66}$ & $11.26_{-7.48}^{7.81}$ & 1.21 & 1137.76 & 1079.9 & -57.86 \\
\hline
\end{tabular}
\end{center}
\end{scriptsize}
\end{table*}

\begin{table*}[!ht]
\caption{The time-resolved spectral analysis of GRB 160325A was conducted using physical model, namely the \sw{Synchrotron}. The flux values (in erg $\rm cm^{-2}$ $\rm s^{-1}$) reported in this study were calculated within the energy range of 8 \keV to 40 MeV.}
\label{Tab:syn}
\setlength{\tabcolsep}{10pt}
\flushleft
\begin{scriptsize}
\setlength\tabcolsep{12.0pt} 
\begin{center}
\begin{tabular}{|c|c|cccc|c|c|}\hline
T$_{\rm start}$ & T$_{\rm stop}$ & \multicolumn{4}{c|}{\bf Synchrotron}& &\\ [1ex]
\bf (s) & \bf (s)  & \bf B (G) & \boldmath $p$ &\boldmath $\gamma_{cool}$ $\times$ $10^4$ ($\mathbf{\keV}$) &\bf Flux ($\times 10^{-06}$) &\bf DIC & \bf $\Delta$DIC \\ [1ex] \hline
1.605 & 4.684 & $2.24_{-0.33}^{0.33}$ & $4.90_{-0.74}^{0.72}$ & $278.36_{-211.47}^{295.90}$ & 1.18 & 2017.57& -82.67\\
4.817 & 6.911 & $1.64_{-0.28}^{0.28}$ & $4.81_{-0.77}^{0.78}$ & $295.88_{-228.85}^{301.89}$ & 0.94 & 1623.3& -307.55\\
6.911 & 10.594 & $1.57_{-0.28}^{0.29}$ & $3.69_{-0.52}^{0.53}$ & $254.06_{-193.23}^{248.57}$ & 2.07 & 2275.61& -15.04\\
10.594 & 14.505 & $2.28_{-0.29}^{0.30}$ & $5.09_{-0.62}^{0.60}$ & $247.48_{-186.56}^{221.99}$ & 1.4 & 2356.47& -74.91\\
44.13 & 45.471 & $1.60_{-0.24}^{0.25}$ & $5.01_{-0.71}^{0.69}$ & $269.59_{-204.95}^{259.69}$ & 1.18 & 1137.83& -365.31\\
\hline
\end{tabular}
\end{center}
\end{scriptsize}
\end{table*}

\begin{table*}[!ht]
\caption{Similar to \ref{TRS:cpl} but for GRB 160802A.}
\label{TRS:cpl_GRB160802A}
\setlength{\tabcolsep}{4pt}
\flushleft
\begin{scriptsize}
\setlength\tabcolsep{4.0pt} 
\begin{center}
\begin{tabular}{|c|c|ccc|cccc|ccc|} \hline
 & &\multicolumn{3}{c|}{\bf CPL}&\multicolumn{4}{c|}{\bf CPL+BB}& {\bf CPL} & {\bf CPL+BB} &\\
T$_{\rm start}$ & T$_{\rm stop}$ &\boldmath $\Gamma$ &\boldmath E$_{\bf c}$  & \bf Flux $\times 10^{-06}$  & {\boldmath $\Gamma$} & {\boldmath E$_{\bf c}$} &\bf kT & \bf Flux $\times 10^{-06}$&\bf DIC & \bf DIC & \bf $\Delta$DIC\\
\bf (s)&  \bf (s) & & (\boldmath$\keV$)& &  & (\boldmath$\keV$) & (\boldmath$\keV$) & & & &  \\[1ex] \hline
0.238 & 0.962 & $-0.39_{-0.03}^{0.03}$ & $319.54_{-15.87}^{15.39}$ & 21.5 & $-0.38_{-0.04}^{0.04}$ & $324.64_{-18.78}^{19.40}$ & $18.25_{-12.54}^{12.56}$ & 21.4 & 1542.86 & 557.25 & -985.61 \\
0.962 & 1.171 & $-0.23_{-0.05}^{0.05}$ & $221.61_{-14.64}^{14.89}$ & 26.0 & $-0.21_{-0.05}^{0.05}$ & $227.45_{-17.96}^{18.46}$ & $18.65_{-12.43}^{12.27}$ & 25.7 & 286.19 & 237.36 & -48.83 \\
1.171 & 1.695 & $-0.30_{-0.04}^{0.03}$ & $215.94_{-9.73}^{9.94}$ & 20.8 & $-0.28_{-0.06}^{0.06}$ & $244.00_{-17.81}^{17.99}$ & $28.98_{-3.26}^{3.40}$ & 20.1 & 1211.21 & 1180.81 & -30.4 \\
1.695 & 2.16 & $-0.29_{-0.05}^{0.05}$ & $158.19_{-9.11}^{9.01}$ & 12.2 & $-0.30_{-0.05}^{0.06}$ & $163.54_{-13.34}^{10.40}$ & $15.28_{-10.63}^{12.61}$ & 12.3 & 958.5 & 929.28 & -29.22 \\
2.16 & 2.504 & $-0.35_{-0.06}^{0.06}$ & $147.31_{-10.87}^{11.13}$ & 8.66 & $-0.34_{-0.07}^{0.07}$ & $148.92_{-12.46}^{12.65}$ & $13.62_{-8.88}^{8.08}$ & 8.6 & 620.06 & 414.34 & -205.72 \\
\hline
\end{tabular}
\end{center}
\end{scriptsize}
\end{table*}

\begin{table*}
\caption{Similar to \ref{TRS:band} but for GRB 160802A.}
\label{TRS:band_GRB160802A}
\setlength{\tabcolsep}{2pt}
\flushleft
\begin{scriptsize}
\setlength\tabcolsep{2.0pt} 
\begin{center}
\begin{tabular}{|c|c|cccc|ccccc|ccc|}\hline
T$_{\rm start}$ & T$_{\rm stop}$ & \boldmath $\it \alpha_{\rm pt}$ & \boldmath $\it \beta_{\rm pt}$ & \boldmath \Ep & \bf Flux &\boldmath $\it \alpha_{\rm pt}$ & \boldmath $\it \beta_{\rm pt}$ & \boldmath \Ep &   kT  & \bf Flux & \bf DIC$_{\rm Band}$ & \bf DIC$_{\rm Band+BB}$ &  \bf $\Delta$DIC\\  \hline
(s) &(s) &  & & \boldmath (\keV) & $\times 10^{-06}$& &  & \boldmath (\keV) &  (\keV)  & $\times 10^{-06}$&  &  &  \\ \hline
0.238 & 0.962 & $-0.35_{-0.04}^{0.03}$ & $-3.05_{-0.22}^{0.22}$ & $482.34_{-18.64}^{19.03}$ & 24.5 & $-0.35_{-0.04}^{0.04}$ & $-3.09_{-0.22}^{0.23}$ & $484.82_{-19.31}^{19.65}$ & $12.79_{-8.44}^{8.40}$ & 24.3 & 1547.48 & 1520.39 & -27.09 \\
0.962 & 1.171 & $-0.06_{-0.07}^{0.07}$ & $-2.53_{-0.14}^{0.14}$ & $322.74_{-19.73}^{19.62}$ & 35.4 & $-0.05_{-0.08}^{0.07}$ & $-2.54_{-0.14}^{0.14}$ & $324.63_{-18.48}^{18.45}$ & $12.45_{-7.72}^{7.42}$ & 35.4 & 271.48 & 258.1 & -13.38 \\
1.171 & 1.695 & $-0.24_{-0.04}^{0.04}$ & $-2.98_{-0.21}^{0.21}$ & $337.61_{-13.44}^{13.48}$ & 24.1 & $-0.22_{-0.05}^{0.06}$ & $-3.14_{-0.24}^{0.24}$ & $367.64_{-26.02}^{25.85}$ & $22.64_{-6.57}^{6.00}$ & 23.0 & 1210.93 & 1194.12 & -16.81 \\
1.695 & 2.16 & $-0.20_{-0.06}^{0.06}$ & $-2.93_{-0.21}^{0.21}$ & $244.21_{-10.98}^{10.86}$ & 14.3 & $-0.19_{-0.06}^{0.06}$ & $-2.94_{-0.21}^{0.21}$ & $244.74_{-11.06}^{10.94}$ & $11.44_{-7.51}^{7.86}$ & 14.3 & 957.81 & 934.19 & -23.62 \\
2.16 & 2.504 & $-0.26_{-0.08}^{0.08}$ & $-2.84_{-0.24}^{0.24}$ & $218.85_{-13.02}^{13.06}$ & 10.5 & $-0.26_{-0.08}^{0.08}$ & $-2.87_{-0.25}^{0.25}$ & $221.21_{-13.11}^{12.79}$ & $12.02_{-7.44}^{7.25}$ & 10.5 & 621.01 & 609.89 & -11.12 \\
2.504 & 3.065 & $-0.41_{-0.08}^{0.08}$ & $-2.81_{-0.23}^{0.23}$ & $188.82_{-12.52}^{12.36}$ & 6.77 & $-0.42_{-0.08}^{0.08}$ & $-2.85_{-0.25}^{0.25}$ & $191.20_{-13.23}^{12.56}$ & $12.08_{-8.24}^{8.42}$ & 6.73 & 1104.57 & 1072.86 & -31.71 \\
\hline
\end{tabular}
\end{center}
\end{scriptsize}
\end{table*}

\begin{table*}
\caption{Similar to \ref{Tab:syn} but for GRB 160802A.}
\label{Tab:syn_160802A}
\setlength{\tabcolsep}{5pt}
\flushleft
\begin{scriptsize}
\setlength\tabcolsep{5.0pt} 
\begin{center}
\begin{tabular}{|c|c|cccc|c|c|}\hline
T$_{\rm start}$ & T$_{\rm stop}$ & \multicolumn{4}{c|}{\bf Synchrotron}& & \\ [1ex]
\bf (s) & \bf (s)  & \bf B (G) & \boldmath $p$ &\boldmath $\gamma_{cool}$ $\times$ $10^4$ ($\mathbf{\keV}$) &\bf Flux ($\times 10^{-06}$) &\bf DIC & \bf $\Delta$DIC \\ [1ex] \hline
0.238 & 0.962 & $4.18_{-0.25}^{0.26}$ & $5.83_{-0.15}^{0.13}$ & $1194.47_{-1060.91}^{1399.32}$ & 28.9 & 567.17&  953.22\\
0.962 & 1.171 & $3.44_{-0.25}^{0.26}$ & $5.67_{-0.28}^{0.26}$ & $1144.14_{-1023.01}^{733.54}$ & 36.4 & 172.61& 85.49\\
1.171 & 1.695 & $2.97_{-0.15}^{0.15}$ & $5.86_{-0.11}^{0.11}$ & $678.58_{-533.58}^{158.49}$ & 27.3 & 1184.84& 9.28\\
1.695 & 2.16 & $2.06_{-0.12}^{0.12}$ & $5.81_{-0.16}^{0.15}$ & $560.14_{-424.78}^{342.60}$ & 15.6 & 1018.2& -84.01\\
2.16 & 2.504 & $1.73_{-0.14}^{0.14}$ & $5.64_{-0.29}^{0.28}$ & $1620.00_{-1505.19}^{964.20}$ & 10.9 & 314.9& 294.99\\
2.504 & 3.065 & $1.36_{-0.12}^{0.12}$ & $5.64_{-0.31}^{0.29}$ & $869.15_{-781.19}^{348.88}$ & 6.7 & 989.67& 83.19\\
\hline
\end{tabular}
\end{center}
\end{scriptsize}
\end{table*}

\begin{table*}
\caption{Similar to \ref{TRS:cpl} but for GRB 160821A.}
\label{TRS:cpl_GRB160821A}
\setlength{\tabcolsep}{2pt}
\flushleft
\begin{scriptsize}
\setlength\tabcolsep{1.0pt} 
\begin{center}
\begin{tabular}{|c|c|ccc|cccc|ccc|} \hline
 & &\multicolumn{3}{c|}{\bf CPL}&\multicolumn{4}{c|}{\bf CPL+BB}& {\bf CPL} & {\bf CPL+BB} &\\
T$_{\rm start}$ & T$_{\rm stop}$ &\boldmath $\Gamma$ &\boldmath E$_{\bf c}$  & \bf Flux $\times 10^{-06}$  & {\boldmath $\Gamma$} & {\boldmath E$_{\bf c}$} &\bf kT & \bf Flux $\times 10^{-06}$&\bf DIC & \bf DIC & \bf $\Delta$DIC\\
\bf (s)&  \bf (s) & & (\boldmath$\keV$)& &  & (\boldmath$\keV$) & (\boldmath$\keV$) & & & &  \\[1ex] \hline
113.502 & 116.61 & $-1.08_{-0.04}^{0.04}$ & $413.39_{-49.64}^{48.83}$ & 1.42 & $-1.13_{-0.04}^{0.03}$ & $519.49_{-36.26}^{104.58}$ & $12.57_{-8.26}^{6.89}$ & 1.63 & 2985.64 & 2699.94 & -285.7 \\
116.61 & 117.57 & $-1.08_{-0.03}^{0.03}$ & $565.55_{-59.82}^{58.97}$ & 2.98 & $-1.13_{0.02}^{0.23}$ & $733.21_{-158.28}^{-2.58}$ & $13.48_{-7.60}^{-4.28}$ & 3.42 & 1820.7 & 1741.92 & -78.78 \\
117.57 & 118.04 & $-1.04_{-0.04}^{0.04}$ & $556.07_{-61.96}^{63.98}$ & 3.97 & $-1.07_{0.02}^{0.19}$ & $769.92_{-117.99}^{34.55}$ & $2.46_{1.26}^{5.21}$ & 5.3 & 1097.2 & 941.09 & -156.11 \\
118.04 & 121.932 & $-0.99_{-0.01}^{0.01}$ & $1013.94_{-9.80}^{7.01}$ & 9.47 & $-0.87_{0.01}^{0.02}$ & $1103.65_{-65.64}^{0.30}$ & $11.02_{-0.30}^{0.50}$ & 10.9 & 3555.41 & 3504.48 & -50.93 \\
121.932 & 124.931 & $-0.92_{-0.01}^{0.01}$ & $996.76_{-23.64}^{24.19}$ & 13.9 & $-0.93_{-0.00}^{0.02}$ & $1108.74_{-79.31}^{-4.65}$ & $4.79_{0.40}^{18.21}$ & 14.6 & 3297.32 & 3265.54 & -31.78 \\
\hline
\end{tabular}
\end{center}
\end{scriptsize}
\end{table*}

\begin{table*}
\caption{Similar to \ref{TRS:band} but for GRB 160821A.}
\label{TRS:band_GRB160821A}
\setlength{\tabcolsep}{2pt}
\flushleft
\begin{scriptsize}
\setlength\tabcolsep{1.0pt} 
\begin{center}
\begin{tabular}{|c|c|cccc|ccccc|ccc|}\hline
T$_{\rm start}$ & T$_{\rm stop}$ & \boldmath $\it \alpha_{\rm pt}$ & \boldmath $\it \beta_{\rm pt}$ & \boldmath \Ep & \bf Flux &\boldmath $\it \alpha_{\rm pt}$ & \boldmath $\it \beta_{\rm pt}$ & \boldmath \Ep &   kT  & \bf Flux & \bf DIC$_{\rm Band}$ & \bf DIC$_{\rm Band+BB}$ &  \bf $\Delta$DIC\\  \hline
(s) &(s) &  & & \boldmath (\keV) & $\times 10^{-06}$& &  & \boldmath (\keV) &  (\keV)  & $\times 10^{-06}$&  &  &  \\ \hline
113.502 & 116.61 & $-1.09_{-0.05}^{0.05}$ & $-2.47_{-0.34}^{0.33}$ & $363.47_{-50.12}^{50.31}$ & 2.01 & $-1.09_{-0.05}^{0.05}$ & $-2.53_{-0.34}^{0.34}$ & $369.79_{-48.55}^{48.92}$ & $10.53_{-7.03}^{7.42}$ & 1.93 & 2982.03 & 2891.22 & -90.81 \\
116.61 & 117.57 & $-1.12_{-0.04}^{0.04}$ & $-2.63_{-0.32}^{0.32}$ & $534.22_{-58.89}^{58.48}$ & 3.81 & $-1.03_{-0.11}^{0.14}$ & $-2.62_{-0.33}^{0.33}$ & $507.26_{-59.38}^{60.33}$ & $8.80_{-3.74}^{5.17}$ & 3.82 & 1814.18 & 1765.9 & -48.28 \\
117.57 & 118.04 & $-1.07_{-0.05}^{0.05}$ & $-2.31_{-0.29}^{0.28}$ & $498.24_{-70.21}^{70.42}$ & 5.97 & $-1.01_{-0.09}^{0.11}$ & $-2.36_{-0.32}^{0.31}$ & $493.90_{-73.68}^{75.00}$ & $8.62_{-4.94}^{6.31}$ & 5.87 & 1090.81 & 783.94 & -306.87 \\
118.04 & 121.932 & $-0.97_{-0.01}^{0.01}$ & $-1.93_{-0.03}^{0.03}$ & $758.51_{-19.71}^{19.61}$ & 17.8 & $-0.82_{-0.03}^{0.03}$ & $-1.92_{-0.02}^{0.03}$ & $701.64_{-26.73}^{24.98}$ & $8.85_{-0.60}^{0.55}$ & 18.3 & 3490.81 & 3470.81 & -20.0 \\
121.932 & 124.931 & $-0.85_{-0.01}^{0.01}$ & $-2.08_{-0.04}^{0.03}$ & $736.18_{-16.51}^{17.92}$ & 21.1 & $-0.84_{-0.03}^{0.03}$ & $-2.14_{-0.05}^{0.05}$ & $807.42_{-30.73}^{29.42}$ & $7.88_{-2.53}^{1.86}$ & 20.8 & 3270.87 & 3242.14 & -28.73 \\
124.931 & 126.547 & $-0.81_{-0.01}^{0.01}$ & $-2.36_{-0.03}^{0.02}$ & $632.87_{-12.32}^{12.04}$ & 18.4 & $-0.87_{-0.01}^{0.01}$ & $-2.32_{-0.08}^{0.08}$ & $717.15_{-22.08}^{22.88}$ & $9.52_{-4.98}^{4.13}$ & 19.9 & 2634.44 & 2603.95 & -30.49 \\
\hline
\end{tabular}
\end{center}
\end{scriptsize}
\end{table*}

\begin{table*}[!ht]
\caption{Similar to \ref{Tab:syn} but for GRB 160821A.}
\label{Tab:syn_160821A}
\setlength{\tabcolsep}{8pt}
\flushleft
\begin{scriptsize}
\setlength\tabcolsep{8.0pt} 
\begin{center}
\begin{tabular}{|c|c|cccc|c|c|}\hline
T$_{\rm start}$ & T$_{\rm stop}$ & \multicolumn{4}{c|}{\bf Synchrotron}& &\\ [1ex]
\bf (s) & \bf (s)  & \bf B (G) & \boldmath $p$ &\boldmath $\gamma_{cool}$ $\times$ $10^4$ ($\mathbf{\keV}$) &\bf Flux ($\times 10^{-06}$) &\bf DIC & \bf $\Delta$DIC \\ [1ex] \hline

113.502 & 116.61 & $0.43_{0.08}^{0.08}$ & $2.02_{0.29}^{0.29}$ & $20.78_{9.11}^{181.07}$ & 3.37 & 2945.62 & -54.40 \\
116.61 & 117.57 & $0.20_{0.04}^{0.04}$ & $2.00_{0.08}^{0.08}$ & $86.02_{3.74}^{116.96}$ & 8.93 & 1797.94 & -32.04 \\
117.57 & 118.04 & $0.24_{0.05}^{0.05}$ & $2.01_{0.06}^{0.06}$ & $127.32_{-5.85}^{247.21}$ & 15.1 & 1103.45 & -319.51 \\
118.04 & 121.932 & $0.60_{0.06}^{0.06}$ & $2.00_{0.03}^{0.03}$ & $56.55_{-6.05}^{12.66}$ & 21.0 & 3549.22 & -78.41 \\
121.932 & 124.931 & $2.06_{0.20}^{0.20}$ & $2.37_{0.10}^{0.10}$ & $20.02_{0.44}^{6.71}$ & 23.2 & 3238.28 & 3.86 \\
124.931 & 126.547 & $2.21_{0.23}^{0.23}$ & $2.89_{0.18}^{0.18}$ & $20.25_{0.27}^{6.58}$ & 20.9 & 2609.42 & -5.47 \\
\hline
\end{tabular}
\end{center}
\end{scriptsize}
\end{table*}

\begin{table*}
\caption{The Pearson correlation (r) between parameters obtained using time-resolved spectral analysis of GRB 160325A, GRB 160802A, and GRB 160821A, respectively.}
\label{tab:correlation}
\setlength{\tabcolsep}{5pt}
\flushleft
\begin{scriptsize}
\setlength\tabcolsep{5.0pt} 
\begin{center}
\begin{tabular}{|c|c|c|c|c|c|c|c|c|} 
\hline
   \multicolumn{9}{|c|}{\textbf{GRB 160325A}} \\
\hline\hline

\textbf{Model} & \multicolumn{2}{c|}{\textbf{log (Flux)-log (\Ep)}} & \multicolumn{2}{c|}{\textbf{log (Flux)-$\alpha_{\rm pt}$}} & \multicolumn{2}{c|}{\textbf{log (\Ep)-$\alpha_{\rm pt}$}}& \multicolumn{2}{c|}{\textbf{log (Flux)-k$T$}}  \\ \hline
 & \bf r & \bf Probability of null hypothesis (p) &  \bf r & \bf p &  \bf r & \bf p & \bf r & \bf p \\ \hline
\sw{Band}& 0.70 & 0.29& -0.90 & 0.10& -0.72 & 0.28& -&- \\
\sw{Band+BB}& 0.75 & 0.25& -0.93 & 0.07 & -0.56 & 0.44 & -0.31 & 0.68\\ \hline
& \multicolumn{3}{c|}{\textbf{log (Flux)-log (B)}} & \multicolumn{3}{c|}{\textbf{log (Flux)-$p$}} & \multicolumn{2}{c|}{\textbf{log (B)-$p$}} \\ \hline
 & \bf r & \multicolumn{2}{c|}{\textbf{p}} &  \bf r & \multicolumn{2}{c|}{\textbf{p}} & \bf r & \bf p  \\ \hline
\sw{Synchrotron}& -0.22 & \multicolumn{2}{c|}{0.78} & -0.77 & \multicolumn{2}{c|}{0.23} & 0.74 & 0.26  \\
\hline
\end{tabular}
\end{center}
\end{scriptsize}
\end{table*}

\end{document}